\documentclass[acmsmall]{acmart}

\usepackage[utf8]{inputenc} % allow utf-8 input
\usepackage[T1]{fontenc}    % use 8-bit T1 fonts
\usepackage{hyperref}       % hyperlinks
\usepackage{url}            % simple URL typesetting
\usepackage{booktabs}       % professional-quality tables
\usepackage{amsfonts}       % blackboard math symbols
\usepackage{nicefrac}       % compact symbols for 1/2, etc.
\usepackage{microtype}      % microtypography
\usepackage{xcolor}         % colors
\usepackage{subfigure}
\usepackage{xspace}
\usepackage{multirow}
\usepackage{makecell}
\usepackage{xcolor}
\usepackage{amsmath}
\usepackage{amsfonts}
\usepackage{bm}
\usepackage{soul}
\usepackage{makecell}
\usepackage[ruled]{algorithm2e}
\usepackage{balance}
\usepackage{tcolorbox}
\usepackage[a-2b]{pdfx}
\usepackage{enumitem}

\usepackage{graphicx}
\usepackage{tabularx} % for 'tabularx' env. and 'X' col. type
\usepackage{ragged2e} % for \RaggedRight macro

\usepackage{microtype}
\usepackage{pifont}
\usepackage{transparent}
\usepackage{threeparttable}
\usepackage{longtable}
\usepackage{threeparttablex}
\usepackage{adjustbox}
\usepackage{tikz}
\usepackage{pgfplots}
\pgfplotsset{compat=1.18}
\usepackage[edges]{forest}

\AtBeginDocument{%
  }

\setcopyright{acmlicensed}
\copyrightyear{2018}
\acmYear{2018}
\acmDOI{XXXXXXX.XXXXXXX}

\acmJournal{JACM}
\acmVolume{37}
\acmNumber{4}
\acmArticle{111}
\acmMonth{8}

\usepackage{xcolor} % 可选：用于高亮

\newcommand{\cmark}{\ding{51}} % check mark
\newcommand{\xmark}{\ding{55}} % cross mark

\newcommand{\eg}{\emph{e.g.},\xspace}
\newcommand{\ie}{\emph{i.e.},\xspace}

\newcommand{\eat}[1]{}

\begin{document}

\title{Large Language Model Powered Intelligent Urban Agents: Concepts, Capabilities, and Applications}

\author{Jindong Han}
\affiliation{%
  \institution{Shandong University}
  \city{Jinan}
  \country{China}}
\email{jindong.han@sdu.edu.cn}

\author{Yansong Ning, Zirui Yuan, Hang Ni, Fan Liu, Tengfei Lyu}
\affiliation{%
  \institution{The Hong Kong University of Science and Technology (Guangzhou)}
  \city{Guangzhou}
  \country{China}}
\email{
{yning092, zyuan779, hni017, fliu236, tlyu077}@connect.hkust-gz.edu.cn}

% \author{Zirui Yuan}
% \affiliation{%
%   \institution{The Hong Kong University of Science and Technology (Guangzhou)}
%   \city{Guangzhou}
% \email{zyuan779@connect.hkust-gz.edu.cn}

% \author{Hang Ni}
% \affiliation{%
%   \institution{The Hong Kong University of Science and Technology (Guangzhou)}
%   \city{Guangzhou}
% \email{hni017@connect.hkustgz.edu.cn}

% \author{Fan Liu}
% \affiliation{%
%   \institution{The Hong Kong University of Science and Technology (Guangzhou)}
%   \city{Guangzhou}
% \email{fliu236@connect.hkust-gz.edu.cn}

% \author{Tengfei Lyu}
% \affiliation{%
%   \institution{The Hong Kong University of Science and Technology (Guangzhou)}
%   \city{Guangzhou}
% \email{tlyu077@connect.hkust-gz.edu.cn}

\author{Hao Liu}
\affiliation{%
  \institution{The Hong Kong University of Science and Technology (Guangzhou), The Hong Kong University of Science and Technology}
  \city{Guangzhou}
  \country{China}}
\email{liuh@ust.hk}

\renewcommand{\shortauthors}{Trovato et al.}

\begin{abstract}
The long-standing vision of intelligent cities is to create efficient, livable, and sustainable urban environments using big data and artificial intelligence technologies. Recently, the advent of Large Language Models (LLMs) has opened new ways toward realizing this vision. With powerful semantic understanding and reasoning capabilities, LLMs can be deployed as intelligent agents capable of autonomously solving complex problems across domains. In this article, we focus on \emph{Urban LLM Agents}, which are LLM-powered agents that are semi-embodied within the hybrid cyber-physical-social space of cities and used for system-level urban decision-making. First, we introduce the concept of urban LLM agents, discussing their unique capabilities and features. Second, we survey the current research landscape from the perspective of agent workflows, encompassing urban sensing, memory management, reasoning, execution, and learning. Third, we categorize the application domains of urban LLM agents into five groups: urban planning, transportation, environment, public safety, and urban society, presenting representative works in each group. Finally, we discuss trustworthiness and evaluation issues that are critical for real-world deployment, and identify several open problems for future research. This survey aims to establish a foundation for the emerging field of urban LLM agents and to provide a roadmap for advancing the intersection of LLMs and urban intelligence. A curated list of relevant papers and open-source resources is maintained and continuously updated at https://github.com/usail-hkust/Awesome-Urban-LLM-Agents.
\end{abstract}


\keywords{Do, Not, Us, This, Code, Put, the, Correct, Terms, for,
  Your, Paper}

\received{20 February 2007}
\received[revised]{12 March 2009}
\received[accepted]{5 June 2009}

\maketitle

\section{Introduction}
With rapid urbanization, modern cities are facing growing challenges, such as traffic congestion, environmental degradation, and energy sustainability. Effectively tackling these issues has become a pressing global priority. In recent years, breakthroughs in Machine Learning (ML) have driven progress toward the vision of the \emph{intelligent city}~\cite{zheng2014urban,bibri2017smart}, which seeks to address these urban challenges through data-driven analytics~\cite{jin2023spatio} and decision-making~\cite{ullah2020applications} with minimal human assistance. By integrating advanced ML technologies into urban systems, intelligent city services allow stakeholders to efficiently analyze massive and heterogeneous urban data, empowering applications like traffic optimization~\cite{zhang2022multi}, energy management~\cite{zhan2022deepthermal}, and policy formulation~\cite{rolnick2022tackling} at various spatial and temporal scales. This data-driven computing paradigm paves the way for more efficient and sustainable urban development.

However, existing intelligent urban systems are still far from ideal due to limitations in flexibility and scalability. Most ML-based approaches are restricted to performing tasks within predefined domains~\cite{goodge2025spatio}, such as those related to specific regions, modalities, and spatio-temporal granularities. When applied to previously unseen urban contexts (\eg new cities or emerging services), these models struggle to provide reliable analysis and decision-making results due to significant disparities in data distribution~\cite{jin2023transferable,sun2024crosslight}. As a result, there is a pressing need to enhance the models' generalization capabilities to support a broader and more flexible range of urban tasks. Moreover, these systems often lack the ability for natural language interaction, reasoning, and autonomous task execution. These capabilities are critical to addressing the diverse demands of urban decision-making and achieving higher levels of intelligence.

The recent revolution in Large Language Models (LLMs), such as GPT-4~\cite{achiam2023gpt} and DeepSeek-R1~\cite{guo2025deepseek}, offers new opportunities to rethink the development of urban intelligence. LLMs are pre-trained on massive open-domain corpora and subsequently refined via post-training techniques~\cite{kumar2025llm}, endowing them with exceptional capabilities for natural language understanding, instruction following, zero-shot generalization, and multimodal integration. More importantly, when provided with sufficient context, LLMs can emulate a human-like reasoning process and solve complex tasks by making executable plans and invoking external tools such as search engines and third-party APIs. As a result, LLMs are increasingly deployed as autonomous agents that can respond and adapt to rapidly changing environments~\cite{wang2024survey, xi2025rise}.

\begin{figure}[t]
    \centering
    \includegraphics[width=1\linewidth]{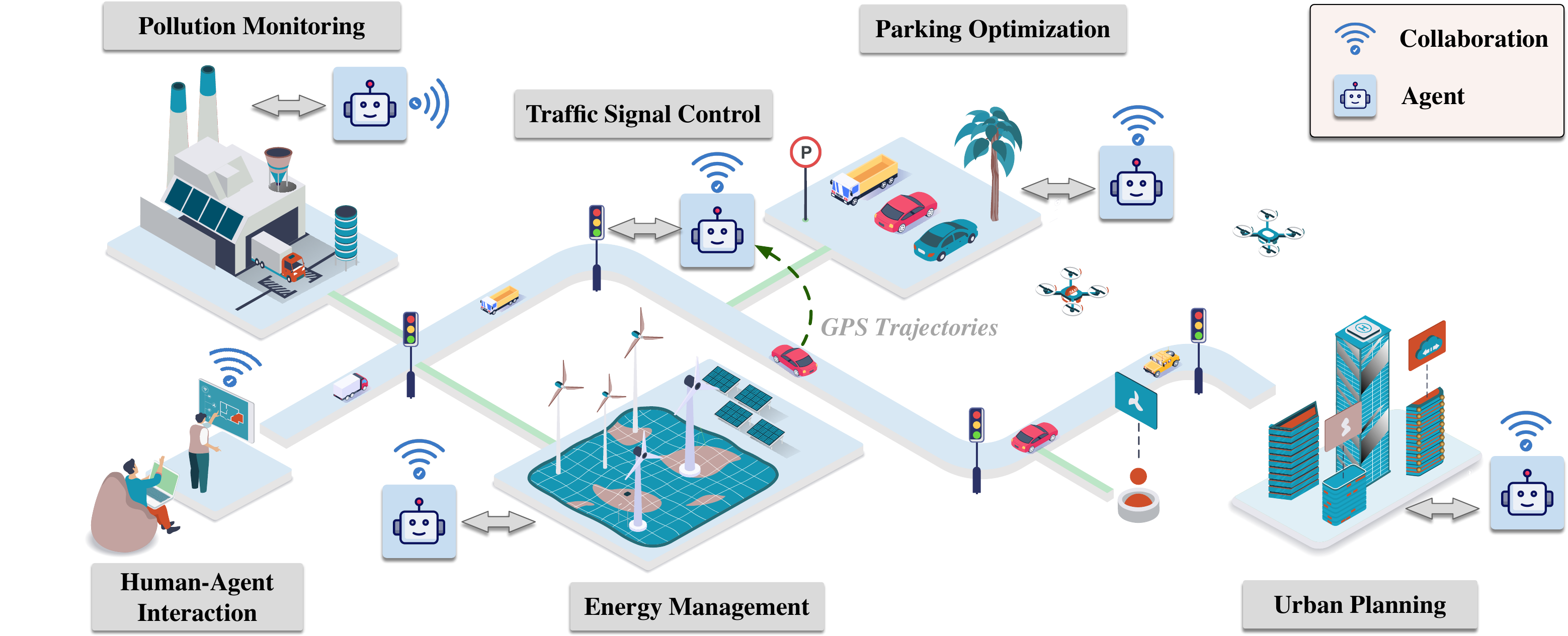}
    \caption{An illustrative example of an intelligent city powered by urban LLM agents. The agents collaborate across various domains. Each agent operates autonomously while maintaining communication with others to coordinate decisions. The system also supports interactions with humans, \ie city officials and citizens.}
    \label{fig:introduction}
    \vspace{-0.6cm}
\end{figure}

Building on the capabilities of LLMs, we envision a new class of intelligent systems for urban operations, referred to as \emph{Urban LLM Agents}. Compared to generic LLM agents, urban LLM agents are deeply integrated with both the virtual and physical infrastructures of cities. They are not physically embodied like robots but semi-embodied (\ie virtually embodied) and interface with urban systems through APIs, databases, and interactive platforms. As illustrated in Figure \ref{fig:introduction}, this involves integrating data from geo-distributed sensors/devices, retrieving regulatory documents, reasoning over space and time, participating in collective urban decision-making workflows, and interacting with humans via interpretable language. Rather than functioning merely as task-specific solvers, these agents serve as \emph{cognitive intermediaries} between city stakeholders and the vast and heterogeneous urban ecosystem that governs human life. In doing so, they can help cities become more efficient, resilient, and responsive. Therefore, we anticipate that urban LLM agents will emerge as a foundational paradigm for intelligent cities in the era of AI.

Although urban LLM agents hold great promise, the field is still in its early stages, facing significant complexity and challenges. 
In light of this, this paper takes a first step to coin urban LLM agents, emphasizing their distinctive capabilities related to \emph{spatio-temporal} aspects. We then present a systematic survey from both agent and application perspectives. From the agent's perspective, we analyze the design principles required for grounding LLM agents in urban scenarios. Specifically, we discuss five key components: urban sensing, memory management, reasoning, execution, and learning. From an application perspective, we categorize the use cases of urban LLM agents into five groups, presenting representative examples in each group. Finally, we explore issues of trustworthiness and evaluation within urban LLM agents and identify open research problems to stimulate further investigation in this burgeoning field.
The major contributions are as follows:
\begin{itemize}
    \item \textbf{Conceptual framework}: We introduce and formalize the concept of urban LLM agents, discussing their core capabilities and unique features that differentiate them from existing agent paradigms. This establishes the scope and roadmap for urban LLM agent research, helping the community understand and engage with this emerging field.
    \item \textbf{Systematic and up-to-date survey}: We provide a survey of the current research landscape on urban LLM agents from two complementary perspectives. Specifically, we categorize existing literature based on agent workflows and target scenarios, presenting the relationships, strengths, and limitations across different subcategories.
    \item \textbf{Trustworthiness, evaluation, and research outlook}: We discuss key issues related to trustworthiness and evaluation in urban LLM agents. Additionally, we give a research outlook on the future of urban LLM agents, identifying several open problems that deserve further investigation.
\end{itemize}

\textbf{Paper organization}: The rest of this paper is organized as follows. In Section 2, we introduce the essential background and basic concepts of urban LLM agents. Section 3 reviews related work from the agent perspective, while Section 4 discusses representative applications. Section 5 delves into the trustworthiness concerns associated with urban LLM agents. Section 6 elaborates on the evaluation protocol. Section 7 discusses potential research problems and provides insights for future exploration. Finally, we conclude this paper in Section 8.

\section{Background and Preliminaries}
\subsection{A Brief History of Urban Agents}
\begin{figure}[t]
    \centering
    \includegraphics[width=1\linewidth]{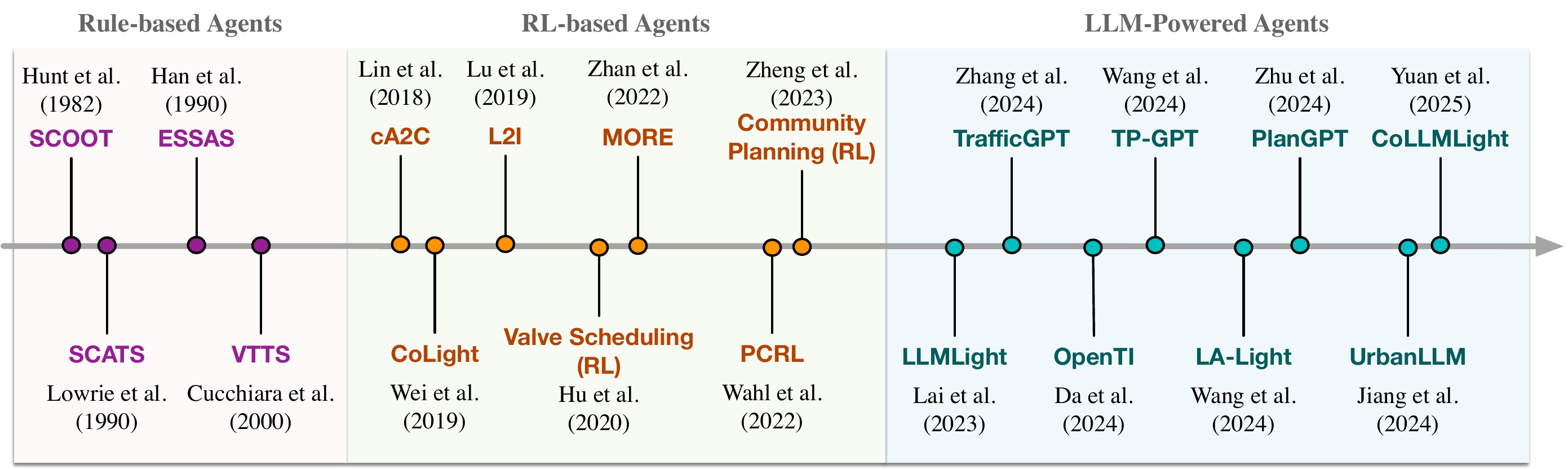}
    \caption{Major milestones in the history of urban agents.}
    \label{fig:urban agent history}
    \vspace{-0.4cm}
\end{figure}

This subsection outlines the evolution of intelligent agents in urban environments, from early rule-based systems to reinforcement learning (RL)-based agents, and most recently to LLM-powered agents. An overview of key milestones in this progression is illustrated in Figure~\ref{fig:urban agent history}.

\subsubsection{Rule-Based Systems}
Rule-based urban systems emerged in the early 1980s, supporting decision-making through predefined rules and incorporating various statistical tools and algorithms. Due to their transparency, such systems were widely adopted in urban planning and traffic management. 
In urban planning, a representative example is ESSAS \cite{han1990essas}, an expert system for site analysis and selection, developed to assist planners in evaluating land parcel suitability by integrating spatial data with planning regulations. 
In the transportation domain, rule-based adaptive traffic signal systems such as SCOOT \cite{hunt1982scoot} and SCATS \cite{lowrie1990scats} were designed to adjust signal timings in real time based on preconfigured logic tied to sensor inputs, and have been widely implemented in cities worldwide. Moreover, Cucchiara et al. \cite{cucchiara2000image} proposed VTTS, a vision-based traffic monitoring system that assesses traffic conditions through rule-based reasoning.
These systems represent early efforts to automate decision-making in urban environments. However, their reliance on static rules, limited scalability, and inability to learn from data make them insufficient for addressing the dynamic, large-scale, and multi-objective challenges of modern urban systems.

\subsubsection{Reinforcement Learning Agents}
Reinforcement learning (RL) is a paradigm for sequential decision-making, where agents learn to optimize long-term cumulative rewards through interactions with dynamic and uncertain environments. Since the emergence of deep RL methods around 2015, RL-based agents have gained significant traction across various domains, including urban computing.
A representative application is traffic signal control, which is typically formulated as a Markov decision process \cite{wiering2000multi, wiering2004intelligent, salkham2008collaborative, wei2019colight}. In this setting, RL agents learn policies that adapt to evolving traffic conditions with the objective of minimizing vehicle delays and queue lengths. Wiering \cite{wiering2000multi} proposed an early multi-agent RL framework that incorporated information from neighboring intersections to estimate cumulative waiting times. Building on this, CoLight \cite{wei2019colight} introduced graph attention networks to enable dynamic communication between intersections, further improving coordination and responsiveness.
Beyond traffic control, RL has also been applied to urban mobility. Lin et al. \cite{lin2018efficient} proposed a contextual multi-agent RL system for fleet management that incorporates both geographic and collaborative contexts. In vehicle routing, Lu et al. \cite{lu2019learning} introduced the L2I framework, where an RL agent selects among multiple local search operators to iteratively improve routing plans, aiming to minimize total travel distance.
RL agents have also been employed in environmental and energy management. For instance, Hu et al. \cite{hu2020deep} developed a DQN-based valve scheduling system for pollution isolation in water distribution networks. The agent uses sensor data as state inputs and schedules valve and hydrant actions to minimize contaminant spread and residual concentration. Zhan et al. \cite{zhan2022deepthermal} proposed MORE, an RL-based controller for thermal power generation units that maximizes combustion efficiency while reducing pollutant emissions. In urban planning, RL has been used to optimize the spatial layout of infrastructure. Wahl et al. \cite{von2022reinforcement} proposed PCRL for charging station placement, enabling agents to observe factors such as demand distribution, existing infrastructure, and land cost, and make decisions about locating and resizing facilities. Moreover, Zheng et al. \cite{zheng2023spatial} designed a deep RL-based agent for community layout planning. The agent selects nodes and edges in a contiguity graph to incrementally place functional components and roads, optimizing spatial efficiency metrics such as service accessibility, ecological quality, and traffic flow.

Despite these advancements, RL agents often rely heavily on task-specific training environments, struggle to generalize across diverse scenarios, lack interpretability, and show limited robustness to unexpected events. In contrast, LLM agents—with their strong generalization, multimodal understanding, tool-use capabilities, and natural language interaction—are increasingly gaining attention as more adaptable and interpretable agents for urban applications.

\subsubsection{LLM-Powered Agents}
Since the release of ChatGPT in 2022, large language models (LLMs) have rapidly become a foundational technology in artificial intelligence. These models demonstrate strong generalization across diverse tasks such as question answering, summarization, code generation, and logical reasoning, without requiring task-specific supervision \cite{kaddour2023challenges}. This versatility has spurred growing interest in deploying LLMs as autonomous agents capable of perceiving dynamic environments, reasoning over multi-step problems, and supporting complex decision-making in data-rich scenarios.

Modern LLMs are built upon the Transformer architecture \cite{vaswani2017attention}, which facilitates efficient modeling of long-range dependencies in text. Their development typically follows a multi-stage pipeline. In the pretraining phase, models are exposed to massive, diverse text corpora and trained using self-supervised objectives like next-token prediction. This equips them with broad linguistic fluency and implicit knowledge about the world \cite{radford2019language, raffel2020exploring}. In the subsequent post-training phase, the pre-trained models are refined to better align with human preferences. Two critical techniques are commonly used: instruction tuning, where the model is fine-tuned to follow natural language instructions using curated prompt-response examples \cite{wei2021finetuned}; and reinforcement learning from human feedback (RLHF), which trains a reward model based on human evaluations to guide the LLM’s outputs toward helpful, safe, and aligned responses \cite{ouyang2022training}.
Following these training phases, the capabilities of LLMs can be further enhanced at inference time. First, LLMs can easily generalize to new tasks simply by conditioning on provided examples in the prompt \cite{wei2021finetuned} through their powerful in-context learning ability. Additionally, their reasoning can be further strengthened using Chain-of-Thought (COT) prompting, which guides models to solve complex tasks in a step-by-step manner \cite{wei2022chain}. To further improve output reliability and problem-solving depth, inference-time computation techniques such as best-of-N sampling, majority voting, and Monte Carlo Tree Search (MCTS) explore diverse reasoning paths and select robust answers \cite{snell2024scaling}. Iterative self-refinement strategies, exemplified by DeepSeek-R1, allow models to critique and revise their outputs, improving reasoning quality \cite{guo2025deepseek}. When domain knowledge or functional capabilities are limited, retrieval-augmented generation (RAG) enables models to incorporate relevant external documents during inference \cite{lewis2020retrieval}. Additionally, access to external tools such as code execution environments and domain-specific APIs can further enhance the output reliability of LLMs \cite{schick2023toolformer}.

Building on the aforementioned technologies, LLMs have emerged as intelligent agents capable of reasoning, planning, memorizing, and interacting with external tools \cite{wooldridge1995intelligent, russell2016artificial, park2023generative}, and have increasingly been applied to urban domains.
For instance, traffic management agents such as TrafficGPT \cite{zhang2024trafficgpt}, OpenTI \cite{da2024open}, and TP-GPT \cite{wang2024traffic} leverage LLMs to retrieve and analyze traffic data, provide interactive decision support, and generate real-time reports. LLMLight \cite{lai2023llmlight}, CoLLMLight \cite{yuan2025collmlight}, and LA-Light \cite{wang2024llm} apply LLMs to traffic signal control, reasoning over real-time observations and selecting signal phases aimed at improving traffic efficiency. In the domain of urban planning and governance, PlanGPT \cite{zhu2024plangpt} integrates local databases and tool-calling capabilities to support policy drafting and zoning evaluation, while UrbanLLM \cite{jiang2024urbanllm} decomposes urban service queries into subtasks and coordinates external AI models to enable autonomous urban activity planning.

\begin{table}[t]
\centering
\small
\newcolumntype{C}{>{\centering\arraybackslash}X}
\caption{Comparison of related surveys with respect to their coverage of agent perspective (Agent), application scenarios (App.), trustworthiness (Trust.), and evaluation (Eval.).}
\begin{tabularx}{\textwidth}{l|l|l|C|C|C|C}
\toprule
\textbf{Article} & \textbf{Domain} & \textbf{Method} & \textbf{Agent} & \textbf{App.} & \textbf{Trust.} & \textbf{Eval.} \\
\midrule
Guo et al.~\cite{guo2024large}        & General & Multi-Agent LLM           & \cmark & \cmark & \xmark & \xmark \\
\midrule
Wang et al.~\cite{wang2024survey}     & General & LLM-based Agents          & \cmark & \cmark & \xmark & \cmark \\
\midrule
Xi et al.~\cite{xi2025rise}           & General & LLM-based Agents          & \cmark & \cmark & \cmark & \cmark \\
\midrule
Li et al.~\cite{li2024personal}       & General & LLM-based Agents          & \cmark & \xmark & \cmark & \cmark \\
\midrule
Zhang et al.~\cite{zhang2024towards}  & Urban   & Foundation Models         & \cmark & \cmark & \cmark & \xmark \\
\midrule
Liang et al.~\cite{liang2025foundation} & Urban   & Foundation Models         & \xmark & \cmark & \xmark & \xmark \\
\midrule
Fang et al.~\cite{fang2025unraveling} & Urban   & Foundation Models         & \xmark & \cmark & \xmark & \xmark \\
\midrule
Li et al.~\cite{li2025urban}          & Urban   & LLMs                      & \cmark & \cmark & \xmark & \cmark \\
\midrule
\textbf{Ours}                         & Urban   & LLM-based Agents          & \cmark & \cmark & \cmark & \cmark \\
\bottomrule
\end{tabularx}

\label{tab:survey-comparison-dimensions}
\vspace{-0.2cm}
\end{table}

\subsection{Related Surveys}
Recent surveys have discussed the rise of LLM-based agents in the general domain. For instance, Wang et al. \cite{wang2024survey} and Xi et al. \cite{xi2025rise} provide comprehensive reviews of general LLM agents, covering their architectures, planning capabilities, and applications, while Guo et al. \cite{guo2024large} focuses specifically on the progress and challenges of LLM-powered multi-agent systems. Meanwhile, Li et al. \cite{li2024personal} discuss the design of lightweight, personalized LLM agents, with emphasis on efficiency, long-term memory, and user-specific customization.

In the urban domain, Zhang et al. \cite{zhang2024towards}, Liang et al. \cite{liang2025foundation}, and Fang et al. \cite{fang2025unraveling} surveyed urban and spatio-temporal foundation models, in which LLMs are regarded as a promising approach for generalizing across heterogeneous spatio-temporal inputs.
Li et al. \cite{li2025urban} further explored the transformative potential of LLMs in urban computing, highlighting applications such as traffic management and spatial decision support.

While these works provide valuable background, our survey is, to the best of our knowledge, the first to focus specifically on LLM-powered agents designed for urban tasks. In contrast to these surveys, we highlight unique spatio-temporal aspects from the agent perspective, including urban sensing, memory management, reasoning, execution, and learning. We then systematically examine how these capabilities are being leveraged to support urban-specific applications such as urban planning, transportation systems, environmental sustainability, public safety, and urban society. Moreover, we also discuss the trustworthiness and evaluation methods of LLM-powered urban agents. Table \ref{tab:survey-comparison-dimensions} provides a comparison with related surveys.

\subsection{Generic LLM Agent Framework}
\subsubsection{Fundamental Components of LLM Agents}
At the core of LLM-powered agents lies a modular architecture designed to emulate goal-oriented behavior \cite{wang2024survey, guo2024large, park2023generative}. While implementations vary, most systems follow a perception--cognition--action loop, enabling agents to perceive input, reason over internal or external context, and perform actions accordingly. This loop is supported by five essential components: perception, memory, planning, action, and learning. Together, these modules allow LLM agents to operate autonomously in dynamic environments, interact with external tools or humans, and evolve over time.

\begin{itemize}
\item \textbf{Perception:} This module is responsible for interpreting inputs from the environment, which can take the form of natural language prompts, structured data (\eg traffic tables, maps), sensor feeds, or multimodal inputs such as images. Perception involves parsing, summarizing, or transforming raw input into a format suitable for reasoning \cite{schick2023toolformer}.

\item \textbf{Memory:} Memory mechanisms allow agents to retain relevant information across interactions or over time. This includes short-term memory (within a single session, via the context window of an LLM) and long-term memory (external databases, vector stores, or document retrieval systems). Effective memory enables continuity in reasoning, retrieval of prior knowledge, and support for personalized or historical context \cite{park2023generative, li2024personal}.

\item \textbf{Planning:} Planning refers to the agent’s ability to interpret goals and formulate a structured strategy to achieve them. This involves breaking down high-level objectives into intermediate subgoals, sequencing them coherently, and maintaining consistency throughout execution. Techniques such as COT and self-refinement can enhance this process by supporting step-by-step reasoning and iterative plan improvement \cite{wei2022chain, guo2025deepseek}.

\item \textbf{Action:} The action module enables the agent to interact with the external world. This includes a broad range of behaviors such as generating textual outputs (\eg reports or recommendations), executing code for simulation or computation, calling APIs or external services, or issuing control commands to affect real-world systems (\eg switching traffic signals) \cite{li2025urban, schick2023toolformer}.

\item \textbf{Learning:} Learning allows the agent to quickly adapt to new tasks and improve its behavior based on feedback or experience. This includes fine-tuning model weights, updating memories, or refining decision-making strategies through self-reflection \cite{shinn2023reflexion}. 
\end{itemize}

\subsubsection{Multi-agent LLM Systems}
Multi-agent systems (MAS) composed of LLM-based agents extend single-agent capabilities by enabling distributed, interactive problem-solving across multiple intelligent agents \cite{guo2024large}. Each agent may take on specialized roles and collaborate with others through natural language, enabling flexible coordination and division of labor. Compared to traditional multi-agent systems, LLM-based agents communicate more naturally, adapt to broader contexts with minimal domain-specific tuning, and demonstrate emergent cooperative behavior through shared prompts and memory \cite{guo2024large, wang2024survey}. Specifically, two fundamental aspects of LLM-based multi-agent systems are agent profiling and inter-agent communication:
\begin{itemize}
\item \textbf{Profiling:} Agent profiling enables role specialization within the agent population. In LLM-based systems, agents can be instantiated with different system prompts to simulate diverse personalities, expertise domains, or behavioral policies. For instance, one agent may be designated as a transportation planner, another as a policy analyst, and a third as a sustainability advisor. These profiles can be manually designed or automatically learned through few-shot examples and instruction tuning. Profiling not only improves task decomposition and division of labor but also facilitates more realistic role-based interaction and negotiation among agents. Advanced frameworks like CAMEL \cite{li2023camel} demonstrate how agents with assigned roles can engage in multi-turn dialogues to collaboratively solve complex tasks.
 
\item \textbf{Communication:} LLM agents communicate through natural language, offering a flexible and interpretable medium for information exchange. Their communication can be structured in various configurations, including centralized frameworks with a coordinator agent, decentralized peer-to-peer systems, and hierarchical organizations. In addition to structural design, the interaction paradigm is shaped by the agents' underlying goals. Agents may engage in cooperative scenarios (where they work toward shared objectives), competitive settings (where they pursue conflicting interests), or hybrid forms that combine both. These factors collectively influence how agents negotiate, share information, and coordinate their actions within complex multi-agent environments.
\end{itemize}

\subsection{Urban LLM Agents}
\subsubsection{Conceptual Framework}
We define \emph{Urban LLM Agents} as a specialized type of LLM-powered agents, specifically designed to operate in complex, ever-changing, and interconnected urban environments. Urban LLM agents are envisioned as \emph{hybrid intelligence} spanning the virtual and physical worlds, working at the intersection of physical infrastructure, digital platforms, and human society. Unlike agents that focus on isolated tasks, urban LLM agents act as intelligent mediators, bridging various urban subsystems with human operators to support scheduling, optimization, management, and planning in intelligent cities. To fulfill these roles effectively, urban LLM agents are expected to possess several essential capabilities that align closely with the spatio-temporal characteristics of urban systems:

% We argue that \emph{Urban LLM Agents} represent an emerging class of LLM-powered agents, specifically designed to operate in complex, dynamic, and interconnected spatio-temporal urban contexts. Urban LLM agents are envisioned as \emph{hybrid intelligence} spanning the virtual and physical worlds, situated at the intersection of physical infrastructure, digital platforms, and the human society. Rather than performing isolated tasks, their central function is to act as intelligent mediators, bridging diverse urban subsystems with human stakeholders to support scheduling, optimization, management, and planning in intelligent cities. To fulfill this role effectively, urban LLM agents are expected to possess a set of core capabilities that are closely aligned with the spatio-temporal characteristics of urban systems. These include:

\begin{itemize}
    \item \textbf{Spatio-temporal data integration}: Urban environments generate vast volumes of heterogeneous data rich in spatial and temporal information, including geovectors, time series, trajectories, geo-tagged images, social media feeds, and regulatory documents. Urban LLM agents need to extract, align, and synthesize these multimodal data sources to create a unified and up-to-date understanding of the city. This integration differs from traditional multimodal fusion, requiring fine-grained spatio-temporal alignment across multiple scales (\eg from street-level to city-wide), which enables agents to maintain situation awareness and respond to fast-changing urban dynamics.
    \item \textbf{Spatio-temporal reasoning}: Reasoning in urban environments requires not only commonsense knowledge but also the ability to understand, infer, and predict across complex spatial topologies (\eg road networks, zoning rules) and temporal patterns (\eg rush-hour patterns, event-driven changes). To operate effectively, urban LLM agents must be capable of leveraging these spatio-temporal structures to identify causal relationships, detect emerging anomalies, and anticipate future developments. Achieving this often requires specialized knowledge and tools in domains such as route planning and resource allocation, which exceed the reasoning capabilities of standard LLMs.
    \item \textbf{Spatio-temporal collaboration}: Cities involve many stakeholders with different goals, responsibilities, and interests. Urban LLM agents must work within this complexity by acting as mediators between groups such as governments, service providers, and citizens. Unlike agents optimized for individual utility, their goal is to support coordinated decisions that balance local needs with broader system-level goals. These agents are required to reason under uncertainty, support fair outcomes, and help align competing goals over space and time. This level of collaboration raises unique challenges, including negotiation, coordination, and joint optimization across different parts of the city.
\end{itemize}

The above three capabilities form the core structure of urban LLM agents. However, they are not natively supported in general-purpose LLMs. Bridging this gap requires careful integration of heterogeneous urban data, utilization of domain-specific reasoning tools, and support for distributed decision-making. In this sense, urban LLM agents should be viewed not as standalone systems, but as complex ecosystems made up of multiple components working together to reason and coordinate across space and time.

% The above three capabilities form the cognitive architecture of urban LLM agents. Importantly, they are not natively supported by off-the-shelf LLMs. Bridging this gap requires the systematic integration of LLMs with heterogeneous urban data sources, specialized spatio-temporal tools, and distributed optimization strategies. In this sense, urban LLM agents should be viewed not as isolated entities, but as composite, multi-agent ecosystems capable of collective reasoning and adaptive coordination across both spatial and temporal dimensions.

\begin{table}[ht]
\centering
\small
\caption{Urban LLM agents vs. related agent paradigms.}
\begin{tabular}{l|c|c|c}
\toprule
\textbf{Dimension} & \textbf{Generic LLM Agents} & \textbf{Embodied Agents} & \textbf{Urban LLM Agents} \\
\midrule
\textbf{Data Sources} & Text/Image & RGB-D/Sensorimotor & Spatio-Temporal Data \\
\textbf{Interaction} & Cyber Space & Physical Space & Cyber-Physical-Social Space \\
\textbf{Embodiment} & Disembodied & Fully Embodied & Semi-Embodied \\
\textbf{Objective} & Individual Task & Individual Task & System-Level Optimization \\
\textbf{Decision Horizon} & Short-Term & Immediate & Short-to-Long Term \\
\textbf{Agency Type} & Language-Mediated & Manipulative & Decision-Supportive \\
\bottomrule
\end{tabular}
\label{tab:agent-comparison}
\vspace{-0.8cm}
\end{table}

\subsubsection{Comparison with Existing Paradigms}
Table~\ref{tab:agent-comparison} compares urban LLM agents with two common types of agents, \ie generic LLM agents~\cite{wang2024survey,xi2025rise} and embodied agents~\cite{liu2024aligning,liu2025embodied}. Generic LLM agents mainly operate in text-based cyberspace, where they are effective at language understanding, reasoning, and single-user interactions. Embodied agents, on the other hand, interact directly with the physical world through sensors and actions, and are typically deployed in well-defined settings such as homes or factories. 
In contrast, urban LLM agents stand at a unique point between these two types of agents. They are designed to interact with the \emph{cyber-physical-social} systems of cities~\cite{zhou2019cyber}, where digital infrastructure (\eg IoT devices, control systems), physical systems (\eg traffic networks, energy grids), and human behavior (\eg mobility patterns, social norms) are deeply intertwined. While these agents are not physically embodied like robots, they are \emph{semi-embodied}: they can affect the physical world through digital interfaces (\eg traffic signal APIs, urban digital twins), or indirectly support human decisions in real-world urban management.

In addition, urban LLM agents differ in the scope and goal of their decision-making. Rather than focusing on short-term, individual tasks, they aim to support \emph{system-level optimization} across various interdependent urban subsystems. This includes both immediate responses to real-time events (\eg managing traffic congestion) and long-term strategic tasks (\eg urban planning or climate adaptation). These broader, more complex goals are rarely addressed by current agent designs and require new approaches for multi-scale coordination and cross-domain reasoning.

\section{Agent-Centric Perspective}
\begin{figure}[t]
    \centering
    \includegraphics[width=0.8\linewidth]{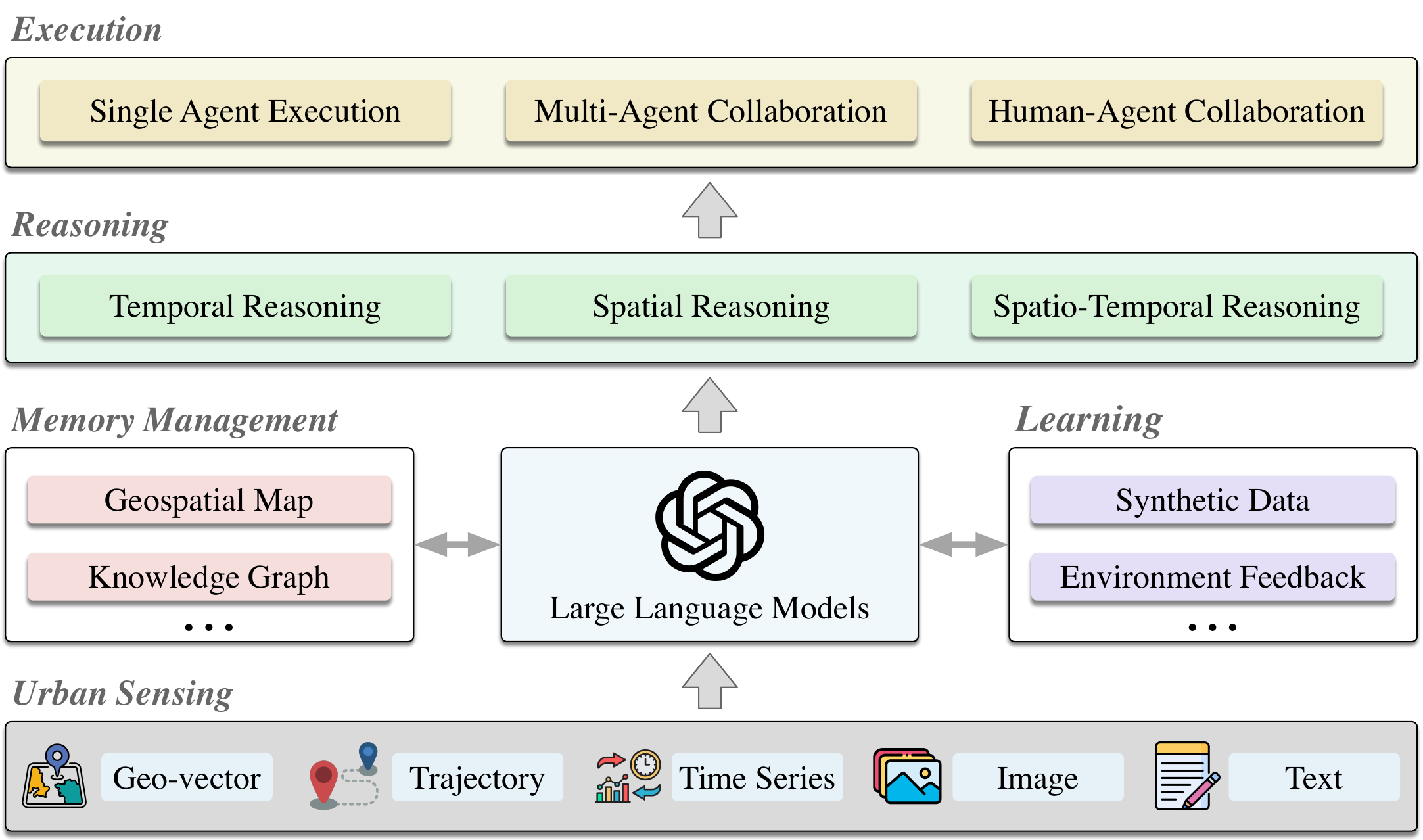}
    \caption{The major components of urban LLM agents.}
    \label{fig:agent-centric perspective}
\end{figure}

As illustrated in Figure~\ref{fig:agent-centric perspective}, the core of urban LLM agents is a large language model, serving as the central controller for interpreting inputs, coordinating internal modules, and interacting with external systems. Building upon this foundation, we identify five key modules required to support the full operational spectrum of intelligent urban agents: (1) \emph{urban sensing}, which enables agents to collect and interpret spatio-temporal urban signals; (2) \emph{memory management}, which organizes and retrieves knowledge across spatial, temporal, and semantic dimensions; (3) \emph{reasoning}, which empowers agents to simulate potential outcomes and generate executable plans; (4) \emph{execution}, which converts linguistic outputs into concrete actions through tool usage, inter-agent coordination, or human-agent interaction; and (5) \emph{learning}, which ensures that agents can continuously improve and adapt to evolving urban environments through synthetic or real-world feedback. In the following sections, we review existing literature that advances these modules, emphasizing design principles, methodologies, and open challenges specific to urban LLM agents.

\subsection{Urban Sensing}
Urban LLM agents can actively collect and integrate diverse data modalities from external APIs, databases, or interactive platforms, enabling a continually updated understanding of urban dynamics. Unlike traditional data pipelines~\cite{zheng2014urban}, these agents operate as adaptive observers, selectively interacting with urban environments based on task requirements. In this section, we first introduce the major sensing modalities within the context of cities and then discuss semantic integration strategies for transforming multimodal inputs into actionable knowledge.

\subsubsection{Sensing Modalities}
Urban LLM agents interact with a wide range of sensing modalities that span the spatial, temporal, and social dimensions of urban life~\cite{zou2025deep}. We categorize them into five representative types: geovector, time series, trajectory, visual, and textual inputs~\cite{zhang2024towards}, each contributing to different aspects of the agents’ perceptual capabilities:
\begin{itemize}[left=0pt]
    \item \textbf{Geovector input}: Static spatial representations such as points, lines, and polygons serve as the foundational geometry of urban environments. Data sources include points of interest (POIs), road networks, and land use~\cite{zheng2014urban}. For urban LLM agents, geovector data provides essential spatial context for tasks such as routing~\cite{liu2020multi}, land-use simulation~\cite{parker2003multi}, and infrastructure analysis~\cite{mao2023detecting}. Accurate alignment and interpretation of geovector inputs allow agents to spatially ground dynamic observations and perform location-aware reasoning.
    \item \textbf{Time series input}: Time series data captures the temporal evolution of urban phenomena and is often collected from environmental sensors and smart infrastructure. Typical examples include traffic flows~\cite{han2024bigst}, air quality~\cite{han2021joint,han2022semi}, energy consumption~\cite{zhou2021informer,yi2024filternet}, and noise levels~\cite{zheng2014diagnosing}. These signals are critical for reasoning about urban dynamics but pose significant challenges due to non-stationarity~\cite{fan2023dish}, complex periodic structures~\cite{fan2022depts}, and spatial correlations.
    \item \textbf{Trajectory input}: Trajectory data records the movement of individuals and vehicles, often sourced from GPS traces, mobile apps, and transit logs~\cite{zheng2015trajectory}. Such data enables agents to understand mobility behaviors and supports tasks such as transportation planning~\cite{wu2024lade}, last-mile logistics~\cite{wu2024lade}, and crowd monitoring~\cite{zhang2017deep}. When combined with geovector data, trajectories can help agents infer causality, identify anomalies~\cite{zhou2024can}, and anticipate future trends~\cite{li2024urbangpt}.
    \item \textbf{Visual input}: Urban environments are increasingly equipped with visual sensors, including satellites, smartphones, and vehicle-mounted cameras~\cite{ito2024understanding}. These data sources provide rich perceptual information for assessing infrastructure conditions, traffic states, construction activities, and more. However, processing visual inputs requires robust computer vision pipelines and, more importantly, effective cross-modal alignment to connect visual content with spatial coordinates and textual semantics~\cite{yan2024urbanclip,hao2024urbanvlp}.
    \item \textbf{Textual input}: Unstructured textual content, such as policy documents, regulatory codes, and social media posts, provides high-level semantic and social context~\cite{zhao2016towards,chen2021location}. These sources inform agents about civic priorities, legal constraints, and collective sentiment, which are often missing from sensor-based inputs. Processing textual content requires advanced language understanding to extract structured, decision-relevant information from noisy and ambiguous sources.
\end{itemize}

\subsubsection{Semantic Integration}
To transform heterogeneous sensing inputs into unified urban knowledge, urban LLM agents require performing semantic integration, \ie aligning and fusing data from different modalities into coherent representations that support downstream reasoning and decision-making. We hereafter discuss three integration strategies: modality-to-text translation, cross-modal alignment, and tool-assisted processing.

\begin{itemize}[left=0pt]
    \item \textbf{Modality-to-text translation}: Given the language-centric nature of LLMs, one straightforward approach is to translate structured and numerical inputs into natural language prompts. The modality-to-text translation can be achieved through hand-crafted templates~\cite{li2024large,beneduce2024large,mai2024opportunities} or specialized tokenization strategies~\cite{gruver2023large}. For example, LLM4POI~\cite{li2024large} translates raw trajectories into prompt sequences using predefined templates, while LLMTime~\cite{gruver2023large} tokenizes time series inputs digit-by-digit to improve forecasting performance. However, such methods often suffer from scalability issues when applied to high-dimensional inputs and are generally less effective for complex modalities such as images or videos.
    \item \textbf{Cross-modal alignment}:  Cross-modal alignment is more efficient than direct translation. This strategy leverages dedicated encoders (\eg small neural networks) for each modality to project diverse signals into a shared embedding space before feeding them into the LLM~\cite{yin2024survey}. For example, UrbanCLIP~\cite{yan2024urbanclip} aligns satellite imagery with textual descriptions using contrastive learning to enhance geographic understanding. Recent works further extend cross-modal alignment to other modalities, including POIs~\cite{xiao2024refound}, street-view imagery~\cite{hao2024urbanvlp}, time series~\cite{li2024urbangpt,zhong2025time}, and trajectories~\cite{xu2025mm}, improving the agent's ability to sense multimodal urban data. 
    \item \textbf{Tool-assisted processing}: Beyond passive fusion, urban LLM agents can dynamically invoke external analytical tools, \eg GIS platforms, physical simulators, or code interpreters, to process complex data streams. These tools act as extensions of the agent's perception capability, enabling advanced operations like spatial transformations~\cite{akinboyewa2024gis} and traffic simulations~\cite{zhang2024trafficgpt}. Tool-augmented strategies enhance the agent’s interpretability and analytical precision beyond what purely language-based processing can achieve.
\end{itemize}

Going forward, several promising directions are emerging for advancing urban sensing in LLM-powered agents. First, handling multimodal uncertainty remains a critical challenge, as urban data is often noisy, incomplete, and inconsistently structured across modalities. Second, it is essential to explore how agents can incorporate novel data sources (\eg LiDAR inputs) and update their knowledge in real-time. Finally, a paradigm shift is underway in which urban LLM agents are not only interpreters of data but also capable of actively generating and contributing new data (\eg crowd-sourced citizen feedback) ~\cite{hou2025urban,liang2025foundation}, \ie LLM agents as sensors.

\subsection{Memory Management}
Memory management enables urban LLM agents to effectively store, organize, and leverage both real-time signals and city-scale knowledge in a task- and context-aware manner. As cognitive intermediaries, these agents require memory not merely as a storage facility but as a dynamic and adaptive system, supporting perception, reasoning, and planning under diverse spatio-temporal constraints. We term this capability agentic memory, which spans three interrelated processes: \emph{memory acquisition}, \emph{memory retrieval}, and \emph{memory utilization}.

\subsubsection{Memory Acquisition}
Memory acquisition refers to how agents gather and structure knowledge about their operating environment. Urban LLM agents acquire memory through both direct sensing of the physical world and indirect extraction from various urban data sources. We summarize four major forms of memory covered in the literature:
\begin{itemize}[left=0pt]
    \item \textbf{Operational state memory}: This type of memory logs real-time, high-frequency signals derived from user interactions and sensed context, typically indexed by geospatial coordinates and timestamps. For example, LLMLight~\cite{lai2023llmlight} obtains the real-time traffic conditions at target intersections, encoding them into readable textual format to guide traffic signal control. CoLLMLight~\cite{yuan2025collmlight} extends the memory by maintaining recent traffic state histories from neighboring intersections for multi-agent coordination. In mobility applications, TrajAgent~\cite{du2024trajagent} stores diverse trajectories and user interaction history in a unified memory structure for trajectory-related tasks. Similarly, AgentMove~\cite{feng2024agentmove} develops a temporal memory to capture users’ recent and long-term mobility patterns in key-value pairs. 
    \item \textbf{Geospatial map memory}: Geospatial maps encapsulate the static spatial layout of the city, including POIs, road networks, and zoning boundaries. Early works primarily incorporate geospatial maps by manually embedding relevant information (\eg nearby POIs) into the prompts or by invoking tools. For example, CityGPT~\cite{feng2024citygpt} manually associates user queries with relevant POIs or road segments for spatial reasoning tasks. TrafficGPT~\cite{zhang2024trafficgpt} processes road networks through simulation tools, enabling visualizations and diagnostic analysis in traffic scenarios. However, such methods usually lack flexible retrieval capabilities. More recent designs treat maps as external structured memory retrievable via natural language queries. Spatial-RAG~\cite{yu2025spatial}, for instance, integrates maps as spatial databases and supports compositional queries like “Find a bar within walking distance from my office—must have live jazz.”
    \item \textbf{Vector database memory}: Vector databases encode unstructured content (\eg urban planning documents, street-view images) into dense embedding vectors for semantic search~\cite{pan2024survey}. ITINERA~\cite{tang2024itinera} constructs a POI-level memory from user travel blogs, using LLMs to extract descriptions and encode them into dense embeddings. The resulting POI embeddings are stored in a continuously updated database that supports fine-grained itinerary generation. PlanGPT~\cite{zhu2024plangpt}, on the other hand, develops a domain-specific vector database by using Plan-Emb, a custom embedding model pre-trained on general Chinese corpora and then fine-tuned on curated urban planning documents. To build the database, PlanGPT preprocesses urban planning texts into semantic chunks and encodes each chunk using Plan-Emb.
    \item \textbf{Knowledge graph memory}: In addition, agents also require structured and symbolic representations of persistent urban knowledge. Knowledge Graph (KG) captures symbolic relations between diverse urban entities (\eg POIs, road segments, and administrative boroughs), thereby supporting a wide range of urban tasks. Traditional KG-based systems, such as UrbanKG~\cite{liu2023urbankg} and UUKG~\cite{ning2023uukg}, rely on pre-defined schemas and manual annotation pipelines to extract entities and relations from urban data. However, manual or rule-based KG construction usually suffers from scalability and flexibility. To address this, UrbanKGent~\cite{ning2024urbankgent} proposes an LLM-powered agent for open-domain KG construction. By fine-tuning a general-purpose LLM and equipping it with tool invocation capabilities, UrbanKGent automatically extracts entities and their relations from urban geographic and text data sources.
\end{itemize}

\subsubsection{Memory Retrieval}
Memory retrieval allows urban LLM agents to selectively access relevant information from previously acquired memory, based on the current spatio-temporal context and task intent. We classify existing approaches into three categories:

\begin{itemize}[left=0pt]
    \item \textbf{Spatio-temporal retrieval}: This retrieval paradigm focuses on indexing and querying information that varies across spatial locations and temporal windows, such as sensor streams and GPS trajectories. Traditional systems like PostGIS and Oracle Spatial employ spatial index structures (\eg K-D tree, R-tree) but often lack efficient temporal access and distributed scalability. To address this, systems like ST-Hadoop~\cite{alarabi2018st}, GeoSpark~\cite{yu2015geospark}, and LocationSpark~\cite{tang2016locationspark} integrate spatio-temporal range queries with big data frameworks. In parallel, NoSQL-based approaches~\cite{hughes2015geomesa} employ multi-level indexing schemes and key-encoding strategies (\eg space-filling curves) to store and retrieve spatio-temporal data at scale. Recently, JUST~\cite{li2020just} further advances this line by supporting online updates in a scalable retrieval engine. We refer to Zheng et al.~\cite{zheng2014urban} and Alam et al.~\cite{alam2022survey} for more details on spatio-temporal retrieval.
    \item \textbf{Semantic retrieval}: Semantic retrieval retrieves relevant memory entries based on task intent and natural language queries, instead of relying on spatial or temporal proximity. It is particularly useful when accessing abstract or high-level knowledge, such as urban plans, policy documents, or user preferences. For example, PlanGPT~\cite{zhu2024plangpt} introduces a hierarchical retrieval framework that combines keyword indexing with cross-attention-based re-ranking, enabling LLMs to locate semantically relevant textual chunks from large-scale planning archives. In contrast, ITINERA~\cite{tang2024itinera} adopts a preference-aware POI retrieval module, where user requests are decomposed into fine-grained intents and encoded into embedding vectors. POIs are retrieved and ranked based on their alignment with positive and negative preference vectors. However, these methods often lack spatio-temporal awareness, which may retrieve semantically relevant but temporally outdated or geographically irrelevant content.
    \item \textbf{Hybrid retrieval}: Hybrid strategies combine spatio-temporal constraints with semantic similarity to enable multi-faceted retrieval. Spatial-RAG~\cite{yu2025spatial} achieves this by unifying sparse spatial retrieval (\ie SQL queries over spatial databases) with dense semantic retrieval based on text embeddings. Spatial candidates are first selected using spatial constraints (\eg proximity, containment), while semantic relevance is assessed through cosine similarity between query and object descriptions. A hybrid scoring function linearly combines spatial and semantic scores, and candidates on the Pareto frontier are reranked by an LLM to balance geometric accuracy and contextual fit. This hybrid strategy enables robust and adaptive retrieval for spatial reasoning tasks beyond the reach of unimodal methods.
\end{itemize}

\subsubsection{Memory Utilization}
Urban LLM agents should also effectively utilize memory for reasoning, planning, and execution. This involves how agents integrate retrieved knowledge into ongoing decision processes. The most direct approach is retrieval-augmented prompting, where the retrieved memory is injected into the prompt as contextual information. This enables the agent to make decisions by incorporating both historical knowledge and the current state of the urban environment. For example, AgentMove~\cite{feng2024agentmove} integrates personalized trajectory histories into the prompt for destination prediction. Similarly, PlanGPT~\cite{zhu2024plangpt} utilizes retrieved documents to inform constraint-aware urban planning. The LLM combines retrieved information with real-time goals to synthesize plans that balance feasibility and policy compliance. Additionally, urban LLM agents can store execution traces (\eg failures, exceptions) in memory for reflective use. This episodic memory allows agents to avoid past mistakes and refine their actions in future tasks. For instance, if an agent encounters traffic congestion during a real-time routing task, it can store this information in memory to avoid the same path in future queries.

In summary, these mechanisms form a closed-loop memory system that encompasses acquisition, retrieval, and utilization, empowering urban LLM agents to function as memory-augmented decision-makers. Future research could explore the unification of symbolic and sub-symbolic memory representations, the lifelong evolution of memory under urban dynamics, and memory sharing across multi-agent systems.

\subsection{Reasoning}
Reasoning is the cognitive core of urban LLM agents, which plays a vital role in decomposing complex tasks and formulating executable plans. In this section, we review existing research from three aspects: \emph{temporal reasoning}, \emph{spatial reasoning}, and \emph{spatio-temporal reasoning}.

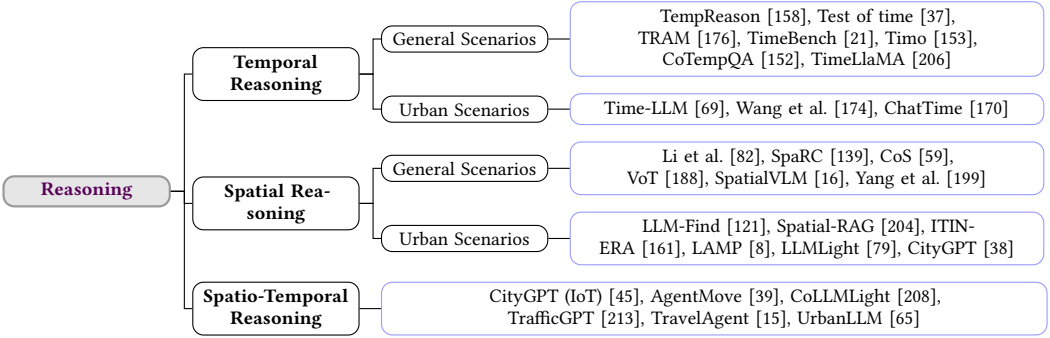
\begin{figure*}[!ht]
\scriptsize
    \begin{adjustbox}{width=\textwidth}
        \begin{forest}
        for tree={
            forked edges,
            grow'=0,
            draw,
            rounded corners,
            node options={align=center},
            text width=3.1cm,
            s sep=6pt,
            calign=edge midpoint,
        },
        [\textbf{Reasoning}, fill=gray!20, draw=gray!80, thick, text=violet!70!black, text width=2cm
            [\textbf{Temporal Reasoning}, text width=2cm
                [General Scenarios, text width=2cm
                    [{TempReason~\cite{tan2023towards}, 
                    Test of time~\cite{fatemi2024test},
                    TRAM~\cite{wang2024tram}, 
                    TimeBench~\cite{chu2024timebench},
                    Timo~\cite{sutimo}, CoTempQA~\cite{su2024living}, TimeLlaMA~\cite{yuan2024back}}, text width=6cm, draw=blue!40, rounded corners]
                ]
                [Urban Scenarios, text width=2cm
                    [{Time-LLM~\cite{jin2024time}, Wang et al.~\cite{wang2024news}, ChatTime~\cite{wang2025chattime}}, text width=6cm, draw=blue!40, rounded corners]
                ]
            ]
            [\textbf{Spatial Reasoning}, text width=2cm
                [General Scenarios, text width=2cm
                    [{Li et al.~\cite{li2024advancing}, SpaRC~\cite{rizvi2024sparc}, CoS~\cite{hu2024chain}, VoT~\cite{wu2024mind}, SpatialVLM~\cite{chen2024spatialvlm}, Yang et al.~\cite{yang2024thinking}}, text width=6cm, draw=blue!40, rounded corners]
                ]
                [Urban Scenarios, text width=2cm
                    [{LLM-Find~\cite{ning2024llm}, Spatial-RAG~\cite{yu2025spatial}, ITINERA~\cite{tang2024itinera}, 
                    LAMP~\cite{balsebre2024lamp},
                    LLMLight~\cite{lai2023llmlight}, CityGPT~\cite{feng2024citygpt}}, text width=6cm, draw=blue!40, rounded corners]
                ]
            ]
            [\textbf{Spatio-Temporal Reasoning}, text width=2cm
                [{CityGPT (IoT)~\cite{guan2024citygpt}, AgentMove~\cite{feng2024agentmove}, CoLLMLight~\cite{yuan2025collmlight}, TrafficGPT~\cite{zhang2024trafficgpt}, TravelAgent~\cite{chen2024travelagent}, UrbanLLM~\cite{jiang2024urbanllm}}, text width=8.5cm, draw=blue!40, rounded corners]
            ]
        ]
        \end{forest}
    \end{adjustbox}
    \caption{Taxonomy of reasoning in urban LLM agents.}
    \label{fig:reasoning_tree}
\end{figure*}

% \begin{figure}
%     \centering
%     \includegraphics[width=1\linewidth]{figure/Reasoning.pdf}
%     \caption{Taxonomy of reasoning in urban LLM agents.}
%     \label{fig:reasoning_tree}
% \end{figure}

\subsubsection{Temporal Reasoning}
Temporal reasoning allows urban LLM agents to interpret events, understand relationships such as order and duration, and make predictions about future trends.

\begin{itemize}[left=0pt]
    \item In \textbf{general scenarios}, recent benchmarks~\cite{tan2023towards,fatemi2024test,wang2024tram,chu2024timebench} have revealed that while LLMs show promising results in basic tasks like event ordering, they often struggle with more complex reasoning involving temporal logic and implicit relations. To address these challenges, researchers have proposed a series of enhancements. For instance, Timo~\cite{sutimo} augments LLMs with mathematical knowledge for arithmetic-based temporal reasoning, while CoTempQA~\cite{su2024living} focuses on improving reasoning capabilities on overlapping and co-occurring events. More recently, TimeLlaMA~\cite{yuan2024back} emphasizes explainability by requiring LLMs to not only produce time-sensitive answers but also justify their reasoning steps. 
    \item In \textbf{urban scenarios}, temporal reasoning empowers urban LLM agents to understand, anticipate, and simulate dynamic real-world processes. For example, Time-LLM~\cite{jin2024time} introduces a reprogramming strategy that transforms raw time series into structured textual prototypes, enabling few-shot generalization under data-constrained scenarios. Subsequently, Wang et al.~\cite{wang2024news} integrate textual event knowledge with temporal data streams, enriching the semantic context of time series forecasting in domains such as transportation and energy. To further enhance the LLM’s ability to reason over complex urban signals, ChatTime~\cite{wang2025chattime} proposes to scale and quantize numerical time series, converting them into discrete tokens that can be directly processed alongside text. Despite fruitful progress, there are also several open issues, such as the ability to reason under temporal uncertainty and support for long-horizon planning. 
\end{itemize}

\subsubsection{Spatial Reasoning}
Spatial reasoning allows agents to interpret and reason about the location, orientation, and spatial relationships of entities in their environment. 
\begin{itemize}[left=0pt]
    \item In \textbf{general scenarios}, recent studies have enhanced the spatial capabilities of LLMs through three main approaches. Structured prompting techniques~\cite{li2024advancing, rizvi2024sparc} help organize spatial relationships into interpretable templates, enabling more systematic reasoning. Another line of research focuses on symbolic abstraction~\cite{li2024reframing, hu2024chain}, which represents spatial concepts in a compact format that supports step-by-step reasoning. For example, Chain-of-Symbol (CoS) prompting~\cite{hu2024chain} encodes textually formatted spatial relations as concise, discrete symbols, improving both interpretability and reasoning efficiency. More recently, several studies leveraged internal visual imagination to enhance spatial reasoning. For example, Visualization-of-Thought (VoT) prompting~\cite{wu2024mind} encourages LLMs to simulate scenes mentally, which has shown promising results on spatial puzzles and navigation tasks. Beyond pure language, SpatialVLM~\cite{chen2024spatialvlm} integrates textual inputs with visual and 3D spatial information, providing enhanced spatial awareness. Yang et al.~\cite{yang2024thinking} further demonstrate that generating cognitive maps can facilitate complex spatial reasoning, particularly in video-based benchmarks. 
    \item In \textbf{urban scenarios}, spatial reasoning is particularly challenging due to the need to ground language in large-scale geographic contexts. General-purpose LLMs often lack access to geospatial knowledge and experience in urban environments, which limits their capacity to reason effectively about urban space. To address this gap, three major strategies have emerged for improving spatial reasoning in urban LLM agents. First, retrieval-augmented approaches such as LLM-Find~\cite{ning2024llm} and Spatial-RAG~\cite{yu2025spatial} integrate external geospatial knowledge bases to retrieve spatial facts and constraints, resulting in higher precision and interpretability. However, their effectiveness depends on the quality of the external data sources. Second, tool-augmented agents extend LLMs with access to spatial tools for solving real-world tasks such as travel planning~\cite{ning2024urbankgent,tang2024itinera,jin2024large}. For instance, ITINERA~\cite{tang2024itinera} combines LLM-generated plans with spatial solvers to ensure routes are contextually appropriate and spatially feasible. Third, instruction tuning in simulated environments offers a way to inject spatial knowledge directly into LLM agents~\cite{lai2023llmlight,balsebre2024lamp,feng2024citygpt}. For example, LLMLight~\cite{lai2023llmlight} trains agents via reinforcement learning in traffic simulators, enhancing their ability to adaptively control traffic signals. CityGPT~\cite{feng2024citygpt} immerses LLMs in city-scale interactive environments, where agents learn to navigate and reason under realistic spatial constraints. These approaches enable agents to internalize complex spatial structures and facilitate generalization across diverse urban environments.
\end{itemize}

\subsubsection{Spatio-Temporal Reasoning}
Urban systems are constantly changing across both space and time. Events such as traffic congestion or public emergencies often start in one location and gradually affect other parts of the city. Thus, urban LLM agents need to reason jointly over spatial and temporal dynamics. This capability is essential for understanding how local changes evolve, interact, and lead to broader consequences.

Recent research explores several promising directions to equip LLM agents with spatio-temporal reasoning capabilities. The first involves decomposing complex urban tasks into smaller, manageable subtasks. For example, CityGPT (IoT)~\cite{guan2024citygpt} splits user queries into separate spatial and temporal components, assigns them to specialized agents, and merges the outputs using a coordination module. Similarly, AgentMove~\cite{feng2024agentmove} divides mobility prediction into individual behaviors, spatial distributions, and shared movement patterns, with each handled by a dedicated module. These designs improve scalability and allow agents to focus on specific aspects of the problem before integrating results. The second direction leverages structured representations to capture urban spatio-temporal dependencies. CoLLMLight~\cite{yuan2025collmlight} constructs a spatio-temporal graph of the road network to model dependencies between intersections over time. It further introduces a complexity-aware mechanism that dynamically adjusts the reasoning depth based on real-time traffic conditions, helping reduce unnecessary computations. The third line of work focuses on hybrid reasoning by combining LLMs’ internal capabilities with external tools. TrafficGPT~\cite{zhang2024trafficgpt} enhances ChatGPT with traffic simulators and predictive models, enabling it to analyze and interpret numerical traffic data. TravelAgent~\cite{chen2024travelagent} utilizes LLMs to reason about time, distance, and scheduling constraints, then employs APIs and arithmetic tools to generate feasible travel plans. UrbanLLM~\cite{jiang2024urbanllm} solves complex urban problems by decomposing them into tractable subtasks, selecting tailored spatio-temporal models for handling subtasks, and synthesizing their results into coherent outputs.

Despite recent progress, most existing methods still rely heavily on observational data or simulated environments, which may not fully capture the causal mechanisms behind urban phenomena. Moving forward, spatio-temporal reasoning in urban LLM agents could benefit from integrating causal reasoning tools, such as causal graphs, structural equation modeling, and physics-informed priors, to improve robustness, interpretability, and generalization in real-world deployments.

\subsection{Execution}
Execution is the operational core of urban LLM agents, translating high-level reasoning into concrete actions. Achieving this goal requires a robust framework capable of functioning under real-world constraints. We categorize existing research into three execution modes: (1) \emph{single-agent execution}, where an individual agent perceives and acts autonomously; (2) \emph{multi-agent collaboration}, where multiple agents coordinate across distributed urban environments; and (3) \emph{human-agent collaboration}, where agents interact with human stakeholders through dialogue, feedback, and shared decision-making.

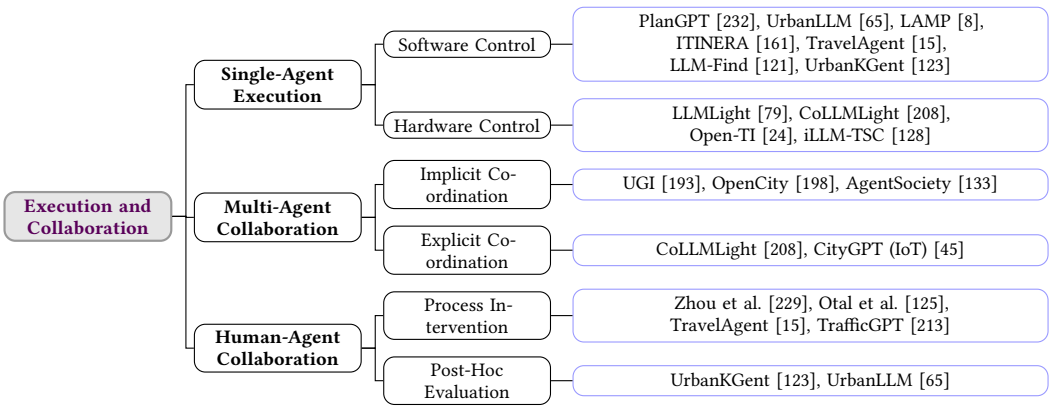
\begin{figure*}[!ht]
\scriptsize
    \begin{adjustbox}{width=\textwidth}
        \begin{forest}
        for tree={
            forked edges,
            grow'=0,
            draw,
            rounded corners,
            node options={align=center},
            text width=3.1cm,
            s sep=6pt,
            calign=edge midpoint,
        },
        [\textbf{Execution and Collaboration}, fill=gray!20, draw=gray!80, thick, text=violet!70!black, text width=2cm
            [\textbf{Single-Agent Execution}, text width=2cm
                [Software Control, text width=2cm
                    [{PlanGPT~\cite{zhu2024plangpt}, UrbanLLM~\cite{jiang2024urbanllm}, LAMP~\cite{balsebre2024lamp}, ITINERA~\cite{tang2024itinera}, TravelAgent~\cite{chen2024travelagent},  LLM-Find~\cite{ning2024llm}, UrbanKGent~\cite{ning2024urbankgent}}, text width=6cm, draw=blue!40, rounded corners]
                ]
                [Hardware Control, text width=2cm
                [{LLMLight~\cite{lai2023llmlight}, CoLLMLight~\cite{yuan2025collmlight}, Open-TI~\cite{da2024open}, iLLM-TSC~\cite{pang2024illm}}, text width=6cm, draw=blue!40, rounded corners]
                ]
            ]
            [\textbf{Multi-Agent Collaboration}, text width=2cm
                [Implicit Coordination, text width=2cm
                    [{UGI~\cite{xu2023urban}, OpenCity~\cite{yan2024opencity}, AgentSociety~\cite{piao2025agentsociety}}, text width=6cm, draw=blue!40, rounded corners]
                ]
                [Explicit Coordination, text width=2cm
                    [{CoLLMLight~\cite{yuan2025collmlight}, CityGPT (IoT)~\cite{guan2024citygpt}}, text width=6cm, draw=blue!40, rounded corners]
                ]
            ]
            [\textbf{Human-Agent Collaboration}, text width=2cm
                [Process Intervention, text width=2cm
                    [{Zhou et al.~\cite{zhou2024large}, Otal et al.~\cite{otal2024llm}, TravelAgent~\cite{chen2024travelagent}, TrafficGPT~\cite{zhang2024trafficgpt}}, text width=6cm, draw=blue!40, rounded corners]
                ]
                [Post-Hoc Evaluation, text width=2cm
                    [{UrbanKGent~\cite{ning2024urbankgent}, UrbanLLM~\cite{jiang2024urbanllm}}, text width=6cm, draw=blue!40, rounded corners]
                ]
            ]
        ]
        \end{forest}
    \end{adjustbox}
    \caption{Taxonomy of execution in urban LLM agents.}
    \label{fig:execution_collaboration_tree}
\end{figure*}

% \begin{figure}
%     \centering
%     \includegraphics[width=1\linewidth]{figure/Execution andCollaboration.pdf}
%     \caption{Taxonomy of execution in urban LLM agents.}
%     \label{fig:execution_collaboration_tree}
% \end{figure}

\subsubsection{Single Agent Execution}
In the single-agent setting, an urban LLM agent operates independently to understand its environment, interpret user instructions, reason over goals and constraints, and perform actions. These agents act as autonomous interfaces between humans and urban systems, converting natural language queries into executable plans or system-level decisions. Depending on how the agent interacts with the environment, we further divide this category into two forms: software control and hardware control.
\begin{itemize}[left=0pt]
    \item \textbf{Software control}: Agents in this category interact with digital platforms to accomplish tasks such as report generation, geographic data retrieval, and structured knowledge synthesis. Several works focus on urban planning and service recommendations. For example, PlanGPT~\cite{zhu2024plangpt} generates structured planning reports to support land use and city development decisions. UrbanLLM~\cite{jiang2024urbanllm} responds to user queries (\eg finding a parking spot) by producing activity plans based on spatio-temporal contexts. Travel agents like LAMP~\cite{balsebre2024lamp}, ITINERA~\cite{tang2024itinera}, and TravelAgent~\cite{chen2024travelagent} provide personalized itineraries by integrating user preferences with transportation constraints.  Additionally, LLM-Find~\cite{ning2024llm} enables agents to retrieve geographic data by writing and executing code, while UrbanKGent~\cite{ning2024urbankgent} automatically constructs urban knowledge graphs by extracting entities and relationships from heterogeneous sources such as text and maps.
    \item \textbf{Hardware control}: In contrast, hardware control allows agents to directly influence physical infrastructure by issuing commands to control devices or systems. A representative example is traffic signal control. LLMLight~\cite{lai2023llmlight} employs a fine-tuned LLM to optimize signal phase timings at intersections through APIs in a simulation environment, aiming to reduce vehicle waiting times. CoLLMLight~\cite{yuan2025collmlight} extends this framework by enabling coordinated control across multiple intersections, each managed by a local LLM agent. Open-TI~\cite{da2024open} adopts a hierarchical structure where a central agent oversees global planning while local agents handle low-level control. iLLM-TSC~\cite{pang2024illm} combines Reinforcement Learning (RL) with LLM-based verification: actions proposed by the RL policy are reviewed and approved by an LLM agent through natural language reasoning, enhancing interpretability and safety.
    
\end{itemize}

\subsubsection{Multi-Agent Collaboration}
Urban environments are inherently multi-agent systems, composed of diverse actors and subsystems operating in parallel. Many urban tasks, such as traffic optimization, emergency coordination, and energy balancing, require agents to operate collaboratively across spatially distributed and functionally heterogeneous domains. In this context, multi-agent collaboration refers to the ability of multiple LLM-based agents to share information, align plans, and jointly execute actions to fulfill collective urban goals. These agents often operate with partial observability and local objectives, making coordination a central technical challenge. We categorize existing approaches into two paradigms: implicit coordination, where collaboration emerges through shared environments or behavioral heuristics, and explicit coordination, where agents communicate directly to align plans and decisions.
\begin{itemize}[left=0pt]
    \item \textbf{Implicit coordination}: Implicit coordination relies on the principle that complex group behaviors can emerge from the independent actions of many agents. This approach is especially useful for modeling social or behavioral dynamics in urban environments. For example, UGI~\cite{xu2023urban} places LLM agents in a simulated city where each agent acts independently based on local observations. Despite the lack of communication protocol or centralized oversight, emergent patterns such as neighborhood clustering or traffic congestion still arise over time. OpenCity~\cite{yan2024opencity} scales this idea by allowing thousands of agents to operate in a shared environment, where local memory and feedback mechanisms guide behavior. AgentSociety~\cite{piao2025agentsociety} further enhances realism by equipping agents with internal traits such as beliefs, goals, and emotions, facilitating emergent social behaviors like norm formation and group polarization. These systems demonstrate the potential of implicit coordination for simulating large-scale, realistic urban interactions. However, the absence of shared goals or communication also makes such systems difficult to control or predict, making them more appropriate for exploratory simulations than tasks demanding consistency, precision, or rapid response.
    \item \textbf{Explicit coordination}: In contrast, explicit coordination involves direct communication among agents to share information, align plans, and jointly optimize performance. This approach is particularly suitable for structured urban tasks, such as traffic management or emergency response. For example, CoLLMLight~\cite{yuan2025collmlight} assigns an LLM agent to each traffic intersection in a city, linking them via a spatiotemporal graph based on geographic proximity. Agents exchange local traffic states and collaboratively generate signal control policies, enabling globally consistent decisions. Unlike implicit frameworks, this system allows for fine-grained control and network-wide optimization. CityGPT~\cite{guan2024citygpt} also employs explicit coordination, albeit in a modular fashion. It adopts a role-based architecture in which agents specialize in tasks such as interpreting user intent, analyzing spatial and temporal dimensions, and synthesizing final outputs. These specialized modules interact through structured message passing, supporting interpretable and composable workflows.
\end{itemize}

Overall, multi-agent collaboration enables urban LLM agents to expand their capabilities from local decision-making to distributed collective intelligence. Implicit strategies offer scalability and behavioral realism, while explicit protocols support precision and optimization. A promising research direction lies in developing hybrid approaches that combine the adaptability of emergent coordination with the controllability of structured interaction, particularly for tasks demanding both social fidelity and operational robustness~\cite{ni2024planning}.

\subsubsection{Human-Agent Collaboration}
Despite the increasing autonomy of urban LLM agents, real-world deployments still require meaningful human involvement. Urban decision-making is deeply embedded in complex social, regulatory, and institutional environments, where transparency, accountability, and human oversight are as important as technical performance. Existing research on human-agent collaboration in urban contexts generally falls into two categories: process intervention during task execution and post-hoc evaluation after task completion.
\begin{itemize}[left=0pt]
    \item \textbf{Process intervention}: This category refers to scenarios in which humans actively intervene to guide or adjust agent behaviors, leveraging human judgment, domain knowledge, or evolving preferences. For instance, Zhou et al.~\cite{zhou2024large} introduce a participatory planning framework where an LLM "planner" interacts with multiple LLM "residents" agents that simulate community feedback. The planner proposes a land-use plan, receives critiques from residents, and then revises the plan accordingly. Otal et al.~\cite{otal2024llm} introduce an LLM agent for emergency response that communicates with citizens and dispatchers, while being overseen by human officials to ensure clarity and appropriateness in high-stakes scenarios. TravelAgent~\cite{chen2024travelagent} supports personalized trip planning via an interactive loop where users iteratively refine travel preferences and constraints, while TrafficGPT~\cite{zhang2024trafficgpt} analyzes traffic data and delegates final decision-making to human operators, thereby incorporating expert oversight into the process.
    \item \textbf{Post-hoc evaluation}: This category focuses on human assessment after task execution, evaluating outputs for accuracy, completeness, and alignment with real-world constraints. UrbanKGent~\cite{ning2024urbankgent} constructs urban knowledge graphs using LLM agents, with human reviewers verifying extracted relationships and facts to safeguard the accuracy of the knowledge base. Similarly, UrbanLLM~\cite{jiang2024urbanllm} engages domain experts to evaluate AI-generated urban activity plans, identifying omissions and inconsistencies relative to professional standards. Looking ahead, a critical frontier is to develop agents capable of continual adaptation based on human feedback, closing the loop between intervention, evaluation, and autonomous learning. Such interactive learning frameworks are crucial for building trustworthy urban LLM agents that evolve alongside dynamic urban systems.
\end{itemize}

In summary, execution constitutes the bridge between reasoning and real-world impact for urban LLM agents. By synthesizing capabilities across single-agent autonomy, distributed collaboration, and human-aligned oversight, execution frameworks could support flexible, scalable, and trustworthy operations in complex, evolving city environments. While existing systems typically specialize in one mode, a major challenge and opportunity lies in designing unified execution architectures that dynamically integrate across modes. Such adaptability is essential for scaling urban LLM agents from task-specific prototypes to robust actors embedded within the cyber-physical-social fabric of future cities.

\subsection{Learning}
Learning plays a central role in enabling urban LLM agents to operate across diverse city environments, adapt to evolving urban dynamics, and improve over time. In this section, we discuss two primary paradigms that support learning in urban LLM agents: learning from synthetic data and learning from environmental feedback.

\begin{figure*}[!ht]
\scriptsize
    \begin{adjustbox}{width=\textwidth}
        \begin{forest}
        for tree={
            forked edges,
            grow'=0,
            draw,
            rounded corners,
            node options={align=center},
            text width=3.1cm,
            s sep=6pt,
            calign=edge midpoint,
        },
        [\textbf{Learning}, fill=gray!20, draw=gray!80, thick, text=violet!70!black, text width=2cm
            [\textbf{Learning from Synthetic Data}, text width=2.5cm
                [Vanilla Instruction Generation, text width=2cm
                    [{\protect\parbox{5cm}{\centering PlanGPT~\protect\cite{zhu2024plangpt}, UrbanLLM~\protect\cite{jiang2024urbanllm}, CoPB~\protect\cite{shao2024chain}}}, text width=5cm, draw=blue!40, rounded corners]
                ]
               [Advanced Reasoning Augmentation, text width=2cm
                    [{\protect\parbox{5cm}{\centering UrbanKGent~\protect\cite{ning2024urbankgent}, UGI~\protect\cite{xu2023urban}, TrajAgent~\protect\cite{du2024trajagent}}}, text width=5cm, draw=blue!40, rounded corners]
                ]
            ]
            [\textbf{Learning from Environmental Feedback}, text width=2.5cm
                [Simulation-Based Feedback, text width=2cm
                    [{\protect\parbox{5cm}{\centering iLLM-TSC~\protect\cite{pang2024illm}, LLMLight~\protect\cite{lai2023llmlight}, CoLLMLight~\protect\cite{yuan2025collmlight}, DiMA~\protect\cite{ning2025dima}}}, text width=5cm, draw=blue!40, rounded corners]
                ]
               [Real-World Feedback, text width=2cm
                    [{\protect\parbox{5cm}{\centering WatchOverGPT~\protect\cite{shahid2024watchovergpt}}}, text width=5cm, draw=blue!40, rounded corners]
                ] 
            ]
        ]
        \end{forest}
    \end{adjustbox}
    \caption{Taxonomy of learning in urban LLM Agents.}
    \label{fig:learning}
\end{figure*}
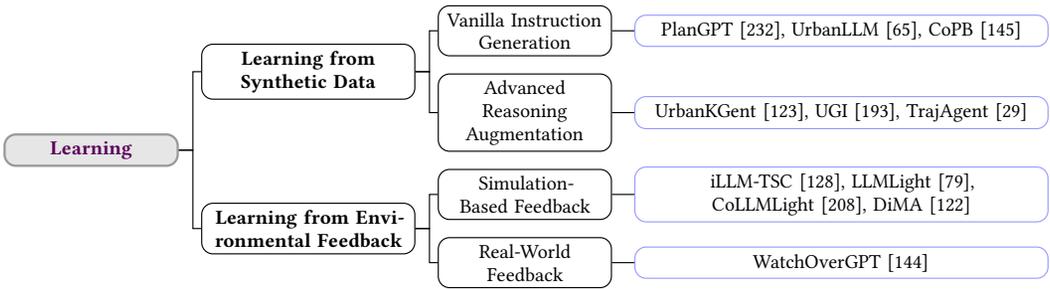

% \begin{figure}
%     \centering
%     \includegraphics[width=1\linewidth]{figure/Learning.pdf}
%     \caption{Taxonomy of learning in urban LLM agents.}
%     \label{fig:learning}
% \end{figure}

\subsubsection{Learning from Synthetic Data}
Unlike general-purpose LLMs trained on large-scale open-domain text, urban LLM agents require domain-specific knowledge to reason within structured, goal-oriented environments such as transportation systems and public service delivery.
Recent research has demonstrated that training on synthetic data is a practical and scalable way to teach agents the rules, workflows, and reasoning patterns that underpin urban systems. 

\begin{itemize}[left=0pt]
\item \textbf{Vanilla instruction generation:}
In urban planning, synthetic datasets are often constructed from formal sources such as zoning codes, development guidelines, and regulatory documents. PlanGPT~\cite{zhu2024plangpt} exemplifies this approach by fine-tuning LLMs on instruction-style data extracted from these materials, allowing the agent to internalize the logic and structure of urban planning policies. In contrast, UrbanLLM~\cite{jiang2024urbanllm} focuses on autonomous activity planning and orchestration rather than primarily learning policy logic from formal documents. 
For longer-term planning tasks, CoPB~\cite{shao2024chain} decomposes decision-making into a chain of intention-driven steps, allowing the agent to simulate human-like planning behavior across spatial and temporal contexts. 

\item \textbf{Advanced reasoning augmentation:}
In addition to the vanilla method, existing studies also invoke external tools or multi-turn agent discussion to obtain more reliable, diverse reasoning traces~\cite{madaan2023self}.
For example, UrbanKGent~\cite{liu2023urbankg} synthesizes tool-use instructions to guide the construction of urban knowledge graphs, enabling LLM agents to interface with spatial databases through structured queries. 
Other efforts focus on agent-level behavior modeling and coordination. UGI~\cite{xu2023urban}, for instance, employs a curriculum learning strategy where agents are trained on progressively more complex tasks, starting with single-step execution and gradually advancing to multi-agent coordination in urban environments.
This staged approach allows agents to build foundational capabilities before handling more complex scenarios. In the domain of mobility, synthetic data has been used to train agents for tasks such as routing and behavior modeling. 
TrajAgent~\cite{du2024trajagent} derives a self-reflection mechanism to generate a large-scale dataset of agent-environment interactions under urban mobility constraints, which is used to train LLMs to generate feasible trajectories and respond to changing urban contexts. 

\end{itemize}

Overall, synthetic data provides a controlled, richly annotated approach for developing urban LLM agents. 
It enables structured supervision and offers a valuable foundation for grounding agents in domain-specific reasoning and policy alignment.

\subsubsection{Learning from Environmental Feedback}
While synthetic data offers a strong starting point for injecting urban knowledge into LLM agents, real-world urban environments are inherently dynamic, uncertain, and context-dependent. To operate effectively in such settings, agents are required to learn through interaction, \ie adapting their behavior in response to changing conditions, user preferences, and unexpected events. This interactive learning is often achieved through feedback-driven interaction with simulation platforms or direct engagement with real-world data streams. 
\begin{itemize}[left=0pt]
\item \textbf{Simulation-Based Feedback:}
In the area of traffic control, simulation-based closed-loop learning has become a widely adopted strategy. Agents such as iLLM-TSC~\cite{pang2024illm} and LLMLight~\cite{lai2023llmlight} integrate LLMs with traffic simulators to iteratively refine control policies. These agents receive feedback signals, such as average delay, queue length, or congestion levels, which guide policy updates and improve performance over time. Building on this approach, CoLLMLight~\cite{yuan2025collmlight} introduces multi-agent collaboration, where intersection-level agents share environmental feedback and coordinate their actions to achieve global traffic optimization across a city. In mobility service applications, interactive learning is also critical. 
DiMA~\cite{ning2025dima} presents a ride-hailing assistant trained through continual fine-tuning in a simulated role-playing environment. This setup enables the agent to adapt to evolving user preferences and operational constraints by interacting with simulated passengers, drivers, and service operators. Such simulation-driven fine-tuning helps the model align with real-world decision-making patterns in ride-hailing scenarios. 

\item \textbf{Real-World Feedback:}
Beyond structured simulations, some studies also incorporate real-time sensory inputs and live reports to support decision-making in complex urban scenarios. For example, WatchOverGPT~\cite{shahid2024watchovergpt} processes data from surveillance sensors, citizen reports, and external alerts to monitor and respond to emergency events. 
Given the uncertainty and urgency of emergencies, WatchOverGPT needs to quickly adjust its predictions and actions based on the latest information, further underscoring the importance of interactive adaptation in real-world urban contexts.  

\end{itemize}

Overall, urban feedback is often sparse, delayed, or biased, especially in underrepresented or underserved regions where sensor coverage and user engagement may be limited. 
Without proper handling, such disparities can reinforce existing inequalities in urban service delivery. 
Moreover, the interdependent nature of urban systems demands that agents not only learn individual policies but also coordinate with others in a trust-aware, socially responsible manner.

\section{Application-Centric Perspective}
\begin{figure}
    \centering
    \includegraphics[width=1 \linewidth]{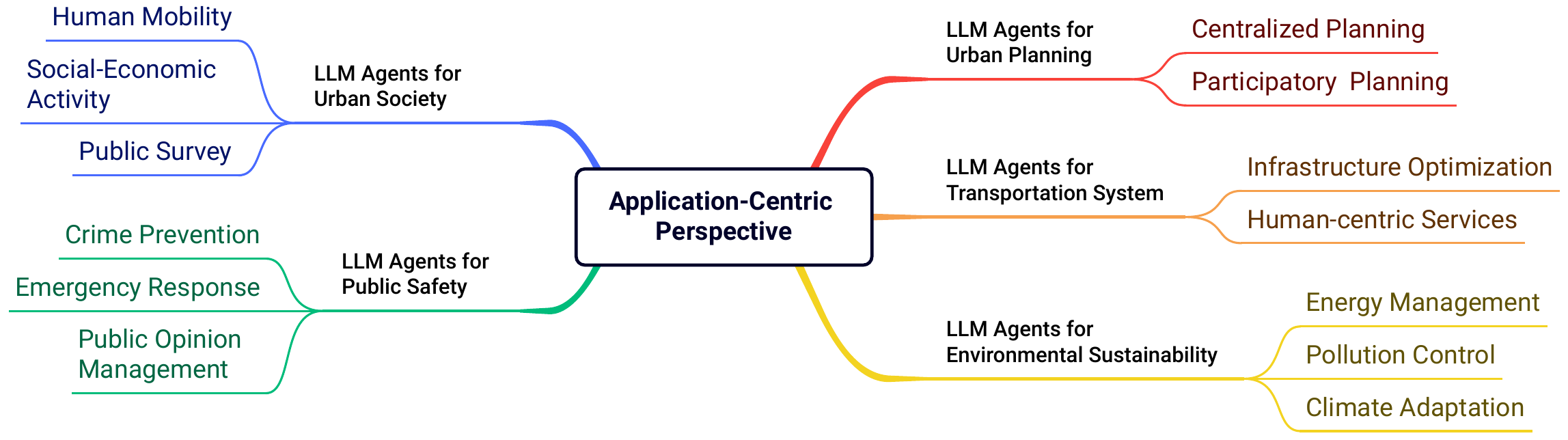}
    \caption{An overview of urban LLM agent from application-centric perspective.}
    \label{fig:application-overview}
    \vspace{-0.4cm}
\end{figure}

The integration of LLMs into urban applications is reshaping how cities tackle complex challenges. 
With their advanced natural language understanding, reasoning, and generation capabilities, LLM-powered agents enable intelligent interaction with urban systems, providing solutions in areas such as urban planning, transportation, environmental sustainability, public safety, and urban society. 
As research in this field progresses, a comprehensive survey of existing work is crucial for understanding current trends and future directions.
% To this end, as shown in Figure \song{TODO}, we summarize urban applications powered by LLM and propose a novel classification framework to organizes the literature accordingly.
To this end, as shown in Figure \ref{fig:application-overview}, we summarize LLM-powered urban applications and present a taxonomy to organize the literature.

\subsection{LLM Agents for Urban Planning}
In this section, we investigate the use of LLM agents in urban planning processes, which can be categorized into centralized planning and participatory planning.

\subsubsection{Centralized Planning}
Centralized urban planning involves government-led initiatives to optimize urban infrastructure and management using LLM agents for efficient decision-making and regulatory compliance.
For example, to ensure adaptable parking facilities that accommodate various types of vehicles and urban settings, Jin et al.~\cite{jin2024large} devise an LLM-based parking planning agent to evaluate and optimize current parking infrastructures in an efficient and flexible way.
Beyond spatial planning, LLM agents are also adopted to improve the urban decision-making process and promote smart city management, due to their remarkable language processing and problem-solving capabilities.
For instance, UrbanPlanBench~\cite{zhengurbanplanbench} lays the foundational investigation into the acquisition of planning knowledge among LLMs, spanning various aspects of the urban planning task, including fundamental principles, professional knowledge, and management regulations, which showcases the proficiency of LLMs in understanding regulations.
To generate, translate, or evaluate professional urban planning documents, PlanGPT~\cite{zhu2024plangpt} crafts an LLM agent that can strategically utilize various data sources representing the latest urban information.
In addition, it is aligned with the style of governmental documents through domain-specific instruction fine-tuning.
To effectively incorporate immense references of urban planning texts, a tailored embedding model and hierarchical retrieval system are devised to address the challenge of low signal-to-noise ratio.
To foster city management, City-LEO~\cite{jiao2024city} synergizes LLMs' logical reasoning abilities to effectively scope down the prior knowledge, which aims to customize the user requirements, and an end-to-end optimizer to derive decisions under uncertain environments.
Besides, Kalyuzhnaya et al.~\cite{kalyuzhnaya2025llm} design a multi-agent system that integrates retrieval-augmented generation (RAG) approaches, demonstrating superior accuracy in answering queries regarding urban management.

\begin{table}[t]
\centering
\small
\caption{Existing LLM-based agents for urban planning.}

\resizebox{\textwidth}{!}{
\begin{tabular}{c|c|c|c|c|c|c}
\toprule
\textbf{Domain} & \textbf{Article} & \textbf{\makecell{Urban Sensing}} & \textbf{\makecell{Memory \\ Management}} & \textbf{\makecell{Planning and \\ Reasoning}} & \textbf{\makecell{Execution and \\ Collaboration}} & \textbf{Learning} \\
\midrule
\multirow{5}{*}{\makecell{Central \\ Planning}} & Jin et al.~\cite{jin2024large} & Geo-Vector &  Operational State & Spatial & Single-Agent & - \\
& UrbanPlanBench~\cite{zhengurbanplanbench} & Text & Vector Database & Spatio-Temporal & Single-Agent & Synthetic Data \\
& PlanGPT~\cite{zhu2024plangpt} & Text & Vector Database & Spatio-Temporal & Single-Agent & Synthetic Data \\
& City-LEO~\cite{jiao2024city} & Text & Vector Database & Spatio-Temporal & Single-Agent & - \\
& Kalyuzhnaya et al.~\cite{kalyuzhnaya2025llm} & Text & Vector Database & Spatio-Temporal & Multi-Agent & - \\
\midrule
\multirow{3}{*}{\makecell{Participatory \\ Planning}} & Zhou et al.~\cite{zhou2024large} & Geo-Vector, Image & - & Spatial & Multi-Agent & - \\
& Singla et al.~\cite{singla2024adaptive} & Geo-Vector, Image & - & Spatial & Multi-Agent & - \\
& Ni et al.~\cite{ni2024planning} & Geo-Vector, Image & Operational State & Spatio-Temporal & Multi-Agent & - \\
\bottomrule
\end{tabular}
}
\label{tab:urban-planning-comparison-dimensions}
% \vspace{-0.8cm}
\end{table}

\subsubsection{Participatory Planning}
Participatory urban planning engages diverse stakeholders, including residents and planners, through LLM agents to collaboratively shape urban development.
For example, Zhou et al.~\cite{zhou2024large} propose a participatory land use planning framework empowered by LLMs.
Specifically, it crafts LLM agents to emulate the planner and residents, and engage them in a multi-agent discussion to balance their divergent needs.
To reduce time and token costs, a fishbowl discussion technique is used, in which only a subset of residents is included in the interaction, avoiding overlong context length for LLMs.
To achieve balanced and area-specific land use layouts, Singla et al.~\cite{singla2024adaptive} harness four specialized LLM agents to manage the development of different sub-areas, targeting the local-regional land-use requirements.
Furthermore, to provide a fine-grained assessment of the urban plans and facilitate continuous plan enhancement, Ni et al.~\cite{ni2024planning} propose the cyclical urban planning paradigm.
It is characterized by the iteration of urban planning, living simulation of residents, and resident interviews, providing nuanced insights into residents’ lived experiences for plan evaluation and regeneration.

\subsection{LLM Agents for Transportation System}
This section reviews the emerging applications of LLM agents in transportation systems, which fall into two major categories: infrastructure optimization and human-centric services. These categories reflect the primary ways LLMs are utilized in this field: to enhance the system’s architecture and operations, and to improve services directly for users.

\subsubsection{Infrastructure Optimization}
This category discusses the use of LLM agents to analyze, predict, manage, and optimize transportation infrastructure and its operations.

\begin{itemize}[left=0pt]
\item \textbf{Traffic analytics}: Large Language Models (LLMs) are playing an increasingly significant role in enhancing predictive analytics in transportation systems. Current applications extend beyond semantic analysis, showcasing their potential in various predictive tasks \cite{ning2023uukg}. A key research focus involves leveraging LLMs to integrate unstructured textual data, such as news reports and social media feeds, into traditional traffic forecasting models, thereby extracting contextual information to enhance forecast accuracy \cite{huang2024enhancing}. Additionally, LLMs are applied to process complex, unstructured geospatial data for tasks like city-wide delivery demand estimation and forecasting \cite{nie2025joint}. Beyond feature enhancement, augmented LLM frameworks, such as Open-TI \cite{da2024open}, are emerging to support advanced traffic analysis, including simulations and external tool integration for tasks like demand optimization and traffic signal control. Furthermore, LLMs also function as intelligent interfaces, simplifying access to complex transportation datasets and models. For example, TransitGPT \cite{devunuri2024transitgpt} translates natural language queries into data requests, thus democratizing transit information access. Similarly, TrafficGPT \cite{zhang2024trafficgpt} and CityGPT-IoT \cite{guan2024citygpt} enable users to interact with transportation data through natural language. These developments underscore the diverse contributions of LLMs to predictive analytics in transportation, ultimately enhancing decision-making in traffic management.
\item \textbf{Traffic control}: LLMs have become transformative tools for decision-making in transportation systems, particularly in generating adaptive strategies under complex operational constraints. One critical application is traffic signal control, a traditionally challenging task often managed by rule-based algorithms or reinforcement learning techniques \cite{koonce2008traffic}. With the emergence of LLMs, researchers have begun exploring their potential in improving traffic signal management.
Early studies \cite{tang2024large, da2023llm} propose diverse strategies for utilizing LLMs. For example, LA-Light \cite{wang2024llm} introduces a hybrid framework where the LLM serves as a central reasoning engine, coordinating specialized tools to gather data and make human-like decisions, particularly for rare events such as sensor failures or emergency vehicles. In contrast, LLMLight \cite{lai2023llmlight} demonstrates the feasibility of using LLMs as independent decision-making agents, interpreting traffic data through prompts and employing Chain-of-Thought reasoning to select signal phases, with a tailored LLM agent fine-tuned for this purpose.
Another approach, iLLM-TSC \cite{pang2024illm}, combines LLMs with reinforcement learning (RL), using the LLM as a supervisor to evaluate and refine RL-generated decisions, addressing RL's limitations in scenarios with imperfect observations or rare events. Building on these efforts, Yuan et al. \cite{yuan2025collmlight} propose CoLLMLight, a cooperative multi-agent system where LLM agents manage traffic signals across a road network. This approach involves constructing spatio-temporal graphs from traffic data and predicting optimal signal configurations. Key innovations include complexity-aware reasoning, which adjusts cooperation based on congestion levels, and simulation-driven fine-tuning, enabling iterative optimization of the LLM agent for better understanding of traffic patterns.

% Similarly, TrafficGPT \cite{zhang2024trafficgpt} fuses LLMs with specialized Traffic Foundation Models (TFMs), allowing users to view, process, and interact with potentially predictive TFMs using natural language, thereby aiding decision support.

% Tang et al. \cite{tang2024large} reviewed early efforts to integrate LLMs into this domain.
% For example, initial systems such as LLM-Assisted-Light \cite{wang2024llm}, iLLM-TSC \cite{pang2024illm}, and LLMLight \cite{lai2023llmlight} demonstrated the capacity of LLMs to function as intelligent traffic signal control agents by interpreting complex traffic patterns and optimizing intersection flow \cite{da2023llm}. 
% Tang et al. \cite{tang2024large} reviewed early efforts to integrate LLMs into the traffic signal control domain.

% Building on these foundational studies, Yuan et al. \cite{yuan2025collmlight} introduced CoLLMLight, a cooperative multi-agent system where multiple LLM-driven agents collaboratively manage traffic signals across an entire network. 
% This system represents a significant step forward, showcasing how LLMs can scale to handle the intricacies of large-scale urban traffic systems through decentralized collaboration.

\end{itemize}

These advancements illustrate the growing role of LLMs in transportation infrastructure optimization, where their ability to reason, adapt, and integrate diverse data sources positions them as powerful tools for optimizing transportation systems.

\subsubsection{Human-centric Services}
This category focuses on utilizing LLMs to enhance public transportation services. By processing and analyzing large-scale data, LLM agents directly interact with human users or provide transportation services (\eg navigation and ride-hailing) to address their mobility needs within the city.

\begin{table}[t]
\centering
\small
\caption{Existing LLM-based agents for transportation systems.}
\resizebox{\textwidth}{!}{
\begin{tabular}{c|c|c|c|c|c|c}
\toprule
\textbf{Domain} & \textbf{Article} & \textbf{\makecell{Urban Sensing}} & \textbf{\makecell{Memory \\ Management}} & \textbf{\makecell{Planning and \\ Reasoning}} & \textbf{\makecell{Execution and \\ Collaboration}} & \textbf{Learning} \\
\midrule
\multirow{6}{*}{\makecell{Traffic \\ Analytics}} & Huang et al.~\cite{huang2024enhancing} & Geovector, Text &  Operational State & Spatio-Temporal & Single-Agent & Synthetic Data \\
& Nie et al.~\cite{nie2025joint} & Trajectory, Text & Geospatial Map & Spatio-Temporal & Single-Agent & Synthetic Data \\
& Open-TI~\cite{da2024open} & Text & Vector Database & Spatial & Single-Agent & Environmental Feedback \\
& TransitGPT~\cite{devunuri2025transitgpt} & Text & Vector Database & Spatial & Multi-Agent & Synthetic Data \\
& TrafficGPT~\cite{zhang2024trafficgpt} & Text & Geospatial Map & Spatio-Temporal & Single-Agent & - \\
& CityGPT-IoT~\cite{guan2024citygpt} & Text & Geospatial Map & Spatial & Multi-Agent & - \\
\midrule
\multirow{4}{*}{\makecell{Traffic \\ Control}} & LA-Light~\cite{zhou2024large} & Text & Operational State & Spatial & Single-Agent &  - \\
& LLMLight~\cite{lai2023llmlight} & Text & Operational State & Spatial & Single-Agent & Environmental Feedback \\
& iLLM-TSC~\cite{pang2024illm} & Text & Operational State & Spatial & Single-Agent & - \\
& CoLLMLight~\cite{yuan2025collmlight} & Text & Operational State & Spatio-Temporal & Multi-Agent & Environmental Feedback \\
\midrule
\multirow{4}{*}{\makecell{Navigation \\ Routing}} & Li et al.~\cite{li2024research} & Geovector, Text & - & Spatial & Single-Agent & Environmental Feedback \\
& TraveLLM~\cite{fang2024travellm} & Image, Text & Geospatial Map & Spatial & Single-Agent & - \\
& LAMP~\cite{balsebre2024lamp} & Text & Geospatial Map & Spatial & Single-Agent & Synthetic Data \\
& Verma et al.~\cite{verma2023generative} & Image, Text & Vector Database & Spatio-Temporal & Single-Agent & Environmental Feedback \\
& TP-RAG~\cite{ni2025tp} & Geovector, Text & Vector Database & Spatio-Temporal & Multi-Agent & - \\
\midrule
\multirow{2}{*}{\makecell{On-demand \\ Mobility}} & GARLIC~\cite{han2024gpt} & Trajectory & Operational State & Spatio-Temporal & Multi-Agent & - \\
& DiMA~\cite{ning2025dima} & Text & Vector Database & Spatio-Temporal & Multi-Agent & Environmental Feedback \\
\bottomrule
\end{tabular}
}
% \vspace{-1cm}
\end{table}

\begin{itemize}[left=0pt]
\item \textbf{Navigation and routing}: LLMs are utilized to provide personalized and context-aware transportation services to urban residents. One significant application is route planning. For example, Li et al. \cite{li2024research} propose an LLM-based framework that dynamically adjusts travel routes by interpreting heterogeneous urban data streams, achieving faster and more effective responses compared to traditional optimization models. Similarly, Liu et al. \cite{liu2023can} develop an LLM-powered system for delivery route optimization, reducing last-mile costs by semantically analyzing urban road network constraints. Another important application is improving user experience and offering intelligent assistance during navigation. 
TraveLLM \cite{fang2024travellm}, for instance, introduces an LLM-driven framework that generates alternative transit plans through conversational interactions, providing real-time support during service disruptions.
TP-RAG \cite{ni2025tp} proposes a spatio-temporal-aware travel planning method, supporting city-scale travel plan generation services.
For location-based recommendations, LAMP \cite{balsebre2024lamp} fine-tunes LLMs with city-specific geospatial knowledge, enabling accurate and context-aware conversational recommendations. Additionally, Verma et al. \cite{verma2023generative} investigates LLM-powered generative agents that interact with urban street view imagery to simulate navigation toward defined destinations.

% To better understand urban environments, Verma et al. \cite{verma2023generative} explore LLM-powered generative agents that simulate interactions with street view images. These agents analyze subjective perceptions of the environment, providing insights into urban design and livability.

\item \textbf{On-demand mobility services}: 
LLMs also support the operation and user interaction in services such as ride-hailing.
Recent work, GARLIC \cite{han2024gpt}, focuses on optimizing vehicle dispatching, a critical task in ride-hailing operations. Specifically, GARLIC first integrates hierarchical traffic state representations using multi-view graphs and introduces a dynamic reward system based on driving behaviors and fare analysis. 
Then, it leverages a GPT-augmented policy learning module with a custom ``GeoLoss'' function to ensure geospatial accuracy and improve dispatching efficiency.
Additionally, frameworks like DiMA \cite{ning2025dima} demonstrate the use of LLM agents for city-scale ride-hailing services. 
DiMA integrates external spatial and temporal tools to enhance the reasoning capabilities of LLMs in understanding user travel intentions, enabling accurate ride-hailing order planning. DiMA also develops a cost-effective dialogue system that assigns LLMs of varying sizes to handle ride-hailing conversations.
% Beyond ride-hailing, other efforts, such as LAUR \cite{wei2025laura}, leverage LLMs to optimize Unmanned Aerial Vehicle (UAV) routing. Initially designed to minimize the age of information in sensor networks, this approach is directly applicable to urban delivery.

% Focusing on smart city integration, CityGPT \cite{guan2024citygpt} introduces a multi-agent framework based on LLMs to interact with spatiotemporal data from urban IoT devices. This framework enhances the ability to learn, analyze, and respond to dynamic urban conditions.

\end{itemize}

As an emerging trend, LLMs are driving human-centric transportation services by interacting with urban data and enhancing the efficiency of city-scale public transportation.

\subsection{LLM Agents for Environmental Sustainability}
% This section examines the application of urban LLM agent in addressing various aspects of environmental sustainability within urban environments. 
This section explores the application of LLM agents in addressing critical aspects of urban environmental sustainability. Specifically, we focus on the studies that utilize LLMs to develop intelligent systems for enhancing energy efficiency, controlling pollution, and facilitating climate adaptation.

% These studies explored how to derive LLMs to develop intelligent systems aimed at improving energy efficiency, waste management, air quality monitoring, and climate change adaptation in cities.

\subsubsection{Energy Management.}  
This section discusses recent academic findings on the application of LLM agents in various aspects of urban energy management, including power grid optimization, renewable energy integration, and emerging urban energy applications \cite{liu2025large}.  
As a critical component of urban infrastructure, power grid optimization faces significant operational challenges. LLMs are increasingly being utilized to address these complexities. For instance, LLM4DistReconfig \cite{christou2025llm4distreconfig} fine-tunes LLMs to optimize network configurations in near real-time, reducing system losses while adhering to operational constraints. This method offers more comprehensive outputs compared to traditional algorithms.  
Beyond specific optimization tasks, LLMs are also being developed as tools to assist in broader power system operations. GAIA \cite{cheng2025large} exemplifies this by supporting human operators in tasks such as operation adjustment, monitoring, and handling complex black start scenarios.  
LLM agents also play a key role in wind and solar energy management. 
Time-LLM \cite{jin2023time}, which adapts general LLMs for time-series tasks, has been widely adopted for forecasting applications in wind, solar, and weather-related contexts.
Similarly, EF-LLM \cite{qiu2024ef} is a framework designed to address challenges in load, photovoltaic (PV), and wind power forecasting. It provides AI-assisted automation for the entire forecasting process, including operational guidance, feature engineering, prediction, and post-forecast decision-making.

% in the realm of energy management, LLMs can democratize energy optimization with llm-assisted optimization autoformalism.
% This simplify solution design including electric vehicle charging and HVAC control, lowering barriers to advanced urban energy management \cite{jin2024democratizing}.

\begin{table}[t]
\centering
\small
\caption{Existing LLM-based agents for environmental sustainability.}
\resizebox{\textwidth}{!}{
\begin{tabular}{c|c|c|c|c|c|c}
\toprule
\textbf{Domain} & \textbf{Article} & \textbf{\makecell{Urban Sensing}} & \textbf{\makecell{Memory \\ Management}} & \textbf{\makecell{Planning and \\ Reasoning}} & \textbf{\makecell{Execution and \\ Collaboration}} & \textbf{Learning} \\
\midrule
\multirow{4}{*}{\makecell{Energy \\ Management}} & LLM4DistReconfig~\cite{christou2025llm4distreconfig} & Time Series, Text &  Operational State & Spatio-Temporal & Single-Agent & Synthetic Data \\
& GAIA~\cite{cheng2025large} & Text & - & Spatial  & Single-Agent & Synthetic Data \\
& Time-LLM~\cite{jin2023time} & Time Series &  Operational State & Temporal & Single-Agent & Synthetic Data \\
& EF-LLM~\cite{qiu2024ef} & Time Series, Text & Operational State & Temporal & Multi-Agent & Synthetic Data \\
\midrule
\multirow{2}{*}{\makecell{Pollution \\ Control}} & Verma et al.~\cite{verma2023generative} & Image, Text & Vector Database &  Spatio-Temporal & Single-Agent & Environmental Feedback \\
& LLMAir~\cite{fan2024llmair} & Time Series & Operational State & Spatio-Temporal & Single-Agent & Synthetic Data \\ \midrule
\multirow{4}{*}{\makecell{Climate \\ Adaptation}} & ClimateBERT~\cite{webersinke2021climatebert} & Text & - & - & Single-Agent & - \\
& ClimateGPT~\cite{thulke2024climategpt} &  Text & Vector Database & Spatio-Temporal & Single-Agent &Synthetic Data \\
& ChatClimate~\cite{vaghefi2023chatclimate} & Text & Vector Database & Spatio-Temporal & Single-Agent & - \\
& ClimaQA~\cite{manivannan2024climaqa} & Text & Vector Database & Spatio-Temporal & Single-Agent & - \\
\bottomrule
\end{tabular}
}
% \vspace{-0.8cm}
\end{table}

\subsubsection{Pollution Control.}
LLM agents are also increasingly recognized for their potential to support urban pollution control across various domains. In urban waste management, LLMs can improve operational efficiency by processing waste pickup requests, managing collection schedules, and optimizing resource allocation \cite{kalyuzhnaya2025llm}. 
Verma et al. \cite{verma2023generative} demonstrate this potential by simulating urban citizens' perceptions of city services, including waste management, underscoring the role of LLMs in understanding and potentially improving public satisfaction. 
Beyond solid waste, LLM agents have also been applied to enhance urban air quality monitoring systems. For example, the LLMAir system \cite{fan2024llmair} employs a multi-agent LLM architecture to analyze air quality data during events such as wildfires. 
Although LLMs show strong performance in the semantic analysis of air quality data, studies indicate that their numerical forecasting capabilities remain limited, requiring further research to improve predictive accuracy \cite{gao2025instructor}. 
These works provide promising solutions for improving the sustainability and livability of urban environments.

% The application of LLM agents extends to the complex domain of urban waste management.
% By effectively processing requests related to waste pickup, managing collection schedules, and coordinating resources, LLM agents could contribute to more efficient waste management operations \cite{kalyuzhnaya2025llm}.
% For example, Verma1 et al., \cite{verma2023generative} simulates urban citizens to perceive various aspects of the city, including waste management infrastructure and overall cleanliness levels.

% LLM agents are also emerging as valuable tools for enhancing urban air quality monitoring systems.
% Specific research efforts are focusing on the development of LLM-based systems for air quality analysis and prediction.
% The LLMAir system \cite{fan2024llmair}, for instance, utilizes a multi-agent LLM architecture to analyze air quality data during events like wildfires.
% While some studies have explored the use of LLMs for time series forecasting of air quality, these models have shown limitations in prediction accuracy compared to their strengths in semantic analysis \cite{gao2025instructor}.
% This indicates a need for further research to adapt and enhance the numerical forecasting capabilities of LLMs for air quality applications

\subsubsection{Climate Adaptation}
Predicting climate change in urban environments remains a multifaceted challenge, with LLM agents emerging as potential tools in this domain.
Early attempts like ClimateBERT \cite{webersinke2021climatebert} adapt general-purpose language models (\eg BERT) through continued training on climate-related corpora to enhance performance on climate-specific NLP tasks such as classification, fact-checking, and claim generation.
Along this line, more recent efforts have introduced specialized models and frameworks to further improve the climate reasoning capabilities of LLMs.
ClimateGPT \cite{thulke2024climategpt} is trained on extensive scientific and climate datasets using the Llama-2 architecture. 
By leveraging retrieval-augmented generation, ClimateGPT enhances the accuracy and reliability of its responses.
In contrast, ChatClimate \cite{vaghefi2023chatclimate} explores the integration of RAG techniques. 
By grounding model responses in authoritative sources such as the IPCC AR6 reports, ChatClimate ensures both scientific validity and up-to-date information, as demonstrated through expert evaluations that show it produces accurate and well-referenced answers.
Complementing these model developments, ClimaQA \cite{manivannan2024climaqa} introduces an automated evaluation framework designed to systematically assess LLMs' understanding of climate science. 
It provides both expert-validated benchmarks and synthetic datasets, enabling comprehensive analysis of model performance on climate-related question answering tasks.

% ClimateGPT \cite{thulke2024climategpt} is a Llama 2 family trained on extensive science and climate data, optimized for reliable RAG. 
% ChatClimate \cite{vaghefi2023chatclimate} uses RAG to ground GPT-4's responses in IPCC AR6, ensuring accurate climate information. 
% ClimaQA \cite{manivannan2024climaqa} offers expert-validated benchmarks and methods to rigorously evaluate LLM understanding of climate science.

\subsection{LLM Agents for Public Safety}
This section discusses the research on the application of LLM agents in improving urban public safety, with a focus on their use in building intelligent systems for crime prevention, emergency response, and public opinion management.

% The focus is on how LLMs are being leveraged to develop intelligent systems for crime prediction and urban emergency management.

\subsubsection{Crime Prevention} 
Emerging research suggests that LLMs can be applied to tasks such as crime classification and generating explanations for crime occurrences \cite{wickramasekara2025exploring}. 
The CriX framework \cite{rezacrix} integrates retrieval-augmented generation with the Mistral AI LLM. CriX dynamically retrieves socio-economic indicators (\eg literacy rates, income levels) and maps them to crime hotspots, producing human-readable explanations that link criminal patterns to underlying societal conditions. 
This approach enables hotspot prediction while providing interpretable outputs for policymakers, addressing the "black-box" limitations of traditional models. 
In addition, WatchOverGPT \cite{shahid2024watchovergpt} demonstrates the potential of LLMs in real-time crime detection. The framework employs YOLOv8 for weapon detection and activity recognition. 
By enabling context-aware communication with law enforcement, WatchOverGPT assists vulnerable individuals (\eg visually impaired users) with minimal human intervention.

% The CriX framework \cite{rezacrix}, for instance, integrates demographic factors with LLMs to offer human-readable explanations linking crime hotspots with relevant societal conditions.
% In addition, WathOverGPT \cite{shahid2024watchovergpt}, utilizes a wearable camera and smartphone integration to continuously monitor the environment for distress or criminal activity by analyzing human actions and weapon presence, combined with location and other smartphone data, enabling a GPT-based application for autonomous decision-making.

\begin{table}[t]
\centering
\small
\caption{Existing LLM-based agents for public safety.}
\resizebox{\textwidth}{!}{
\begin{tabular}{c|c|c|c|c|c|c}
\toprule
\textbf{Domain} & \textbf{Article} & \textbf{\makecell{Urban Sensing}} & \textbf{\makecell{Memory \\ Management}} & \textbf{\makecell{Planning and \\ Reasoning}} & \textbf{\makecell{Execution and \\ Collaboration}} & \textbf{Learning} \\
\midrule
\multirow{2}{*}{\makecell{Crime \\ Prevention}} & CriX~\cite{rezacrix} & Trajectory, Text &  Operational State & Spatio-Temporal & Single-Agent & Synthetic Data \\
& WatchOverGPT~\cite{shahid2024watchovergpt} & Trajectory, Image, Text & - & Spatio-Temporal  & Multi-Agent & Environmental Feedback \\
\midrule
\multirow{2}{*}{\makecell{Emergency \\ Response}} & He et al.~\cite{he2024multi} & Text & Vector Database & Spatial & Multi-Agent &  - \\
& LLM-Assisted Crisis~\cite{otal2024llm} & Text & Knowledge Graph & Spatio-Temporal & Single-Agent & Environmental Feedback \\ \midrule
\multirow{2}{*}{\makecell{Public \\ Opinion}} & RE'EM~\cite{fan2025invisible} & Geovector, Text & Vector Database & Spatial & Multi-Agent & Synthetic Data \\
& \cite{han2024enhanced}, \cite{xia2024question}, \cite{sun2023unleashing}, \cite{durmus2024role} &  Text & - & - & Single-Agent & - \\
\bottomrule
\end{tabular}
}
% \vspace{-0.8cm}
\end{table}

\subsubsection{Emergency Response}
LLM agents demonstrate significant potential in enhancing urban emergency response systems by analyzing live data from various sources during emergencies and coordinating effective responses in real-time.
By improving the simulation capabilities of agent-based models, LLMs can support the intricate decision-making processes involved in coordinating emergency responses and mitigating the impact of urban disasters \cite{he2024multi}.
In addition, researchers are also exploring LLMs, such as Llama-2, to support public safety telecommunicators during large-scale emergencies. 
These models can classify emergency events from 911 calls or social media, assist dispatchers with real-time recommendations, and alert relevant agencies when systems are overwhelmed \cite{otal2024llm}.

\subsubsection{Public Opinion Management}
LLM agents can also be employed to understand and engage with complex public opinion and social dynamics by analyzing public discourse, simulating social interactions, and managing information flow. 
For example, RE'EM ~\cite{fan2025invisible} proposes using LLMs to analyze large-scale public text data, such as social media, to uncover subtle public opinions and social divisions. 
The proposed RE'EM framework enables the processing of nuanced language to capture lived experiences, such as urban segregation. 
Other studies focus on applying LLMs to analyze public sentiment and deliver information during and after disasters, including earthquakes~\cite{han2024enhanced}, typhoons~\cite{xia2024question}, floods~\cite{sun2023unleashing}, and fires~\cite{durmus2024role}, to enhance public safety and situational awareness. 
These approaches leverage LLMs for social media text analysis, question answering, and potentially visual information extraction to understand public reactions and needs. 
Such applications aim to support disaster response and indirectly influence public opinion by providing timely and relevant information.

% \subsubsection{Public Opinion Management}
% LLM agent are used to understand and engage with complex public opinion and social dynamics by analyzing public discourse, simulating social interactions, and managing information flow.
% For example, Fan et al., \cite{fan2025invisible} proposes to use LLMs to analyze extensive public text data, such as social media, to identify subtle public opinions and social divisions. 
% The proposed RE'EM framework enables the processing of nuanced language to understand lived experiences like urban segregation. 
% Other works focus on LLM to analyze public sentiment and provide information during and after disasters like earthquakes \cite{han2024enhanced}, typhoons \cite{xia2024question}, floods \cite{sun2023unleashing}, and fires \cite{durmus2024role} for public safety and situational awareness. 
% They leverage LLMs for social media text analysis, question answering systems, and potentially visual information extraction to understand public reactions and needs. 
% These applications aim to support disaster response and indirectly influence public opinion by providing timely and relevant information.

% \subsection{LLM Agents for Society Simulation}
\subsection{LLM Agents for Urban Society}
This section introduces studies that utilize LLM agents to model and simulate complex societal dynamics within urban contexts. 
They intend to generate synthetic data, replicate and discover societal phenomena, and offer counterfactual or prospective insights.
Existing research can be categorized into three groups: human mobility, social-economic activity, and public survey.

\subsubsection{Human Mobility}
The impressive abilities of LLM agents in general reasoning and role-playing pave the way for more interpretable and generalizable simulation of human mobility behaviors.
LLMob~\cite{jiawei2024large} is one of the pioneering studies, which capitalizes on LLM agents to generate individual mobility patterns.
Based on two principal influential factors of human activities—habitual activity patterns and dynamic motivations—LLM agents first extract the typical movement patterns and preferences from the historical data consistently.
Then the agent derives evolving motivations and situational needs, followed by a final action decision.
Based on LLMob, to incorporate fine-grained collective patterns into mobility generation, MobAgent~\cite{li2024more} applies agent clustering according to individual profiles.
To better align with the real-world data, MobAgent develops a spatial mechanistic model to physically constrain human movements, mapping the human activities to actual locations.
In addition, TrajLLM~\cite{ju2025trajllm} designs a memory module to record the historical activities and preferences of agents, alongside a weighted metric indicating the significance of memory items.
Drawing from behavioral theories, CoPB~\cite{shao2024chain} explicitly models the interweaving of attitudes, subjective norms, and perceived behavioral control within the reasoning process of agents.
The integration of the gravity model further encourages the alignment with real-world mobility distributions.

Another line of studies focuses on mobility modeling, leveraging the reasoning and pattern recognition capabilities of LLMs to analyze or predict human movements with high interpretability. 
LLM-Mob~\cite{wang2023would} introduces a novel framework that formats mobility data into historical and contextual stays, capturing both long-term and short-term dependencies while enabling time-aware predictions through carefully designed prompts. 
Further, AgentMove~\cite{feng2024agentmove} proposes a systematic agentic framework that decomposes mobility prediction into subtasks, including individual pattern mining, urban structure modeling, and collective knowledge extraction, achieving superior performance across diverse datasets. 
Further advancing the field, TrajAgent~\cite{du2024trajagent} unifies trajectory modeling tasks under a single LLM-based agentic framework, integrating data augmentation and parameter optimization to adapt models dynamically. These studies collectively highlight the potential of LLMs to transcend traditional mobility prediction methods by combining interpretability and scalability, while also addressing challenges such as data sparsity and geographical bias.

\begin{table}[t]
\centering
\small
\caption{Existing LLM-based agents for urban society.}
\resizebox{\textwidth}{!}{
\begin{tabular}{c|c|c|c|c|c|c}
\toprule
\textbf{Domain} & \textbf{Article} & \textbf{\makecell{Urban \\ Sensing}} & \textbf{\makecell{Memory \\ Management}} & \textbf{\makecell{Planning and \\ Reasoning}} & \textbf{\makecell{Execution and \\ Collaboration}} & \textbf{Learning} \\
\midrule
\multirow{7}{*}{\makecell{Human Mobility}} & LLMob~\cite{jiawei2024large} & Trajectory & Operational State & Spatio-Temporal & Single-Agent & - \\
& MobAgent~\cite{li2024more} & Trajectory & Operational State & Spatio-Temporal & Single-Agent & - \\
& TrajLLM~\cite{ju2025trajllm} & Trajectory & Operational State & Spatio-Temporal & Single-Agent & - \\
& CoPB~\cite{shao2024chain} & Trajectory & Operational State & Spatio-Temporal & Single-Agent & - \\
& LLM-Mob~\cite{wang2023would} & Trajectory & - & Spatio-Temporal & Single-Agent & - \\
& AgentMove~\cite{feng2024agentmove} & Trajectory & Operational State & Spatio-Temporal & Single-Agent & - \\
& TrajAgent~\cite{du2024trajagent} & Trajectory & Operational State & Spatio-Temporal & Multi-Agent & - \\
\midrule
\multirow{8}{*}{\makecell{Social-Economic \\ Activity}} & Generative Agents~\cite{park2023generative} & Trajectory,Text & Vector Database & Spatio-Temporal & Multi-Agent & - \\
& Humanoid Agents~\cite{wang2023humanoid} & Trajectory,Text & Vector Database & Spatio-Temporal & Multi-Agent & - \\
& D2A~\cite{wang2024simulating} & Trajectory,Text & Vector Database & Spatio-Temporal & Multi-Agent & - \\
& Williams et al.~\cite{williams2023epidemic} & Text & - & Spatio-Temporal & Multi-Agent & - \\
& EconAgent~\cite{li2023econagent} & Text & Operational State & - & Multi-Agent & - \\
& AgentSociety~\cite{piao2025agentsociety} & Trajectory,Text & Vector Database & Spatio-Temporal & Multi-Agent & - \\
& AgentTorch~\cite{chopra2024limits} & Text & - & Spatio-Temporal & Multi-Agent & - \\
& OpenCity~\cite{yan2024opencity} & Trajectory,Text & Vector Database & Spatio-Temporal & Multi-Agent & - \\
\midrule
\multirow{3}{*}{\makecell{Public Survey}} & Bhandari et al.~\cite{bhandari2024urban} & Text & - & Spatio-Temporal & Multi-Agent & - \\
& Park et al.~\cite{park2024generative} & Text & - & Spatio-Temporal & Multi-Agent & - \\
& Yang et al.~\cite{yang2024llm} & Text & - & Spatio-Temporal & Multi-Agent & - \\
\bottomrule
\end{tabular}
}
\label{tab:urban-society-comparison-dimensions}
% \vspace{-0.8cm}
\end{table}

\subsubsection{Social-Economic Activity}
Beyond the analysis of human mobility data, social-economic activity emphasizes more complex and diverse interactive behaviors in social and economic scenarios~\cite{mou2024individual}.
Preliminary efforts focus on behavior simulation in sandbox environments.
For example, Park et al.~\cite{park2023generative} create an interactive simulation of complex human behaviors in a game environment, based on generative agents consisting of modules including planning and reacting, memory and retrieval, and active reflection.
These agents intend to emulate the real lives of humans by moving, working, interacting with the environment, and with each other through natural language.
Based on these agents, Humanoid Agents~\cite{wang2023humanoid} supplement the System 1 thinking process featuring intuitive and instantaneous desires, embracing basic needs for survival, emotions, and the closeness of social relationships, which further consolidate the authenticity of the simulation.
Similarly, inspired by the Theory of Needs~\cite{mcleod2007maslow}, D2A~\cite{wang2024simulating} introduces a desire-driven autonomous agent through an activity generation workflow constrained with a dynamic value system.

In addition to sandbox simulation, another line of research delves into the real-world environment.
For instance, Williams et al.~\cite{williams2023epidemic} adopt LLM agents for epidemic modeling, showcasing that the generative agents can mimic realistic quarantining and self-isolation behaviors during the COVID-19 pandemic.
EconAgent~\cite{li2023econagent} utilizes LLM agents to create a simulation of macroeconomic activities, highlighting labor supply and consumption agent decisions intertwined with dynamics of financial markets and government taxation.
AgentSociety~\cite{piao2025agentsociety} proposes a large-scale simulator, integrating LLM agents, a realistic society environment, and a powerful simulation engine.
The LLM agents are motivated by various psychological states and are associated with inter-dependencies among mobility, socioeconomic behaviors, which are situated in an integral societal environment.
However, the emulation of complex societal phenomena necessitates large-scale agent simulation, which results in challenges of excessive time and token costs.
To solve this issue, Chopra et al.~\cite{chopra2024limits} propose a scalable framework, AgentTorch, which creates archetypes representing unique agent characteristics and avoids the redundant simulation for similar agent behaviors.
A case study of emulating isolation and employment activities during the COVID-19 pandemic demonstrates the balance between the agency and simulation scale.
Furthermore, OpenCity~\cite{yan2024opencity} introduces a group-and-distill strategy that implements a prototype learning paradigm, discovering agents with similar profiles for batch simulation.

\subsubsection{Public Survey}
Public survey aims to efficiently replicate the feedback of agents in response to surveys or interviews, which can be leveraged to reflect and analyze opinions of the masses, potentially promoting urban policy refinement.
As an alternative to traditional travel survey methods, Bhandari et al.~\cite{bhandari2024urban} leverage LLM agents to generate surveyed mobility data, representing the daily movements of people, which avoid privacy concerns, participant noncompliance, and expensive time and labor costs.
The results reveal that LLMs fine-tuned on even limited real data can closely mimic the actual surveys.
Further, Park et al.~\cite{park2024generative} apply qualitative interviews to the agents to mirror their attitudes and behaviors during realistic lives, which are demonstrated to accurately replicate human participants' authentic responses.
In opposition to these works, Yang et al.~\cite{yang2024llm} investigate the voting behaviors of LLMs in response to various urban projects, which exhibit the limitations of LLM agents in simulating diverse and unbiased viewpoints.

\section{Trustworthiness of Urban LLM Agents}
Urban LLM agents interact with diverse data sources, manage critical physical infrastructures, and affect the lives of millions of residents in the city. Therefore, trustworthiness is a foundational requirement for their real-world deployment~\cite{sun2024trustllm}. In this section, we delve into the safety, fairness, accountability, and privacy of urban LLM agents, emphasizing the unique challenges posed in the context of cities.

\subsubsection*{5.1 Safety} 
Urban LLM agents operate in high-stakes environments where errors can quickly propagate through interconnected urban systems. Unlike traditional LLMs that process static text, these agents continuously engage with dynamic and multimodal data (\eg traffic flows, public safety alerts, and citizen reports) and are often embedded in real-time decision loops that affect physical infrastructure. Their partial embodiment in both digital and physical systems introduces distinct safety risks that must be addressed before real-world deployment. We identify three major categories of threats in the urban context.

First, \emph{adversarial attacks}~\cite{sun2022adversarial} can exploit spatial and temporal dependencies in city-scale data. For instance, injecting subtle noise into traffic flows near hospitals can mislead routing systems, potentially delaying emergency vehicles in surrounding
areas~\cite{liu2023robust,liu2024adversarial}. Similarly, falsified construction events during peak hours can trigger traffic re-routing that cascades through an entire road network. These attacks go beyond input-level perturbations and target the spatio-temporal reasoning that underpins agent behavior.
Second, \emph{backdoor attacks}~\cite{li2022backdoor} introduce hidden triggers into the agent's decision process. Many urban agents incorporate community feedback into their reasoning, such as planning documents or social media posts. Malicious actors may embed common urban terms like ``green corridor'' or ``historic preservation'' as semantic triggers to sway outcomes toward specific interests~\cite{alam2021comprehensive}. Because these terms appear legitimate, detecting such backdoors is difficult. However, their influence can distort long-term policy decisions or resource allocations, which could undermine public trust. Third, \emph{prompt injection attacks}~\cite{liu2023prompt} exploit the decentralized and asynchronous nature of urban information flows. Urban LLM agents process updates from multiple stakeholders, including utility providers, transport operators, and emergency responders. Malicious instructions embedded in routine inputs, such as a fake outage notice, may remain dormant until activated by specific conditions, like a weather event or system overload~\cite{liu2024formalizing}. These delayed effects are particularly dangerous in tightly coupled urban systems.

To mitigate these risks, several directions merit further research. One is to enforce spatial and temporal consistency constraints during training and fine-tuning, helping agents align with realistic urban behaviors~\cite{liu2024adversarial}. Another is to shift from single-point anomaly detection to monitoring collective behavior patterns across time and space. For example, coordinated anomalies across districts may signal system-wide manipulation~\cite{dobra2015spatiotemporal,sofuoglu2022gloss}. Finally, ensuring data provenance~\cite{alam2021comprehensive}, which tracks the origin, transformation, and credibility of inputs, can strengthen the trustworthiness of real-time data pipelines. Techniques such as source verification, cross-source redundancy checks, and automated logging can help detect and contain malicious modifications. Overall, these approaches emphasize the need for a system-level perspective on safety, which accounts for not just the model itself, but also its interaction with the broader urban infrastructure landscape.

\subsubsection*{5.2 Fairness}
As urban LLM agents are increasingly used to support decisions in transportation, urban planning, and public service delivery, fairness emerges as a central concern. Unlike general-purpose LLMs, which are primarily scrutinized for demographic bias in language outputs~\cite{liu2023trustworthy,sun2024trustllm}, urban agents interact with spatially distributed data and influence diverse communities. This introduces fairness challenges that span from individual-level disparities to systemic inequities across neighborhoods and cities~\cite{bousquet2018algorithmic, yan2021fairness}.

One key dimension is \emph{spatial fairness}, which arises when data-driven decisions consistently favor or neglect specific geographic areas. Urban LLM agents often rely on inputs such as GPS traces, service requests, and social media activity. However, these data sources are typically uneven~\cite{zheng2014urban}. Affluent districts may generate more civic feedback and ride-hailing data, while low-income or linguistically isolated communities remain underrepresented. This imbalance can lead to skewed resource allocation, where well-represented areas receive more services, and underserved areas are overlooked. Over time, such gaps may widen as data-poor regions become increasingly invisible to algorithmic systems, reinforcing spatial inequities already embedded in urban infrastructure. Another critical challenge is \emph{cross-stakeholder fairness}. Urban decisions often involve competing priorities among residents, businesses, and public agencies. For instance, a traffic control agent might reduce congestion for commuters by diverting traffic through residential zones, increasing noise and pollution. Residents may prioritize safety and air quality, while businesses may care more about logistics and accessibility. In such contexts, fairness cannot be reduced to a single metric. It should reflect trade-offs among competing values and the lived experiences of different groups. Importantly, these preferences can vary significantly across communities, further complicating the design of fair decision-making agents.

Addressing fairness in urban LLM agents requires methods that are sensitive to the spatial nature of cities and the multiplicity of voices involved. One promising approach is to compute geospatial equity scores~\cite{yan2019fairst}, which quantify how equitably an agent’s decisions affect different regions. Another is counterfactual analysis~\cite{lyu2023if}, where synthetic inputs from underrepresented areas are introduced to test whether the agent responds fairly under more balanced conditions. Simulation-based audits using urban digital twins~\cite{zhai2025heterogeneous} can reveal hidden biases. For example, a policy that seems efficient may in fact reduce access for low-demand neighborhoods, reinforcing long-term exclusion. Beyond algorithmic techniques, interactive interfaces could enable communities to express their fairness preferences, while negotiation-based frameworks may help agents mediate between conflicting goals. Ultimately, fairness in urban scenarios is not only a technical issue but a social and political one. Building trustworthy agents means embedding fairness into both their design and their interactions with the diverse communities they serve. 

\subsubsection*{5.3 Accountability}
Accountability is essential in urban environments, where LLM agents operate within complex ecosystems that involve physical infrastructure, digital systems, and diverse human stakeholders. These agents interact with sensors, APIs, other agents, and human operators, making it difficult to pinpoint responsibility when things go wrong. This is known as the many hands problem~\cite{thompson1980moral, dey2025towards}, \ie when multiple entities contribute to an outcome, it becomes unclear who is accountable for specific decisions or failures. 

Ensuring accountability in this context requires mechanisms that address both how decisions are made internally by the agent and how the agent interacts with the broader system. Internally, urban LLM agents should generate interpretable traces of their decision-making processes. These may include the sequence of tools invoked, intermediate goals formed during reasoning, confidence scores, and branching logic when handling uncertainty or exceptions. Such traces can help developers, auditors, or even end-users understand how specific outputs were generated. Externally, transparency should also extend to the agent's interactions with its environment. This includes how it acquires information, ranging from sensors, user reports, or other agents, and how its actions influence physical systems like traffic signals or emergency dispatch platforms. For instance, if an agent reallocates ambulances based on crowd-sourced incident reports, it should be possible to reconstruct which data sources contributed to the decision, what planning logic was used, and which subsystems executed the resulting actions.

We envision a hybrid accountability framework that combines multiple layers of traceability. At the reasoning level, symbolic planning diagrams~\cite{ghallab2004automated} can visualize goal decomposition and tool usage. At the causal level, recent work in causal auditing~\cite{sharkey2024causal} offers ways to trace how specific inputs lead to specific outcomes. System-wide logs, synchronized across modules, can capture contextual dependencies and temporal order, which are especially important in fast-changing urban settings. In some cases, post-hoc rationalization techniques~\cite{ribeiro2016should} can generate simplified explanations of agent behavior based on execution histories, aiding human understanding and debugging. Ultimately, meaningful accountability for urban LLM agents requires end-to-end visibility, ranging from data intake to actuation, and from single-agent behavior to system-level coordination. As these agents are entrusted with decisions that affect public services and infrastructure, transparency must be embedded not only at the model level but across the entire operational pipeline.

\subsubsection*{5.4 Privacy}
Urban LLM agents operate in environments rich with real-time data, much of which contains sensitive personal or community-level information~\cite{zhang2017security}. These include GPS traces, utility usage logs, service requests, and social media posts. Tasks such as ambulance routing~\cite{ji2019real} or housing evaluation~\cite{zhang2024meta} often require access to mobility patterns, health records, or financial data. If not carefully managed, such data can be exposed during both model training and inference, raising serious privacy concerns.

A key challenge is preventing information leakage through agent behavior. This can occur in several ways, \eg models may memorize sensitive details from training data, reveal private facts through reasoning outputs, or infer personal attributes by combining multiple inputs~\cite{sun2024trustllm}. These risks are especially prominent in urban scenarios, where data streams are continuous and tied to physical locations. In such cases, even indirect outputs like an agent's recommendation or routing decision can unintentionally expose personal information. Protecting privacy requires safeguards across multiple levels. At the data governance level, agents should operate under strict access controls based on task type, user role, and geographic or legal boundaries (\eg school districts or health zones). Role-based permissions and data use agreements can help limit unnecessary exposure. At the algorithmic level, techniques such as differential privacy~\cite{ahuja2023neural} and federated learning~\cite{yang2019federated} reduce risk by avoiding direct access to raw data and minimizing the retention of individual-level information. For long-term deployments, mechanisms like privacy budgets~\cite{steinke2022composition} can track cumulative exposure over time, ensuring that agents remain within acceptable usage limits.

Another critical challenge arises from the integration of heterogeneous data sources. Individually anonymized datasets can become identifiable when combined. For example, linking public transit logs with mobile app usage patterns could reveal someone’s identity or daily routines. To mitigate this, agents should include real-time input inspection modules that assess sensitivity levels and redact or block high-risk data before processing. Techniques such as privacy labeling, red-teaming tests~\cite{perez2022red}, and consent enforcement mechanisms can add additional layers of protection. Finally, privacy safeguards should be embedded into the broader system architecture. Urban LLM agents often support high-stakes services such as public health or emergency response, making regulatory compliance and public trust essential. This includes maintaining audit logs of data usage, supporting external reviews, and enabling rollback mechanisms when violations occur. Agents should also comply with local privacy laws. In highly sensitive domains, human-in-the-loop oversight may be necessary to ensure ethical and lawful use of personal data.

\section{Performance Evaluation}
In this section, we first review existing evaluation approaches and early efforts to build benchmarks tailored for urban LLM agents. Then, we outline directions toward next-generation benchmarks, aiming to enable more realistic, holistic, and trustworthy assessment of urban LLM agents in open-world environments.

\subsubsection*{6.1 Existing Evaluation Approaches}
In this part, we review existing evaluation methods of urban LLM agents from three aspects, including agent-centric behaviors, task-specific effectiveness, and spatio-temporal generalization.
\begin{itemize}[left=0pt]
    \item \textbf{Agent-centric evaluation}:
    Urban LLM agents often maintain continuous operation within dynamic, evolving environments populated by multiple stakeholders and external agents. Therefore, agent-centric evaluation should not only measure task outcomes but also assess the robustness, adaptability, and societal alignment of the agent’s decision-making process. These metrics focus on the intrinsic qualities of the agent’s behavior, independent of any specific downstream application. Existing works~\cite{lin2023mcu, chang2024survey} typically evaluate agent performance from three core dimensions:
    (1) \emph{Utility}: Measures the agent’s ability to achieve intended urban goals, such as minimizing traffic congestion or optimizing public service delivery, relative to predefined success criteria~\cite{chang2024survey}.
    (2) \emph{Efficiency}: Assesses the agent’s resource consumption (\eg computational cost, response time, and communication overhead) when operating under real-world urban constraints~\cite{shinn2023reflexion}.
    (3) \emph{Trustworthiness}: Evaluates the agent’s reliability, safety, ethical compliance, and transparency, especially in applications that directly affect public stakeholders~\cite{amodei2016concrete, chang2024survey}.
    \item \textbf{Task-specific evaluation}:
    It is crucial to assess how well urban LLM agents perform on specific urban tasks. We organize task-specific evaluation across five application domains:
    (1) \emph{Urban planning}:  
    (1.1) \textit{Regulatory compliance}: Measures how accurately the agent can generate or evaluate urban plans that align with legal, regulatory, and professional standards~\cite{zhu2024plangpt, zhengurbanplanbench}.  
    (1.2) \textit{Participatory alignment}: Evaluates the agent’s ability to balance diverse stakeholder needs in participatory planning scenarios~\cite{zhou2024large, singla2024adaptive}.
    (2) \emph{Transportation systems}:  
    (2.1) \textit{Goal-conditioned navigation}: Assesses the agent's ability to generate feasible and efficient travel plans under real-world constraints (\eg traffic congestion, road closures)~\cite{li2024research, fang2024travellm, ni2025tp}.  
    (2.2) \textit{Traffic signal optimization}: Measures improvements in throughput and congestion mitigation through agent-controlled traffic scheduling, typically evaluated by average wait time and queue length~\cite{wang2024llm, lai2023llmlight, yuan2025collmlight}.
    (2.3) \textit{Incident impact mitigation}: Evaluates the agent’s ability to predict, manage, and mitigate disruptions caused by traffic incidents such as accidents or protests~\cite{da2024open, pang2024illm}.
    (3) \emph{Environmental sustainability}:  
    (3.1) \textit{Energy load forecasting accuracy}: Measures the accuracy of energy demand predictions, critical for managing renewable resources (\eg wind, solar) and grid stability~\cite{jin2023time, qiu2024ef}. (3.2) \textit{Pollution event detection recall}: Assesses the recall and precision of the agent in identifying pollution anomalies such as wildfire smoke or air quality deterioration~\cite{fan2024llmair, verma2023generative}.
    (4) \emph{Public safety}:  
    (4.1) \textit{Crime risk explanation quality}: Evaluates the factual grounding and interpretability of crime hotspot analysis or risk assessments generated by the agent~\cite{rezacrix, wickramasekara2025exploring}.  
    (4.2) \textit{Emergency response coordination}: Measures the agent’s effectiveness in facilitating multi-agency coordination during urban emergencies such as natural disasters or public health crises~\cite{he2024multi, otal2024llm}.
    (5) \emph{Urban society}:  
    (5.1) \textit{Mobility pattern realism}: Assesses how closely the agent-generated human mobility trajectories match real-world patterns observed in cities~\cite{jiawei2024large, li2024more, du2024trajagent}.  
    (5.2) \textit{Socioeconomic behavior plausibility}: Evaluates the realism, diversity, and logical consistency of simulated social or economic behaviors modeled by urban LLM agents~\cite{park2023generative, wang2023humanoid, piao2025agentsociety}.
    \item \textbf{Spatio-temporal generalization}:  
    Spatio-temporal generalization is critical for deploying urban LLM agents in real-world settings, where unseen variations are inevitable and often profound. Following the perspective in~\cite{goodge2025spatio}, we discuss four critical dimensions of spatio-temporal generalization: (1) \emph{Domain generalization}:
    Measures the agent’s ability to transfer knowledge across different urban application domains, such as traffic management, public health, and environmental monitoring, while minimizing negative transfer effects~\cite{yuan2024unist}. (2) \emph{Spatial generalization}:  
    Assesses the agent’s adaptability across diverse geographic regions, accounting for variations in infrastructure, demographics, and urban design~\cite{jin2022selective,jin2023transferable}.
    (3) \emph{Temporal generalization}:  
    Evaluates the agent’s robustness over different time periods, including gradual changes such as seasonal shifts and sudden disruptions such as emergencies or policy shifts~\cite{han2023ieta}.
    (4) \emph{Scale generalization}: Tests the agent’s ability to operate effectively across varying spatial and temporal scales, from real-time street-level interventions to long-term metropolitan-scale planning~\cite{lai2023llmlight,zhu2024plangpt}. 
\end{itemize}

% \textbf{Emergent behavioral dimensions}:
% \begin{itemize}
% \item \textbf{Causal coherence}: Evaluates whether the agent’s decisions align with physical laws, temporal constraints, and regulatory frameworks. For instance, traffic rerouting plans violating real-world speed limits or flood zone restrictions would be penalized~\cite{shinn2023reflexion}.
% \item \textbf{Adaptation latency}: Measures the response time and quality of behavioral adjustments following environmental changes such as unexpected congestion, infrastructure failure, or natural disasters~\cite{chang2024survey}.
% \item \textbf{Failure recovery rate}: Captures the proportion of action failures that the agent successfully diagnoses and corrects without requiring human or centralized system intervention~\cite{chang2024survey, amodei2016concrete}.
% \item \textbf{Inter-agent alignment}: Assesses the agent’s ability to coordinate and align intentions with other agents, including both human collaborators and autonomous peers, to accomplish shared urban objectives~\cite{xu2024theagentcompany}.
% \end{itemize}
% This evaluation framework moves beyond static accuracy metrics and focuses on the real-world viability, resilience, and social compatibility of urban LLM agents operating in open, dynamic environments.  

\subsubsection*{6.2 Existing Benchmarks} 
Traditional benchmarks from natural language processing and urban computing fall short in capturing the multifaceted nature of urban tasks, which often involve multi-modality, spatio-temporal dependencies, interactive dynamics, and policy-sensitive decision-making. To address these gaps, several emerging benchmarks have begun to evaluate LLM agents under realistic urban conditions: (1) \emph{CityBench}~\cite{feng2024citybench}:  
A comprehensive benchmark that spans a range of urban tasks, from street-view image localization to traffic signal control. It supports both static and dynamic evaluations and enables comparative analysis across different cities and contexts. (2) \emph{STBench}~\cite{li2024stbench}:  Tailored for spatio-temporal understanding, STBench includes tasks that require fusing geographic and temporal knowledge, emphasizing time-sensitive predictions and reasoning over urban phenomena. (3) \emph{UrBench}~\cite{urbench2025}:  
Designed for evaluating multimodal LLMs in complex urban environments, UrBench contains 11.6K questions across 14 task types, including geolocalization, scene reasoning, and object understanding. It emphasizes multi-view reasoning and fine-grained urban scene comprehension. Evaluations on 21 multimodal LLMs show significant performance gaps compared to humans, particularly on cross-view reasoning tasks. (4) \emph{UrbanPlanBench}~\cite{urbanplanbench2025}:  Focused on professional urban planning tasks, this benchmark evaluates LLMs across fundamental principles, domain-specific expertise, and regulatory compliance. It highlights persistent challenges in aligning with human-level expectations, particularly in regulation interpretation. Accompanied by the large-scale PlanText SFT dataset, UrbanPlanBench aims to support the integration of LLMs into real-world urban planning workflows.

\subsubsection*{6.3 Towards Next-Generation Benchmarks}
Current benchmarks typically rely on curated datasets and single-turn evaluation tasks. While useful for tracking early-stage progress, they are limited in realism, interactivity, and task diversity, failing to evaluate the full operational cycle of urban LLM agents. Recent advances in urban digital twins~\cite{bettencourt2024recent} have shown the feasibility of building high-fidelity replicas of cities. Motivated by this, we envision a new generation of benchmarks grounded in simulation-oriented urban digital twins, which are detailed below:

\begin{itemize}[left=0pt]
    \item \textbf{Core components}:
    (1) \emph{Urban digital twin simulator}: A geospatially accurate and semantically rich virtual city that includes traffic systems, public infrastructure, zoning regulations, citizen behavior models, and real-time emulated sensor streams. This virtual environment provides the contextual foundation for all evaluation tasks; (2) \emph{Urban task generator}: A dynamic task generation engine that produces diverse and realistic tasks (\eg traffic signal control, ambulance redeployment). These tasks encode real-world constraints, multi-agent collaboration, and ethical considerations; (3) \emph{Multimodal perception interface}: Agents receive inputs from multiple modalities such as geospatial vectors, time series, trajectories, geo-tagged imagery, and natural language instructions. These inputs are perturbed with realistic noise, delay, and missing data to assess robustness; (4) \emph{Tool and API execution layer}: A functional interface that allows agents to interact with simulators (\eg SUMO, energy models), public APIs (\eg weather, routing, policy databases), and city-specific digital services to plan and execute actions; (5) \emph{Human-centered interaction module}: A participatory system simulating interactions with diverse urban stakeholders, including citizens, planners, and service providers. Agents are evaluated on their ability to explain decisions, accommodate feedback, and resolve goal conflicts using natural language.

    \item \textbf{End-to-end evaluation}:
    Unlike traditional benchmarks that focus on isolated tasks, this framework enables evaluation across the entire agentic loop, including (1) \emph{Perception}: Can the agent correctly interpret noisy, multimodal, and temporally dynamic urban input? (2) \emph{Reasoning \& planning}: Can it generate goal-oriented and contextually feasible plans using available tools and data? (3) \emph{Execution}: Can it coordinate its actions across systems and stakeholders, including humans and other agents? (4) \emph{Feedback adaptation}: Can it incorporate feedback, detect policy changes, and update its strategy over time? Additionally, we also recommend including urban-specific trustworthiness evaluation, such as (1) \emph{Disaster response}: Include event-driven disruptions (\eg floods, protests, power outages) to test agents' resilience and real-time coordination; (2) \emph{Spatial fairness}: Scenarios should evaluate equitable service delivery and fairness-aware decision-making, especially under demographic disparities; (3) \emph{Multi-agent negotiation}: Assess the ability to align conflicting goals among multiple stakeholders (\eg balancing delivery speed with pedestrian safety); (4) \emph{Regulatory complexity}: Tasks should involve multi-level governance constraints (\eg municipal vs. regional zoning laws) to test legal compliance. In summary, benchmarks based on urban digital twins offer a promising direction for evaluating urban LLM agents in realistic, high-stakes, and evolving settings. Such benchmarks support holistic assessment, ensuring agents are capable not only of solving abstract reasoning problems but also operating effectively in the complex and dynamic systems that characterize modern cities.
\end{itemize}

\section{Future Directions}
Despite the growing potential of urban LLM agents, unlocking their full capabilities requires addressing a series of fundamental research challenges. In this section, we discuss several key directions that can guide future work in this emerging field.

\subsubsection*{7.1 Urban Multimodal Fusion}
Urban environments generate vast amounts of data from heterogeneous and often asynchronous sources, ranging from mobility traces and traffic sensor records to geo-tagged images and social media posts. These data streams are typically noisy, incomplete, and misaligned across space and time. Existing multimodal fusion approaches usually assume clean and semantically aligned inputs~\cite{yin2024survey}, which limits their effectiveness in urban applications. Thus, future research could explore hierarchical and adaptive fusion frameworks that operate across multiple spatial scales (\eg street, neighborhood, city) and temporal resolutions (\eg real-time, daily, seasonal). Important open questions include: How can agents recover meaningful signals from fragmented or uncertain modalities? Can semantic, spatial, and temporal alignments be learned jointly in an end-to-end manner?

\subsubsection*{7.2 Rehearsal-Based Reasoning}
Urban LLM agents operate within highly dynamic and uncertain environments, where fully identifying causal mechanisms is often infeasible. Rehearsal learning~\cite{zhou2022rehearsal} provides a practical alternative by enabling agents to simulate possible interventions and outcomes without relying on complete causal models. Drawing inspiration from human mental rehearsal prior to decision-making, this approach allows agents to explore candidate interventions (\eg adjusting traffic lights, rerouting transit flows, or reallocating energy) by simulating counterfactual outcomes using pseudo-data or virtual environments. Future research could focus on developing rehearsal-based spatio-temporal reasoning frameworks, designing lightweight simulators tailored for decision rehearsal in urban systems, and investigating how agents can integrate considerations of cost, risk, and fairness into their hypothetical planning processes.

\subsubsection*{7.3 Toolchain Ecosystem Building}
Urban LLM agents rely heavily on interacting with external tools such as APIs, digital maps, control platforms, and public databases. However, urban systems are fragmented, with inconsistent protocols, access methods, and update frequencies. These issues pose significant challenges to reliable and scalable agent deployment. We advocate for building a modular and programmable toolchain ecosystem—analogous to an operating system—that can support real-time sensing, dynamic configuration, and secure action execution. Such a system should enable agents to automatically discover, invoke, and adapt to APIs from urban services, such as transit information or public records. It also calls for open standards and collaborative infrastructure to ensure interoperability between agents and urban systems.

\subsubsection*{7.4 Self-Evolution}
Urban environments are constantly evolving due to changes in infrastructure, population behavior, and policy. As a result, static models quickly become outdated. This requires mechanisms for detecting and responding to distribution shifts, especially those that are localized in space or time~\cite{jin2022selective,han2023ieta}. Urban LLM agents should be able to incorporate feedback from the environment, simulate possible futures, and selectively update their models without forgetting past knowledge. Future research could explore several fundamental problems: How can agents detect and localize concept drift over space and time? How can they adapt incrementally without forgetting previously acquired knowledge? And how can feedback from stakeholders be effectively incorporated to guide safe and socially aligned adaptation?

\subsubsection*{7.5 Multi-Agent Multi-Stakeholder Collaboration}
Urban decision-making often involves multiple agents and stakeholders, each with different goals, constraints, and information access. This complexity requires agents to negotiate, coordinate, and act under decentralized and possibly conflicting conditions. Unlike standard multi-agent reinforcement learning~\cite{zheng2024survey}, urban LLM agents should account for asymmetric information, fragmented governance, and real-world regulatory boundaries. Future work could explore methods for aligning local decisions with global system-level goals, balancing trade-offs (\eg minimizing regional traffic congestion while ensuring equitable service access), and building federated decision-making frameworks that respect urban governance structures. These efforts can support more equitable and efficient urban operations.

\subsubsection*{7.6 Agents as Autonomous Urban Scientists}
Beyond task execution, urban LLM agents have the potential to serve as collaborators in scientific discovery across urban domains. They could generate hypotheses, design simulations, and analyze results to derive new insights in urban planning, environmental science, or social policy. Inspired by recent progress in autonomous science~\cite{lu2024ai}, we envision agents that can support the entire workflows of urban scientific discovery. Recent pioneering work such as AutoUrbanCI~\cite{xia2025reimagining} demonstrates this potential by developing LLM-powered agents for urban causal inference. For example, such agents might explore how different land use patterns influence urban heat islands or how transit policies affect social equity. This direction elevates agents from task executors to collaborators in knowledge creation, thereby unlocking new frontiers in AI-driven urban science.

\subsubsection*{7.7 Value Alignment}
Urban decision-making is not only purely technical but also social, political, and value-laden. Decisions made by urban LLM agents, such as optimizing traffic flow or allocating resources, can have far-reaching impacts on equity, privacy, and sustainability. It is essential that urban LLM agents not only perform well technically but also align with the values of affected communities. We advocate for value-sensitive design approaches that incorporate participatory governance, preference learning, and ethical fine-tuning. Future work should address how agents can reason about conflicting values, respond to changing ethical or legal norms, and provide transparent explanations of their decisions. Developing reward models that balance technical performance with social goals is crucial for trustworthy deployment.

\section{Conclusion}
The advent of LLMs holds transformative potential for the development of next-generation intelligent cities, offering new opportunities to reshape urban operations and everyday life. In this paper, we focused on urban LLM agents, an emerging paradigm of LLM-powered systems. We systematically discussed their concepts, capabilities, applications, and future directions based on existing literature. Currently, research in this area is in the early stages. Urban LLM agents still lack sufficient domain-specific knowledge and spatio-temporal reasoning abilities. Their ability to support large-scale, cross-regional, and cross-task collaboration remains underexplored, leaving substantial room for improvement. Moreover, urban LLM agents represent a high-risk, high-demand, and high-value domain. Ensuring the utility, efficiency, and reliability in real-world deployments requires rigorous evaluation frameworks and close attention to trustworthiness issues. Looking forward, we advocate for interdisciplinary collaboration to build a full-stack ecosystem for urban LLM agents and to ensure their deployment reflects technical soundness, ethical responsibility, and societal value.

% \begin{acks}
% \end{acks}

\bibliographystyle{ACM-Reference-Format}
\bibliography{ref}


\begin{thebibliography}{233}

%%% the Web address.

\ifx \showCODEN    \undefined \def \showCODEN     #1{\unskip}     \fi
\ifx \showDOI      \undefined \def \showDOI       #1{#1}\fi
\ifx \showISBNx    \undefined \def \showISBNx     #1{\unskip}     \fi
\ifx \showISBNxiii \undefined \def \showISBNxiii  #1{\unskip}     \fi
\ifx \showISSN     \undefined \def \showISSN      #1{\unskip}     \fi
\ifx \showLCCN     \undefined \def \showLCCN      #1{\unskip}     \fi
\ifx \shownote     \undefined \def \shownote      #1{#1}          \fi
\ifx \showarticletitle \undefined \def \showarticletitle #1{#1}   \fi
\ifx \showURL      \undefined \def \showURL       {\relax}        \fi
% invisible to TeX
\providecommand\bibfield[2]{#2}
\providecommand\bibinfo[2]{#2}
\providecommand\natexlab[1]{#1}
\providecommand\showeprint[2][]{arXiv:#2}

\bibitem[Achiam et~al\mbox{.}(2023)]%
        {achiam2023gpt}
\bibfield{author}{\bibinfo{person}{Josh Achiam}, \bibinfo{person}{Steven Adler}, \bibinfo{person}{Sandhini Agarwal}, \bibinfo{person}{Lama Ahmad}, \bibinfo{person}{Ilge Akkaya}, \bibinfo{person}{Florencia~Leoni Aleman}, \bibinfo{person}{Diogo Almeida}, \bibinfo{person}{Janko Altenschmidt}, \bibinfo{person}{Sam Altman}, \bibinfo{person}{Shyamal Anadkat}, {et~al\mbox{.}}} \bibinfo{year}{2023}\natexlab{}.
\newblock \showarticletitle{Gpt-4 technical report}.
\newblock \bibinfo{journal}{\emph{arXiv preprint arXiv:2303.08774}} (\bibinfo{year}{2023}).
\newblock


\bibitem[Ahuja et~al\mbox{.}(2023)]%
        {ahuja2023neural}
\bibfield{author}{\bibinfo{person}{Ritesh Ahuja}, \bibinfo{person}{Sepanta Zeighami}, \bibinfo{person}{Gabriel Ghinita}, {and} \bibinfo{person}{Cyrus Shahabi}.} \bibinfo{year}{2023}\natexlab{}.
\newblock \showarticletitle{A neural approach to spatio-temporal data release with user-level differential privacy}.
\newblock \bibinfo{journal}{\emph{Proceedings of the ACM on Management of Data}} \bibinfo{volume}{1}, \bibinfo{number}{1} (\bibinfo{year}{2023}), \bibinfo{pages}{1--25}.
\newblock


\bibitem[Akinboyewa et~al\mbox{.}(2024)]%
        {akinboyewa2024gis}
\bibfield{author}{\bibinfo{person}{Temitope Akinboyewa}, \bibinfo{person}{Zhenlong Li}, \bibinfo{person}{Huan Ning}, {and} \bibinfo{person}{M~Naser Lessani}.} \bibinfo{year}{2024}\natexlab{}.
\newblock \showarticletitle{GIS copilot: Towards an autonomous GIS agent for spatial analysis}.
\newblock \bibinfo{journal}{\emph{arXiv preprint arXiv:2411.03205}} (\bibinfo{year}{2024}).
\newblock


\bibitem[Alam et~al\mbox{.}(2022)]%
        {alam2022survey}
\bibfield{author}{\bibinfo{person}{Md~Mahbub Alam}, \bibinfo{person}{Luis Torgo}, {and} \bibinfo{person}{Albert Bifet}.} \bibinfo{year}{2022}\natexlab{}.
\newblock \showarticletitle{A survey on spatio-temporal data analytics systems}.
\newblock \bibinfo{journal}{\emph{Comput. Surveys}} \bibinfo{volume}{54}, \bibinfo{number}{10s} (\bibinfo{year}{2022}), \bibinfo{pages}{1--38}.
\newblock


\bibitem[Alam and Wang(2021)]%
        {alam2021comprehensive}
\bibfield{author}{\bibinfo{person}{Md~Morshed Alam} {and} \bibinfo{person}{Weichao Wang}.} \bibinfo{year}{2021}\natexlab{}.
\newblock \showarticletitle{A comprehensive survey on the state-of-the-art data provenance approaches for security enforcement}.
\newblock \bibinfo{journal}{\emph{arXiv preprint arXiv:2107.01678}} (\bibinfo{year}{2021}).
\newblock


\bibitem[Alarabi et~al\mbox{.}(2018)]%
        {alarabi2018st}
\bibfield{author}{\bibinfo{person}{Louai Alarabi}, \bibinfo{person}{Mohamed~F Mokbel}, {and} \bibinfo{person}{Mashaal Musleh}.} \bibinfo{year}{2018}\natexlab{}.
\newblock \showarticletitle{St-hadoop: A mapreduce framework for spatio-temporal data}.
\newblock \bibinfo{journal}{\emph{GeoInformatica}}  \bibinfo{volume}{22} (\bibinfo{year}{2018}), \bibinfo{pages}{785--813}.
\newblock


\bibitem[Amodei et~al\mbox{.}(2016)]%
        {amodei2016concrete}
\bibfield{author}{\bibinfo{person}{Dario Amodei}, \bibinfo{person}{Chris Olah}, \bibinfo{person}{Jacob Steinhardt}, \bibinfo{person}{Paul Christiano}, \bibinfo{person}{John Schulman}, {and} \bibinfo{person}{Dan Man{\'e}}.} \bibinfo{year}{2016}\natexlab{}.
\newblock \showarticletitle{Concrete problems in AI safety}.
\newblock \bibinfo{journal}{\emph{arXiv preprint arXiv:1606.06565}} (\bibinfo{year}{2016}).
\newblock


\bibitem[Balsebre et~al\mbox{.}(2024)]%
        {balsebre2024lamp}
\bibfield{author}{\bibinfo{person}{Pasquale Balsebre}, \bibinfo{person}{Weiming Huang}, {and} \bibinfo{person}{Gao Cong}.} \bibinfo{year}{2024}\natexlab{}.
\newblock \showarticletitle{LAMP: A language model on the map}.
\newblock \bibinfo{journal}{\emph{arXiv preprint arXiv:2403.09059}} (\bibinfo{year}{2024}).
\newblock


\bibitem[Beneduce et~al\mbox{.}(2024)]%
        {beneduce2024large}
\bibfield{author}{\bibinfo{person}{Ciro Beneduce}, \bibinfo{person}{Bruno Lepri}, {and} \bibinfo{person}{Massimiliano Luca}.} \bibinfo{year}{2024}\natexlab{}.
\newblock \showarticletitle{Large language models are zero-shot next location predictors}.
\newblock \bibinfo{journal}{\emph{arXiv preprint arXiv:2405.20962}} (\bibinfo{year}{2024}).
\newblock


\bibitem[Bettencourt(2024)]%
        {bettencourt2024recent}
\bibfield{author}{\bibinfo{person}{Lu{\'\i}s~MA Bettencourt}.} \bibinfo{year}{2024}\natexlab{}.
\newblock \showarticletitle{Recent achievements and conceptual challenges for urban digital twins}.
\newblock \bibinfo{journal}{\emph{Nature Computational Science}} \bibinfo{volume}{4}, \bibinfo{number}{3} (\bibinfo{year}{2024}), \bibinfo{pages}{150--153}.
\newblock


\bibitem[Bhandari et~al\mbox{.}(2024)]%
        {bhandari2024urban}
\bibfield{author}{\bibinfo{person}{Prabin Bhandari}, \bibinfo{person}{Antonios Anastasopoulos}, {and} \bibinfo{person}{Dieter Pfoser}.} \bibinfo{year}{2024}\natexlab{}.
\newblock \showarticletitle{Urban mobility assessment using llms}. In \bibinfo{booktitle}{\emph{Proceedings of the 32nd ACM International Conference on Advances in Geographic Information Systems}}. \bibinfo{pages}{67--79}.
\newblock


\bibitem[Bibri and Krogstie(2017)]%
        {bibri2017smart}
\bibfield{author}{\bibinfo{person}{Simon~Elias Bibri} {and} \bibinfo{person}{John Krogstie}.} \bibinfo{year}{2017}\natexlab{}.
\newblock \showarticletitle{Smart sustainable cities of the future: An extensive interdisciplinary literature review}.
\newblock \bibinfo{journal}{\emph{Sustainable cities and society}}  \bibinfo{volume}{31} (\bibinfo{year}{2017}), \bibinfo{pages}{183--212}.
\newblock


\bibitem[Bousquet(2018)]%
        {bousquet2018algorithmic}
\bibfield{author}{\bibinfo{person}{Chris Bousquet}.} \bibinfo{year}{2018}\natexlab{}.
\newblock \showarticletitle{Algorithmic fairness: Tackling bias in city algorithms}.
\newblock \bibinfo{journal}{\emph{Data-Smart City Solutions}} (\bibinfo{year}{2018}).
\newblock


\bibitem[Chang et~al\mbox{.}(2024)]%
        {chang2024survey}
\bibfield{author}{\bibinfo{person}{Yupeng Chang}, \bibinfo{person}{Xu Wang}, \bibinfo{person}{Jindong Wang}, \bibinfo{person}{Yuan Wu}, \bibinfo{person}{Linyi Yang}, \bibinfo{person}{Kaijie Zhu}, \bibinfo{person}{Hao Chen}, \bibinfo{person}{Xiaoyuan Yi}, \bibinfo{person}{Cunxiang Wang}, \bibinfo{person}{Yidong Wang}, {et~al\mbox{.}}} \bibinfo{year}{2024}\natexlab{}.
\newblock \showarticletitle{A survey on evaluation of large language models}.
\newblock \bibinfo{journal}{\emph{ACM transactions on intelligent systems and technology}} \bibinfo{volume}{15}, \bibinfo{number}{3} (\bibinfo{year}{2024}), \bibinfo{pages}{1--45}.
\newblock


\bibitem[Chen et~al\mbox{.}(2024a)]%
        {chen2024travelagent}
\bibfield{author}{\bibinfo{person}{Aili Chen}, \bibinfo{person}{Xuyang Ge}, \bibinfo{person}{Ziquan Fu}, \bibinfo{person}{Yanghua Xiao}, {and} \bibinfo{person}{Jiangjie Chen}.} \bibinfo{year}{2024}\natexlab{a}.
\newblock \showarticletitle{TravelAgent: An AI assistant for personalized travel planning}.
\newblock \bibinfo{journal}{\emph{arXiv preprint arXiv:2409.08069}} (\bibinfo{year}{2024}).
\newblock


\bibitem[Chen et~al\mbox{.}(2024b)]%
        {chen2024spatialvlm}
\bibfield{author}{\bibinfo{person}{Boyuan Chen}, \bibinfo{person}{Zhuo Xu}, \bibinfo{person}{Sean Kirmani}, \bibinfo{person}{Brain Ichter}, \bibinfo{person}{Dorsa Sadigh}, \bibinfo{person}{Leonidas Guibas}, {and} \bibinfo{person}{Fei Xia}.} \bibinfo{year}{2024}\natexlab{b}.
\newblock \showarticletitle{Spatialvlm: Endowing vision-language models with spatial reasoning capabilities}. In \bibinfo{booktitle}{\emph{Proceedings of the IEEE/CVF Conference on Computer Vision and Pattern Recognition}}. \bibinfo{pages}{14455--14465}.
\newblock


\bibitem[Chen et~al\mbox{.}(2021)]%
        {chen2021location}
\bibfield{author}{\bibinfo{person}{Zhida Chen}, \bibinfo{person}{Lisi Chen}, \bibinfo{person}{Gao Cong}, {and} \bibinfo{person}{Christian~S Jensen}.} \bibinfo{year}{2021}\natexlab{}.
\newblock \showarticletitle{Location-and keyword-based querying of geo-textual data: a survey}.
\newblock \bibinfo{journal}{\emph{The VLDB Journal}} \bibinfo{volume}{30}, \bibinfo{number}{4} (\bibinfo{year}{2021}), \bibinfo{pages}{603--640}.
\newblock


\bibitem[Cheng et~al\mbox{.}(2025)]%
        {cheng2025large}
\bibfield{author}{\bibinfo{person}{Yuheng Cheng}, \bibinfo{person}{Huan Zhao}, \bibinfo{person}{Xiyuan Zhou}, \bibinfo{person}{Junhua Zhao}, \bibinfo{person}{Yuji Cao}, \bibinfo{person}{Chao Yang}, {and} \bibinfo{person}{Xinlei Cai}.} \bibinfo{year}{2025}\natexlab{}.
\newblock \showarticletitle{A large language model for advanced power dispatch}.
\newblock \bibinfo{journal}{\emph{Scientific Reports}} \bibinfo{volume}{15}, \bibinfo{number}{1} (\bibinfo{year}{2025}), \bibinfo{pages}{8925}.
\newblock


\bibitem[Chopra et~al\mbox{.}(2024)]%
        {chopra2024limits}
\bibfield{author}{\bibinfo{person}{Ayush Chopra}, \bibinfo{person}{Shashank Kumar}, \bibinfo{person}{Nurullah Giray-Kuru}, \bibinfo{person}{Ramesh Raskar}, {and} \bibinfo{person}{Arnau Quera-Bofarull}.} \bibinfo{year}{2024}\natexlab{}.
\newblock \showarticletitle{On the limits of agency in agent-based models}.
\newblock \bibinfo{journal}{\emph{arXiv preprint arXiv:2409.10568}} (\bibinfo{year}{2024}).
\newblock


\bibitem[Christou et~al\mbox{.}(2025)]%
        {christou2025llm4distreconfig}
\bibfield{author}{\bibinfo{person}{Panayiotis Christou}, \bibinfo{person}{Md~Zahidul Islam}, \bibinfo{person}{Yuzhang Lin}, {and} \bibinfo{person}{Jingwei Xiong}.} \bibinfo{year}{2025}\natexlab{}.
\newblock \showarticletitle{LLM4DistReconfig: A Fine-tuned Large Language Model for Power Distribution Network Reconfiguration}.
\newblock \bibinfo{journal}{\emph{arXiv preprint arXiv:2501.14960}} (\bibinfo{year}{2025}).
\newblock


\bibitem[Chu et~al\mbox{.}(2024)]%
        {chu2024timebench}
\bibfield{author}{\bibinfo{person}{Zheng Chu}, \bibinfo{person}{Jingchang Chen}, \bibinfo{person}{Qianglong Chen}, \bibinfo{person}{Weijiang Yu}, \bibinfo{person}{Haotian Wang}, \bibinfo{person}{Ming Liu}, {and} \bibinfo{person}{Bing Qin}.} \bibinfo{year}{2024}\natexlab{}.
\newblock \showarticletitle{TimeBench: A Comprehensive Evaluation of Temporal Reasoning Abilities in Large Language Models}. In \bibinfo{booktitle}{\emph{Proceedings of the 62nd Annual Meeting of the Association for Computational Linguistics (Volume 1: Long Papers)}}. \bibinfo{pages}{1204--1228}.
\newblock


\bibitem[Cucchiara et~al\mbox{.}(2000)]%
        {cucchiara2000image}
\bibfield{author}{\bibinfo{person}{Rita Cucchiara}, \bibinfo{person}{Massimo Piccardi}, {and} \bibinfo{person}{Paola Mello}.} \bibinfo{year}{2000}\natexlab{}.
\newblock \showarticletitle{Image analysis and rule-based reasoning for a traffic monitoring system}.
\newblock \bibinfo{journal}{\emph{IEEE transactions on intelligent transportation systems}} \bibinfo{volume}{1}, \bibinfo{number}{2} (\bibinfo{year}{2000}), \bibinfo{pages}{119--130}.
\newblock


\bibitem[Da et~al\mbox{.}(2023)]%
        {da2023llm}
\bibfield{author}{\bibinfo{person}{Longchao Da}, \bibinfo{person}{Minchiuan Gao}, \bibinfo{person}{Hao Mei}, {and} \bibinfo{person}{Hua Wei}.} \bibinfo{year}{2023}\natexlab{}.
\newblock \showarticletitle{Llm powered sim-to-real transfer for traffic signal control}.
\newblock \bibinfo{journal}{\emph{arXiv preprint arXiv:2308.14284}} (\bibinfo{year}{2023}).
\newblock


\bibitem[Da et~al\mbox{.}(2024)]%
        {da2024open}
\bibfield{author}{\bibinfo{person}{Longchao Da}, \bibinfo{person}{Kuanru Liou}, \bibinfo{person}{Tiejin Chen}, \bibinfo{person}{Xuesong Zhou}, \bibinfo{person}{Xiangyong Luo}, \bibinfo{person}{Yezhou Yang}, {and} \bibinfo{person}{Hua Wei}.} \bibinfo{year}{2024}\natexlab{}.
\newblock \showarticletitle{Open-ti: Open traffic intelligence with augmented language model}.
\newblock \bibinfo{journal}{\emph{International Journal of Machine Learning and Cybernetics}} \bibinfo{volume}{15}, \bibinfo{number}{10} (\bibinfo{year}{2024}), \bibinfo{pages}{4761--4786}.
\newblock


\bibitem[Devunuri and Lehe(2024)]%
        {devunuri2024transitgpt}
\bibfield{author}{\bibinfo{person}{Saipraneeth Devunuri} {and} \bibinfo{person}{Lewis Lehe}.} \bibinfo{year}{2024}\natexlab{}.
\newblock \showarticletitle{TransitGPT: A Generative AI-based framework for interacting with GTFS data using Large Language Models}.
\newblock \bibinfo{journal}{\emph{arXiv preprint arXiv:2412.06831}} (\bibinfo{year}{2024}).
\newblock


\bibitem[Devunuri and Lehe(2025)]%
        {devunuri2025transitgpt}
\bibfield{author}{\bibinfo{person}{Saipraneeth Devunuri} {and} \bibinfo{person}{Lewis Lehe}.} \bibinfo{year}{2025}\natexlab{}.
\newblock \showarticletitle{TransitGPT: A Generative AI-based framework for interacting with GTFS data using Large Language Models}.
\newblock \bibinfo{journal}{\emph{Public Transport}} (\bibinfo{year}{2025}), \bibinfo{pages}{1--27}.
\newblock


\bibitem[Dey et~al\mbox{.}(2025)]%
        {dey2025towards}
\bibfield{author}{\bibinfo{person}{Suvodip Dey}, \bibinfo{person}{Yi-Jyun Sun}, \bibinfo{person}{Gokhan Tur}, {and} \bibinfo{person}{Dilek Hakkani-Tur}.} \bibinfo{year}{2025}\natexlab{}.
\newblock \showarticletitle{Towards Preventing Overreliance on Task-Oriented Conversational AI Through Accountability Modeling}.
\newblock \bibinfo{journal}{\emph{arXiv preprint arXiv:2501.10316}} (\bibinfo{year}{2025}).
\newblock


\bibitem[Dobra et~al\mbox{.}(2015)]%
        {dobra2015spatiotemporal}
\bibfield{author}{\bibinfo{person}{Adrian Dobra}, \bibinfo{person}{Nathalie~E Williams}, {and} \bibinfo{person}{Nathan Eagle}.} \bibinfo{year}{2015}\natexlab{}.
\newblock \showarticletitle{Spatiotemporal detection of unusual human population behavior using mobile phone data}.
\newblock \bibinfo{journal}{\emph{PloS one}} \bibinfo{volume}{10}, \bibinfo{number}{3} (\bibinfo{year}{2015}), \bibinfo{pages}{e0120449}.
\newblock


\bibitem[Du et~al\mbox{.}(2024)]%
        {du2024trajagent}
\bibfield{author}{\bibinfo{person}{Yuwei Du}, \bibinfo{person}{Jie Feng}, \bibinfo{person}{Jie Zhao}, {and} \bibinfo{person}{Yong Li}.} \bibinfo{year}{2024}\natexlab{}.
\newblock \showarticletitle{TrajAgent: An Agent Framework for Unified Trajectory Modelling}.
\newblock \bibinfo{journal}{\emph{arXiv preprint arXiv:2410.20445}} (\bibinfo{year}{2024}).
\newblock


\bibitem[Durmus et~al\mbox{.}(2024)]%
        {durmus2024role}
\bibfield{author}{\bibinfo{person}{Dilan Durmus}, \bibinfo{person}{Alberto Giretti}, \bibinfo{person}{Ori Ashkenazi}, \bibinfo{person}{Alessandro Carbonari}, \bibinfo{person}{Shabtai Isaac}, {et~al\mbox{.}}} \bibinfo{year}{2024}\natexlab{}.
\newblock \showarticletitle{The Role of Large Language Models for Decision Support in Fire Safety Planning}.
\newblock \bibinfo{journal}{\emph{PROCEEDINGS OF THE... ISARC}} (\bibinfo{year}{2024}), \bibinfo{pages}{339--346}.
\newblock


\bibitem[Fan et~al\mbox{.}(2025)]%
        {fan2025invisible}
\bibfield{author}{\bibinfo{person}{Bingbing Fan}, \bibinfo{person}{Lin Chen}, \bibinfo{person}{Songwei Li}, \bibinfo{person}{Jian Yuan}, \bibinfo{person}{Fengli Xu}, \bibinfo{person}{Pan Hui}, {and} \bibinfo{person}{Yong Li}.} \bibinfo{year}{2025}\natexlab{}.
\newblock \showarticletitle{Invisible Walls in Cities: Leveraging Large Language Models to Predict Urban Segregation Experience with Social Media Content}.
\newblock \bibinfo{journal}{\emph{arXiv preprint arXiv:2503.04773}} (\bibinfo{year}{2025}).
\newblock


\bibitem[Fan et~al\mbox{.}(2024)]%
        {fan2024llmair}
\bibfield{author}{\bibinfo{person}{Jinxiao Fan}, \bibinfo{person}{Haolin Chu}, \bibinfo{person}{Liang Liu}, {and} \bibinfo{person}{Huadong Ma}.} \bibinfo{year}{2024}\natexlab{}.
\newblock \showarticletitle{LLMAir: Adaptive Reprogramming Large Language Model for Air Quality Prediction}. In \bibinfo{booktitle}{\emph{2024 IEEE 30th International Conference on Parallel and Distributed Systems (ICPADS)}}. IEEE, \bibinfo{pages}{423--430}.
\newblock


\bibitem[Fan et~al\mbox{.}(2023)]%
        {fan2023dish}
\bibfield{author}{\bibinfo{person}{Wei Fan}, \bibinfo{person}{Pengyang Wang}, \bibinfo{person}{Dongkun Wang}, \bibinfo{person}{Dongjie Wang}, \bibinfo{person}{Yuanchun Zhou}, {and} \bibinfo{person}{Yanjie Fu}.} \bibinfo{year}{2023}\natexlab{}.
\newblock \showarticletitle{Dish-ts: a general paradigm for alleviating distribution shift in time series forecasting}. In \bibinfo{booktitle}{\emph{Proceedings of the AAAI conference on artificial intelligence}}, Vol.~\bibinfo{volume}{37}. \bibinfo{pages}{7522--7529}.
\newblock


\bibitem[Fan et~al\mbox{.}(2022)]%
        {fan2022depts}
\bibfield{author}{\bibinfo{person}{Wei Fan}, \bibinfo{person}{Shun Zheng}, \bibinfo{person}{Xiaohan Yi}, \bibinfo{person}{Wei Cao}, \bibinfo{person}{Yanjie Fu}, \bibinfo{person}{Jiang Bian}, {and} \bibinfo{person}{Tie-Yan Liu}.} \bibinfo{year}{2022}\natexlab{}.
\newblock \showarticletitle{DEPTS: Deep expansion learning for periodic time series forecasting}.
\newblock \bibinfo{journal}{\emph{arXiv preprint arXiv:2203.07681}} (\bibinfo{year}{2022}).
\newblock


\bibitem[Fang et~al\mbox{.}(2024)]%
        {fang2024travellm}
\bibfield{author}{\bibinfo{person}{Bowen Fang}, \bibinfo{person}{Zixiao Yang}, \bibinfo{person}{Shukai Wang}, {and} \bibinfo{person}{Xuan Di}.} \bibinfo{year}{2024}\natexlab{}.
\newblock \showarticletitle{TraveLLM: Could you plan my new public transit route in face of a network disruption?}
\newblock \bibinfo{journal}{\emph{arXiv preprint arXiv:2407.14926}} (\bibinfo{year}{2024}).
\newblock


\bibitem[Fang et~al\mbox{.}(2025)]%
        {fang2025unraveling}
\bibfield{author}{\bibinfo{person}{Yuchen Fang}, \bibinfo{person}{Hao Miao}, \bibinfo{person}{Yuxuan Liang}, \bibinfo{person}{Liwei Deng}, \bibinfo{person}{Yue Cui}, \bibinfo{person}{Ximu Zeng}, \bibinfo{person}{Yuyang Xia}, \bibinfo{person}{Yan Zhao}, \bibinfo{person}{Torben~Bach Pedersen}, \bibinfo{person}{Christian~S Jensen}, {et~al\mbox{.}}} \bibinfo{year}{2025}\natexlab{}.
\newblock \showarticletitle{Unraveling Spatio-Temporal Foundation Models via the Pipeline Lens: A Comprehensive Review}.
\newblock \bibinfo{journal}{\emph{arXiv preprint arXiv:2506.01364}} (\bibinfo{year}{2025}).
\newblock


\bibitem[Fatemi et~al\mbox{.}(2024)]%
        {fatemi2024test}
\bibfield{author}{\bibinfo{person}{Bahare Fatemi}, \bibinfo{person}{Mehran Kazemi}, \bibinfo{person}{Anton Tsitsulin}, \bibinfo{person}{Karishma Malkan}, \bibinfo{person}{Jinyeong Yim}, \bibinfo{person}{John Palowitch}, \bibinfo{person}{Sungyong Seo}, \bibinfo{person}{Jonathan Halcrow}, {and} \bibinfo{person}{Bryan Perozzi}.} \bibinfo{year}{2024}\natexlab{}.
\newblock \showarticletitle{Test of time: A benchmark for evaluating llms on temporal reasoning}.
\newblock \bibinfo{journal}{\emph{arXiv preprint arXiv:2406.09170}} (\bibinfo{year}{2024}).
\newblock


\bibitem[Feng et~al\mbox{.}(2024a)]%
        {feng2024citygpt}
\bibfield{author}{\bibinfo{person}{Jie Feng}, \bibinfo{person}{Yuwei Du}, \bibinfo{person}{Tianhui Liu}, \bibinfo{person}{Siqi Guo}, \bibinfo{person}{Yuming Lin}, {and} \bibinfo{person}{Yong Li}.} \bibinfo{year}{2024}\natexlab{a}.
\newblock \showarticletitle{Citygpt: Empowering urban spatial cognition of large language models}.
\newblock \bibinfo{journal}{\emph{arXiv preprint arXiv:2406.13948}} (\bibinfo{year}{2024}).
\newblock


\bibitem[Feng et~al\mbox{.}(2024b)]%
        {feng2024agentmove}
\bibfield{author}{\bibinfo{person}{Jie Feng}, \bibinfo{person}{Yuwei Du}, \bibinfo{person}{Jie Zhao}, {and} \bibinfo{person}{Yong Li}.} \bibinfo{year}{2024}\natexlab{b}.
\newblock \showarticletitle{Agentmove: Predicting human mobility anywhere using large language model based agentic framework}.
\newblock \bibinfo{journal}{\emph{arXiv preprint arXiv:2408.13986}} (\bibinfo{year}{2024}).
\newblock


\bibitem[Feng et~al\mbox{.}(2024c)]%
        {feng2024citybench}
\bibfield{author}{\bibinfo{person}{Jie Feng}, \bibinfo{person}{Jun Zhang}, \bibinfo{person}{Junbo Yan}, \bibinfo{person}{Xin Zhang}, \bibinfo{person}{Tianjian Ouyang}, \bibinfo{person}{Tianhui Liu}, \bibinfo{person}{Yuwei Du}, \bibinfo{person}{Siqi Guo}, {and} \bibinfo{person}{Yong Li}.} \bibinfo{year}{2024}\natexlab{c}.
\newblock \showarticletitle{Citybench: Evaluating the capabilities of large language model as world model}.
\newblock \bibinfo{journal}{\emph{arXiv preprint arXiv:2406.13945}} (\bibinfo{year}{2024}).
\newblock


\bibitem[Gao et~al\mbox{.}(2025)]%
        {gao2025instructor}
\bibfield{author}{\bibinfo{person}{Kyle Gao}, \bibinfo{person}{Dening Lu}, \bibinfo{person}{Liangzhi Li}, \bibinfo{person}{Nan Chen}, \bibinfo{person}{Hongjie He}, \bibinfo{person}{Linlin Xu}, {and} \bibinfo{person}{Jonathan Li}.} \bibinfo{year}{2025}\natexlab{}.
\newblock \showarticletitle{Instructor-Worker Large Language Model System for Policy Recommendation: a Case Study on Air Quality Analysis of the January 2025 Los Angeles Wildfires}.
\newblock \bibinfo{journal}{\emph{arXiv preprint arXiv:2503.00566}} (\bibinfo{year}{2025}).
\newblock


\bibitem[Ghallab et~al\mbox{.}(2004)]%
        {ghallab2004automated}
\bibfield{author}{\bibinfo{person}{Malik Ghallab}, \bibinfo{person}{Dana Nau}, {and} \bibinfo{person}{Paolo Traverso}.} \bibinfo{year}{2004}\natexlab{}.
\newblock \bibinfo{booktitle}{\emph{Automated Planning: theory and practice}}.
\newblock \bibinfo{publisher}{Elsevier}.
\newblock


\bibitem[Goodge et~al\mbox{.}(2025)]%
        {goodge2025spatio}
\bibfield{author}{\bibinfo{person}{Adam Goodge}, \bibinfo{person}{Wee~Siong Ng}, \bibinfo{person}{Bryan Hooi}, {and} \bibinfo{person}{See~Kiong Ng}.} \bibinfo{year}{2025}\natexlab{}.
\newblock \showarticletitle{Spatio-Temporal Foundation Models: Vision, Challenges, and Opportunities}.
\newblock \bibinfo{journal}{\emph{arXiv preprint arXiv:2501.09045}} (\bibinfo{year}{2025}).
\newblock


\bibitem[Gruver et~al\mbox{.}(2023)]%
        {gruver2023large}
\bibfield{author}{\bibinfo{person}{Nate Gruver}, \bibinfo{person}{Marc Finzi}, \bibinfo{person}{Shikai Qiu}, {and} \bibinfo{person}{Andrew~G Wilson}.} \bibinfo{year}{2023}\natexlab{}.
\newblock \showarticletitle{Large language models are zero-shot time series forecasters}.
\newblock \bibinfo{journal}{\emph{Advances in Neural Information Processing Systems}}  \bibinfo{volume}{36} (\bibinfo{year}{2023}), \bibinfo{pages}{19622--19635}.
\newblock


\bibitem[Guan et~al\mbox{.}(2024)]%
        {guan2024citygpt}
\bibfield{author}{\bibinfo{person}{Qinghua Guan}, \bibinfo{person}{Jinhui Ouyang}, \bibinfo{person}{Di Wu}, {and} \bibinfo{person}{Weiren Yu}.} \bibinfo{year}{2024}\natexlab{}.
\newblock \showarticletitle{CityGPT: Towards Urban IoT Learning, Analysis and Interaction with Multi-Agent System}.
\newblock \bibinfo{journal}{\emph{arXiv preprint arXiv:2405.14691}} (\bibinfo{year}{2024}).
\newblock


\bibitem[Guo et~al\mbox{.}(2025)]%
        {guo2025deepseek}
\bibfield{author}{\bibinfo{person}{Daya Guo}, \bibinfo{person}{Dejian Yang}, \bibinfo{person}{Haowei Zhang}, \bibinfo{person}{Junxiao Song}, \bibinfo{person}{Ruoyu Zhang}, \bibinfo{person}{Runxin Xu}, \bibinfo{person}{Qihao Zhu}, \bibinfo{person}{Shirong Ma}, \bibinfo{person}{Peiyi Wang}, \bibinfo{person}{Xiao Bi}, {et~al\mbox{.}}} \bibinfo{year}{2025}\natexlab{}.
\newblock \showarticletitle{Deepseek-r1: Incentivizing reasoning capability in llms via reinforcement learning}.
\newblock \bibinfo{journal}{\emph{arXiv preprint arXiv:2501.12948}} (\bibinfo{year}{2025}).
\newblock


\bibitem[Guo et~al\mbox{.}(2024)]%
        {guo2024large}
\bibfield{author}{\bibinfo{person}{Taicheng Guo}, \bibinfo{person}{Xiuying Chen}, \bibinfo{person}{Yaqi Wang}, \bibinfo{person}{Ruidi Chang}, \bibinfo{person}{Shichao Pei}, \bibinfo{person}{Nitesh~V Chawla}, \bibinfo{person}{Olaf Wiest}, {and} \bibinfo{person}{Xiangliang Zhang}.} \bibinfo{year}{2024}\natexlab{}.
\newblock \showarticletitle{Large language model based multi-agents: A survey of progress and challenges}.
\newblock \bibinfo{journal}{\emph{arXiv preprint arXiv:2402.01680}} (\bibinfo{year}{2024}).
\newblock


\bibitem[Han et~al\mbox{.}(2023)]%
        {han2023ieta}
\bibfield{author}{\bibinfo{person}{Jindong Han}, \bibinfo{person}{Hao Liu}, \bibinfo{person}{Shui Liu}, \bibinfo{person}{Xi Chen}, \bibinfo{person}{Naiqiang Tan}, \bibinfo{person}{Hua Chai}, {and} \bibinfo{person}{Hui Xiong}.} \bibinfo{year}{2023}\natexlab{}.
\newblock \showarticletitle{iETA: A robust and scalable incremental learning framework for time-of-arrival estimation}. In \bibinfo{booktitle}{\emph{Proceedings of the 29th ACM SIGKDD Conference on Knowledge Discovery and Data Mining}}. \bibinfo{pages}{4100--4111}.
\newblock


\bibitem[Han et~al\mbox{.}(2022)]%
        {han2022semi}
\bibfield{author}{\bibinfo{person}{Jindong Han}, \bibinfo{person}{Hao Liu}, \bibinfo{person}{Haoyi Xiong}, {and} \bibinfo{person}{Jing Yang}.} \bibinfo{year}{2022}\natexlab{}.
\newblock \showarticletitle{Semi-supervised air quality forecasting via self-supervised hierarchical graph neural network}.
\newblock \bibinfo{journal}{\emph{IEEE Transactions on Knowledge and Data Engineering}} \bibinfo{volume}{35}, \bibinfo{number}{5} (\bibinfo{year}{2022}), \bibinfo{pages}{5230--5243}.
\newblock


\bibitem[Han et~al\mbox{.}(2021)]%
        {han2021joint}
\bibfield{author}{\bibinfo{person}{Jindong Han}, \bibinfo{person}{Hao Liu}, \bibinfo{person}{Hengshu Zhu}, \bibinfo{person}{Hui Xiong}, {and} \bibinfo{person}{Dejing Dou}.} \bibinfo{year}{2021}\natexlab{}.
\newblock \showarticletitle{Joint air quality and weather prediction based on multi-adversarial spatiotemporal networks}. In \bibinfo{booktitle}{\emph{Proceedings of the AAAI Conference on Artificial Intelligence}}, Vol.~\bibinfo{volume}{35}. \bibinfo{pages}{4081--4089}.
\newblock


\bibitem[Han et~al\mbox{.}(2024a)]%
        {han2024bigst}
\bibfield{author}{\bibinfo{person}{Jindong Han}, \bibinfo{person}{Weijia Zhang}, \bibinfo{person}{Hao Liu}, \bibinfo{person}{Tao Tao}, \bibinfo{person}{Naiqiang Tan}, {and} \bibinfo{person}{Hui Xiong}.} \bibinfo{year}{2024}\natexlab{a}.
\newblock \showarticletitle{Bigst: Linear complexity spatio-temporal graph neural network for traffic forecasting on large-scale road networks}.
\newblock \bibinfo{journal}{\emph{Proceedings of the VLDB Endowment}} \bibinfo{volume}{17}, \bibinfo{number}{5} (\bibinfo{year}{2024}), \bibinfo{pages}{1081--1090}.
\newblock


\bibitem[Han et~al\mbox{.}(2024c)]%
        {han2024enhanced}
\bibfield{author}{\bibinfo{person}{Jin Han}, \bibinfo{person}{Zhe Zheng}, \bibinfo{person}{Xin-Zheng Lu}, \bibinfo{person}{Ke-Yin Chen}, {and} \bibinfo{person}{Jia-Rui Lin}.} \bibinfo{year}{2024}\natexlab{c}.
\newblock \showarticletitle{Enhanced earthquake impact analysis based on social media texts via large language model}.
\newblock \bibinfo{journal}{\emph{International Journal of Disaster Risk Reduction}}  \bibinfo{volume}{109} (\bibinfo{year}{2024}), \bibinfo{pages}{104574}.
\newblock


\bibitem[Han and Kim(1990)]%
        {han1990essas}
\bibfield{author}{\bibinfo{person}{Sang-Yun Han} {and} \bibinfo{person}{Tschangho~John Kim}.} \bibinfo{year}{1990}\natexlab{}.
\newblock \showarticletitle{ESSAS: expert system for site analysis and selection}.
\newblock In \bibinfo{booktitle}{\emph{Expert systems: applications to urban planning}}. \bibinfo{publisher}{Springer}, \bibinfo{pages}{145--158}.
\newblock


\bibitem[Han et~al\mbox{.}(2024b)]%
        {han2024gpt}
\bibfield{author}{\bibinfo{person}{Xiao Han}, \bibinfo{person}{Zijian Zhang}, \bibinfo{person}{Xiangyu Zhao}, \bibinfo{person}{Guojiang Shen}, \bibinfo{person}{Xiangjie Kong}, \bibinfo{person}{Xuetao Wei}, \bibinfo{person}{Liqiang Nie}, {and} \bibinfo{person}{Jieping Ye}.} \bibinfo{year}{2024}\natexlab{b}.
\newblock \showarticletitle{GPT-Augmented Reinforcement Learning with Intelligent Control for Vehicle Dispatching}.
\newblock \bibinfo{journal}{\emph{arXiv preprint arXiv:2408.10286}} (\bibinfo{year}{2024}).
\newblock


\bibitem[Hao et~al\mbox{.}(2024)]%
        {hao2024urbanvlp}
\bibfield{author}{\bibinfo{person}{Xixuan Hao}, \bibinfo{person}{Wei Chen}, \bibinfo{person}{Yibo Yan}, \bibinfo{person}{Siru Zhong}, \bibinfo{person}{Kun Wang}, \bibinfo{person}{Qingsong Wen}, {and} \bibinfo{person}{Yuxuan Liang}.} \bibinfo{year}{2024}\natexlab{}.
\newblock \showarticletitle{UrbanVLP: Multi-Granularity Vision-Language Pretraining for Urban Socioeconomic Indicator Prediction}.
\newblock \bibinfo{journal}{\emph{arXiv preprint arXiv:2403.16831}} (\bibinfo{year}{2024}).
\newblock


\bibitem[He et~al\mbox{.}(2024)]%
        {he2024multi}
\bibfield{author}{\bibinfo{person}{Lyulong He}, \bibinfo{person}{Hongyuan Zhang}, \bibinfo{person}{Kunxiao Liu}, {and} \bibinfo{person}{Xi Wu}.} \bibinfo{year}{2024}\natexlab{}.
\newblock \showarticletitle{Multi-Agent Enhanced Complex Decision-Making Support Framework: An Urban Emergency Case Study}. In \bibinfo{booktitle}{\emph{2024 6th International Conference on Frontier Technologies of Information and Computer (ICFTIC)}}. IEEE, \bibinfo{pages}{413--419}.
\newblock


\bibitem[Hou et~al\mbox{.}(2025)]%
        {hou2025urban}
\bibfield{author}{\bibinfo{person}{Ce Hou}, \bibinfo{person}{Fan Zhang}, \bibinfo{person}{Yong Li}, \bibinfo{person}{Haifeng Li}, \bibinfo{person}{Gengchen Mai}, \bibinfo{person}{Yuhao Kang}, \bibinfo{person}{Ling Yao}, \bibinfo{person}{Wenhao Yu}, \bibinfo{person}{Yao Yao}, \bibinfo{person}{Song Gao}, {et~al\mbox{.}}} \bibinfo{year}{2025}\natexlab{}.
\newblock \showarticletitle{Urban sensing in the era of large language models}.
\newblock \bibinfo{journal}{\emph{The Innovation}} \bibinfo{volume}{6}, \bibinfo{number}{1} (\bibinfo{year}{2025}).
\newblock


\bibitem[Hu et~al\mbox{.}(2020)]%
        {hu2020deep}
\bibfield{author}{\bibinfo{person}{Chengyu Hu}, \bibinfo{person}{Junyi Cai}, \bibinfo{person}{Deze Zeng}, \bibinfo{person}{Xuesong Yan}, \bibinfo{person}{Wenyin Gong}, {and} \bibinfo{person}{Ling Wang}.} \bibinfo{year}{2020}\natexlab{}.
\newblock \showarticletitle{Deep reinforcement learning based valve scheduling for pollution isolation in water distribution network}.
\newblock \bibinfo{journal}{\emph{Math. Biosci. Eng}} \bibinfo{volume}{17}, \bibinfo{number}{1} (\bibinfo{year}{2020}), \bibinfo{pages}{105--122}.
\newblock


\bibitem[Hu et~al\mbox{.}(2024)]%
        {hu2024chain}
\bibfield{author}{\bibinfo{person}{Hanxu Hu}, \bibinfo{person}{Hongyuan Lu}, \bibinfo{person}{Huajian Zhang}, \bibinfo{person}{Yun-Ze Song}, \bibinfo{person}{Wai Lam}, {and} \bibinfo{person}{Yue Zhang}.} \bibinfo{year}{2024}\natexlab{}.
\newblock \showarticletitle{Chain-of-Symbol Prompting For Spatial Reasoning in Large Language Models}. In \bibinfo{booktitle}{\emph{First Conference on Language Modeling}}.
\newblock


\bibitem[Huang(2024)]%
        {huang2024enhancing}
\bibfield{author}{\bibinfo{person}{Xiannan Huang}.} \bibinfo{year}{2024}\natexlab{}.
\newblock \showarticletitle{Enhancing traffic prediction with textual data using large language models}.
\newblock \bibinfo{journal}{\emph{arXiv preprint arXiv:2405.06719}} (\bibinfo{year}{2024}).
\newblock


\bibitem[Hughes et~al\mbox{.}(2015)]%
        {hughes2015geomesa}
\bibfield{author}{\bibinfo{person}{James~N Hughes}, \bibinfo{person}{Andrew Annex}, \bibinfo{person}{Christopher~N Eichelberger}, \bibinfo{person}{Anthony Fox}, \bibinfo{person}{Andrew Hulbert}, {and} \bibinfo{person}{Michael Ronquest}.} \bibinfo{year}{2015}\natexlab{}.
\newblock \showarticletitle{Geomesa: a distributed architecture for spatio-temporal fusion}. In \bibinfo{booktitle}{\emph{Geospatial informatics, fusion, and motion video analytics V}}, Vol.~\bibinfo{volume}{9473}. SPIE, \bibinfo{pages}{128--140}.
\newblock


\bibitem[Hunt et~al\mbox{.}(1982)]%
        {hunt1982scoot}
\bibfield{author}{\bibinfo{person}{PB Hunt}, \bibinfo{person}{DI Robertson}, \bibinfo{person}{RD Bretherton}, {and} \bibinfo{person}{M~Cr Royle}.} \bibinfo{year}{1982}\natexlab{}.
\newblock \showarticletitle{The SCOOT on-line traffic signal optimisation technique}.
\newblock \bibinfo{journal}{\emph{Traffic Engineering \& Control}} \bibinfo{volume}{23}, \bibinfo{number}{4} (\bibinfo{year}{1982}).
\newblock


\bibitem[Ito et~al\mbox{.}(2024)]%
        {ito2024understanding}
\bibfield{author}{\bibinfo{person}{Koichi Ito}, \bibinfo{person}{Yuhao Kang}, \bibinfo{person}{Ye Zhang}, \bibinfo{person}{Fan Zhang}, {and} \bibinfo{person}{Filip Biljecki}.} \bibinfo{year}{2024}\natexlab{}.
\newblock \showarticletitle{Understanding urban perception with visual data: A systematic review}.
\newblock \bibinfo{journal}{\emph{Cities}}  \bibinfo{volume}{152} (\bibinfo{year}{2024}), \bibinfo{pages}{105169}.
\newblock


\bibitem[Ji et~al\mbox{.}(2019)]%
        {ji2019real}
\bibfield{author}{\bibinfo{person}{Shenggong Ji}, \bibinfo{person}{Yu Zheng}, \bibinfo{person}{Wenjun Wang}, {and} \bibinfo{person}{Tianrui Li}.} \bibinfo{year}{2019}\natexlab{}.
\newblock \showarticletitle{Real-time ambulance redeployment: A data-driven approach}.
\newblock \bibinfo{journal}{\emph{IEEE Transactions on Knowledge and Data Engineering}} \bibinfo{volume}{32}, \bibinfo{number}{11} (\bibinfo{year}{2019}), \bibinfo{pages}{2213--2226}.
\newblock


\bibitem[Jiang et~al\mbox{.}(2024)]%
        {jiang2024urbanllm}
\bibfield{author}{\bibinfo{person}{Yue Jiang}, \bibinfo{person}{Qin Chao}, \bibinfo{person}{Yile Chen}, \bibinfo{person}{Xiucheng Li}, \bibinfo{person}{Shuai Liu}, {and} \bibinfo{person}{Gao Cong}.} \bibinfo{year}{2024}\natexlab{}.
\newblock \showarticletitle{Urbanllm: Autonomous urban activity planning and management with large language models}.
\newblock \bibinfo{journal}{\emph{arXiv preprint arXiv:2406.12360}} (\bibinfo{year}{2024}).
\newblock


\bibitem[Jiao et~al\mbox{.}(2024)]%
        {jiao2024city}
\bibfield{author}{\bibinfo{person}{Zihao Jiao}, \bibinfo{person}{Mengyi Sha}, \bibinfo{person}{Haoyu Zhang}, \bibinfo{person}{Xinyu Jiang}, {and} \bibinfo{person}{Wei Qi}.} \bibinfo{year}{2024}\natexlab{}.
\newblock \showarticletitle{City-LEO: Toward Transparent City Management Using LLM with End-to-End Optimization}.
\newblock \bibinfo{journal}{\emph{arXiv preprint arXiv:2406.10958}} (\bibinfo{year}{2024}).
\newblock


\bibitem[JIAWEI et~al\mbox{.}(2024)]%
        {jiawei2024large}
\bibfield{author}{\bibinfo{person}{WANG JIAWEI}, \bibinfo{person}{Renhe Jiang}, \bibinfo{person}{Chuang Yang}, \bibinfo{person}{Zengqing Wu}, \bibinfo{person}{Ryosuke Shibasaki}, \bibinfo{person}{Noboru Koshizuka}, \bibinfo{person}{Chuan Xiao}, {et~al\mbox{.}}} \bibinfo{year}{2024}\natexlab{}.
\newblock \showarticletitle{Large language models as urban residents: An llm agent framework for personal mobility generation}.
\newblock \bibinfo{journal}{\emph{Advances in Neural Information Processing Systems}}  \bibinfo{volume}{37} (\bibinfo{year}{2024}), \bibinfo{pages}{124547--124574}.
\newblock


\bibitem[Jin et~al\mbox{.}(2023b)]%
        {jin2023spatio}
\bibfield{author}{\bibinfo{person}{Guangyin Jin}, \bibinfo{person}{Yuxuan Liang}, \bibinfo{person}{Yuchen Fang}, \bibinfo{person}{Zezhi Shao}, \bibinfo{person}{Jincai Huang}, \bibinfo{person}{Junbo Zhang}, {and} \bibinfo{person}{Yu Zheng}.} \bibinfo{year}{2023}\natexlab{b}.
\newblock \showarticletitle{Spatio-temporal graph neural networks for predictive learning in urban computing: A survey}.
\newblock \bibinfo{journal}{\emph{IEEE Transactions on Knowledge and Data Engineering}} \bibinfo{volume}{36}, \bibinfo{number}{10} (\bibinfo{year}{2023}), \bibinfo{pages}{5388--5408}.
\newblock


\bibitem[Jin et~al\mbox{.}(2024)]%
        {jin2024time}
\bibfield{author}{\bibinfo{person}{Ming Jin}, \bibinfo{person}{Shiyu Wang}, \bibinfo{person}{Lintao Ma}, \bibinfo{person}{Zhixuan Chu}, \bibinfo{person}{James Zhang}, \bibinfo{person}{Xiaoming Shi}, \bibinfo{person}{Pin-Yu Chen}, \bibinfo{person}{Yuxuan Liang}, \bibinfo{person}{Yuan-fang Li}, \bibinfo{person}{Shirui Pan}, {et~al\mbox{.}}} \bibinfo{year}{2024}\natexlab{}.
\newblock \showarticletitle{Time-LLM: Time Series Forecasting by Reprogramming Large Language Models}. In \bibinfo{booktitle}{\emph{International Conference on Learning Representations}}.
\newblock


\bibitem[Jin et~al\mbox{.}(2023c)]%
        {jin2023time}
\bibfield{author}{\bibinfo{person}{Ming Jin}, \bibinfo{person}{Shiyu Wang}, \bibinfo{person}{Lintao Ma}, \bibinfo{person}{Zhixuan Chu}, \bibinfo{person}{James~Y Zhang}, \bibinfo{person}{Xiaoming Shi}, \bibinfo{person}{Pin-Yu Chen}, \bibinfo{person}{Yuxuan Liang}, \bibinfo{person}{Yuan-Fang Li}, \bibinfo{person}{Shirui Pan}, {et~al\mbox{.}}} \bibinfo{year}{2023}\natexlab{c}.
\newblock \showarticletitle{Time-llm: Time series forecasting by reprogramming large language models}.
\newblock \bibinfo{journal}{\emph{arXiv preprint arXiv:2310.01728}} (\bibinfo{year}{2023}).
\newblock


\bibitem[Jin et~al\mbox{.}(2022)]%
        {jin2022selective}
\bibfield{author}{\bibinfo{person}{Yilun Jin}, \bibinfo{person}{Kai Chen}, {and} \bibinfo{person}{Qiang Yang}.} \bibinfo{year}{2022}\natexlab{}.
\newblock \showarticletitle{Selective cross-city transfer learning for traffic prediction via source city region re-weighting}. In \bibinfo{booktitle}{\emph{Proceedings of the 28th ACM SIGKDD Conference on Knowledge Discovery and Data Mining}}. \bibinfo{pages}{731--741}.
\newblock


\bibitem[Jin et~al\mbox{.}(2023a)]%
        {jin2023transferable}
\bibfield{author}{\bibinfo{person}{Yilun Jin}, \bibinfo{person}{Kai Chen}, {and} \bibinfo{person}{Qiang Yang}.} \bibinfo{year}{2023}\natexlab{a}.
\newblock \showarticletitle{Transferable graph structure learning for graph-based traffic forecasting across cities}. In \bibinfo{booktitle}{\emph{Proceedings of the 29th ACM SIGKDD conference on knowledge discovery and data mining}}. \bibinfo{pages}{1032--1043}.
\newblock


\bibitem[Jin and Ma(2024)]%
        {jin2024large}
\bibfield{author}{\bibinfo{person}{Yuping Jin} {and} \bibinfo{person}{Jun Ma}.} \bibinfo{year}{2024}\natexlab{}.
\newblock \showarticletitle{Large language model as parking planning agent in the context of mixed period of autonomous vehicles and Human-Driven vehicles}.
\newblock \bibinfo{journal}{\emph{Sustainable Cities and Society}}  \bibinfo{volume}{117} (\bibinfo{year}{2024}), \bibinfo{pages}{105940}.
\newblock


\bibitem[Ju et~al\mbox{.}(2025)]%
        {ju2025trajllm}
\bibfield{author}{\bibinfo{person}{Chenlu Ju}, \bibinfo{person}{Jiaxin Liu}, \bibinfo{person}{Shobhit Sinha}, \bibinfo{person}{Hao Xue}, {and} \bibinfo{person}{Flora Salim}.} \bibinfo{year}{2025}\natexlab{}.
\newblock \showarticletitle{TrajLLM: A Modular LLM-Enhanced Agent-Based Framework for Realistic Human Trajectory Simulation}.
\newblock \bibinfo{journal}{\emph{arXiv preprint arXiv:2502.18712}} (\bibinfo{year}{2025}).
\newblock


\bibitem[Kaddour et~al\mbox{.}(2023)]%
        {kaddour2023challenges}
\bibfield{author}{\bibinfo{person}{Jean Kaddour}, \bibinfo{person}{Joshua Harris}, \bibinfo{person}{Maximilian Mozes}, \bibinfo{person}{Herbie Bradley}, \bibinfo{person}{Roberta Raileanu}, {and} \bibinfo{person}{Robert McHardy}.} \bibinfo{year}{2023}\natexlab{}.
\newblock \showarticletitle{Challenges and applications of large language models}.
\newblock \bibinfo{journal}{\emph{arXiv preprint arXiv:2307.10169}} (\bibinfo{year}{2023}).
\newblock


\bibitem[Kalyuzhnaya et~al\mbox{.}(2025)]%
        {kalyuzhnaya2025llm}
\bibfield{author}{\bibinfo{person}{Anna Kalyuzhnaya}, \bibinfo{person}{Sergey Mityagin}, \bibinfo{person}{Elizaveta Lutsenko}, \bibinfo{person}{Andrey Getmanov}, \bibinfo{person}{Yaroslav Aksenkin}, \bibinfo{person}{Kamil Fatkhiev}, \bibinfo{person}{Kirill Fedorin}, \bibinfo{person}{Nikolay~O Nikitin}, \bibinfo{person}{Natalia Chichkova}, \bibinfo{person}{Vladimir Vorona}, {et~al\mbox{.}}} \bibinfo{year}{2025}\natexlab{}.
\newblock \showarticletitle{LLM Agents for Smart City Management: Enhancing Decision Support Through Multi-Agent AI Systems.}
\newblock \bibinfo{journal}{\emph{Smart Cities (2624-6511)}} \bibinfo{volume}{8}, \bibinfo{number}{1} (\bibinfo{year}{2025}).
\newblock


\bibitem[Koonce et~al\mbox{.}(2008)]%
        {koonce2008traffic}
\bibfield{author}{\bibinfo{person}{Peter Koonce} {et~al\mbox{.}}} \bibinfo{year}{2008}\natexlab{}.
\newblock \bibinfo{booktitle}{\emph{Traffic signal timing manual}}.
\newblock \bibinfo{type}{{T}echnical {R}eport}. \bibinfo{institution}{United States. Federal Highway Administration}.
\newblock


\bibitem[Kumar et~al\mbox{.}(2025)]%
        {kumar2025llm}
\bibfield{author}{\bibinfo{person}{Komal Kumar}, \bibinfo{person}{Tajamul Ashraf}, \bibinfo{person}{Omkar Thawakar}, \bibinfo{person}{Rao~Muhammad Anwer}, \bibinfo{person}{Hisham Cholakkal}, \bibinfo{person}{Mubarak Shah}, \bibinfo{person}{Ming-Hsuan Yang}, \bibinfo{person}{Phillip~HS Torr}, \bibinfo{person}{Salman Khan}, {and} \bibinfo{person}{Fahad~Shahbaz Khan}.} \bibinfo{year}{2025}\natexlab{}.
\newblock \showarticletitle{Llm post-training: A deep dive into reasoning large language models}.
\newblock \bibinfo{journal}{\emph{arXiv preprint arXiv:2502.21321}} (\bibinfo{year}{2025}).
\newblock


\bibitem[Lai et~al\mbox{.}(2023)]%
        {lai2023llmlight}
\bibfield{author}{\bibinfo{person}{Siqi Lai}, \bibinfo{person}{Zhao Xu}, \bibinfo{person}{Weijia Zhang}, \bibinfo{person}{Hao Liu}, {and} \bibinfo{person}{Hui Xiong}.} \bibinfo{year}{2023}\natexlab{}.
\newblock \showarticletitle{LLMLight: Large Language Models as Traffic Signal Control Agents}.
\newblock \bibinfo{journal}{\emph{arXiv preprint arXiv:2312.16044}} (\bibinfo{year}{2023}).
\newblock


\bibitem[Lewis et~al\mbox{.}(2020)]%
        {lewis2020retrieval}
\bibfield{author}{\bibinfo{person}{Patrick Lewis}, \bibinfo{person}{Ethan Perez}, \bibinfo{person}{Aleksandra Piktus}, \bibinfo{person}{Fabio Petroni}, \bibinfo{person}{Vladimir Karpukhin}, \bibinfo{person}{Naman Goyal}, \bibinfo{person}{Heinrich K{\"u}ttler}, \bibinfo{person}{Mike Lewis}, \bibinfo{person}{Wen-tau Yih}, \bibinfo{person}{Tim Rockt{\"a}schel}, {et~al\mbox{.}}} \bibinfo{year}{2020}\natexlab{}.
\newblock \showarticletitle{Retrieval-augmented generation for knowledge-intensive nlp tasks}.
\newblock \bibinfo{journal}{\emph{Advances in neural information processing systems}}  \bibinfo{volume}{33} (\bibinfo{year}{2020}), \bibinfo{pages}{9459--9474}.
\newblock


\bibitem[Li et~al\mbox{.}(2024h)]%
        {li2024research}
\bibfield{author}{\bibinfo{person}{Bohang Li}, \bibinfo{person}{Kai Zhang}, \bibinfo{person}{Yiping Sun}, {and} \bibinfo{person}{Jianke Zou}.} \bibinfo{year}{2024}\natexlab{h}.
\newblock \showarticletitle{Research on travel route planning optimization based on large language model}. In \bibinfo{booktitle}{\emph{2024 6th International Conference on Data-driven Optimization of Complex Systems (DOCS)}}. IEEE, \bibinfo{pages}{352--357}.
\newblock


\bibitem[Li et~al\mbox{.}(2024b)]%
        {li2024advancing}
\bibfield{author}{\bibinfo{person}{Fangjun Li}, \bibinfo{person}{David~C Hogg}, {and} \bibinfo{person}{Anthony~G Cohn}.} \bibinfo{year}{2024}\natexlab{b}.
\newblock \showarticletitle{Advancing spatial reasoning in large language models: An in-depth evaluation and enhancement using the stepgame benchmark}. In \bibinfo{booktitle}{\emph{Proceedings of the AAAI Conference on Artificial Intelligence}}, Vol.~\bibinfo{volume}{38}. \bibinfo{pages}{18500--18507}.
\newblock


\bibitem[Li et~al\mbox{.}(2024c)]%
        {li2024reframing}
\bibfield{author}{\bibinfo{person}{Fangjun Li}, \bibinfo{person}{David~C Hogg}, {and} \bibinfo{person}{Anthony~G Cohn}.} \bibinfo{year}{2024}\natexlab{c}.
\newblock \showarticletitle{Reframing spatial reasoning evaluation in language models: a real-world simulation benchmark for qualitative reasoning}. In \bibinfo{booktitle}{\emph{Proceedings of the Thirty-Third International Joint Conference on Artificial Intelligence}}. \bibinfo{pages}{6342--6349}.
\newblock


\bibitem[Li et~al\mbox{.}(2023b)]%
        {li2023camel}
\bibfield{author}{\bibinfo{person}{Guohao Li}, \bibinfo{person}{Hasan Hammoud}, \bibinfo{person}{Hani Itani}, \bibinfo{person}{Dmitrii Khizbullin}, {and} \bibinfo{person}{Bernard Ghanem}.} \bibinfo{year}{2023}\natexlab{b}.
\newblock \showarticletitle{Camel: Communicative agents for" mind" exploration of large language model society}.
\newblock \bibinfo{journal}{\emph{Advances in Neural Information Processing Systems}}  \bibinfo{volume}{36} (\bibinfo{year}{2023}), \bibinfo{pages}{51991--52008}.
\newblock


\bibitem[Li et~al\mbox{.}(2023a)]%
        {li2023econagent}
\bibfield{author}{\bibinfo{person}{Nian Li}, \bibinfo{person}{Chen Gao}, \bibinfo{person}{Mingyu Li}, \bibinfo{person}{Yong Li}, {and} \bibinfo{person}{Qingmin Liao}.} \bibinfo{year}{2023}\natexlab{a}.
\newblock \showarticletitle{Econagent: large language model-empowered agents for simulating macroeconomic activities}.
\newblock \bibinfo{journal}{\emph{arXiv preprint arXiv:2310.10436}} (\bibinfo{year}{2023}).
\newblock


\bibitem[Li et~al\mbox{.}(2024a)]%
        {li2024large}
\bibfield{author}{\bibinfo{person}{Peibo Li}, \bibinfo{person}{Maarten de Rijke}, \bibinfo{person}{Hao Xue}, \bibinfo{person}{Shuang Ao}, \bibinfo{person}{Yang Song}, {and} \bibinfo{person}{Flora~D Salim}.} \bibinfo{year}{2024}\natexlab{a}.
\newblock \showarticletitle{Large language models for next point-of-interest recommendation}. In \bibinfo{booktitle}{\emph{Proceedings of the 47th International ACM SIGIR Conference on Research and Development in Information Retrieval}}. \bibinfo{pages}{1463--1472}.
\newblock


\bibitem[Li et~al\mbox{.}(2020)]%
        {li2020just}
\bibfield{author}{\bibinfo{person}{Ruiyuan Li}, \bibinfo{person}{Huajun He}, \bibinfo{person}{Rubin Wang}, \bibinfo{person}{Yuchuan Huang}, \bibinfo{person}{Junwen Liu}, \bibinfo{person}{Sijie Ruan}, \bibinfo{person}{Tianfu He}, \bibinfo{person}{Jie Bao}, {and} \bibinfo{person}{Yu Zheng}.} \bibinfo{year}{2020}\natexlab{}.
\newblock \showarticletitle{Just: Jd urban spatio-temporal data engine}. In \bibinfo{booktitle}{\emph{2020 IEEE 36th International Conference on Data Engineering (ICDE)}}. IEEE, \bibinfo{pages}{1558--1569}.
\newblock


\bibitem[Li et~al\mbox{.}(2024g)]%
        {li2024stbench}
\bibfield{author}{\bibinfo{person}{Wenbin Li}, \bibinfo{person}{Di Yao}, \bibinfo{person}{Ruibo Zhao}, \bibinfo{person}{Wenjie Chen}, \bibinfo{person}{Zijie Xu}, \bibinfo{person}{Chengxue Luo}, \bibinfo{person}{Chang Gong}, \bibinfo{person}{Quanliang Jing}, \bibinfo{person}{Haining Tan}, {and} \bibinfo{person}{Jingping Bi}.} \bibinfo{year}{2024}\natexlab{g}.
\newblock \showarticletitle{STBench: Assessing the ability of large language models in spatio-temporal analysis}.
\newblock \bibinfo{journal}{\emph{arXiv preprint arXiv:2406.19065}} (\bibinfo{year}{2024}).
\newblock


\bibitem[Li et~al\mbox{.}(2024d)]%
        {li2024more}
\bibfield{author}{\bibinfo{person}{Xuchuan Li}, \bibinfo{person}{Fei Huang}, \bibinfo{person}{Jianrong Lv}, \bibinfo{person}{Zhixiong Xiao}, \bibinfo{person}{Guolong Li}, {and} \bibinfo{person}{Yang Yue}.} \bibinfo{year}{2024}\natexlab{d}.
\newblock \showarticletitle{Be more real: Travel diary generation using llm agents and individual profiles}.
\newblock \bibinfo{journal}{\emph{arXiv preprint arXiv:2407.18932}} (\bibinfo{year}{2024}).
\newblock


\bibitem[Li et~al\mbox{.}(2022)]%
        {li2022backdoor}
\bibfield{author}{\bibinfo{person}{Yiming Li}, \bibinfo{person}{Yong Jiang}, \bibinfo{person}{Zhifeng Li}, {and} \bibinfo{person}{Shu-Tao Xia}.} \bibinfo{year}{2022}\natexlab{}.
\newblock \showarticletitle{Backdoor learning: A survey}.
\newblock \bibinfo{journal}{\emph{IEEE transactions on neural networks and learning systems}} \bibinfo{volume}{35}, \bibinfo{number}{1} (\bibinfo{year}{2022}), \bibinfo{pages}{5--22}.
\newblock


\bibitem[Li et~al\mbox{.}(2024e)]%
        {li2024personal}
\bibfield{author}{\bibinfo{person}{Yuanchun Li}, \bibinfo{person}{Hao Wen}, \bibinfo{person}{Weijun Wang}, \bibinfo{person}{Xiangyu Li}, \bibinfo{person}{Yizhen Yuan}, \bibinfo{person}{Guohong Liu}, \bibinfo{person}{Jiacheng Liu}, \bibinfo{person}{Wenxing Xu}, \bibinfo{person}{Xiang Wang}, \bibinfo{person}{Yi Sun}, {et~al\mbox{.}}} \bibinfo{year}{2024}\natexlab{e}.
\newblock \showarticletitle{Personal llm agents: Insights and survey about the capability, efficiency and security}.
\newblock \bibinfo{journal}{\emph{arXiv preprint arXiv:2401.05459}} (\bibinfo{year}{2024}).
\newblock


\bibitem[Li et~al\mbox{.}(2025)]%
        {li2025urban}
\bibfield{author}{\bibinfo{person}{Zhonghang Li}, \bibinfo{person}{Lianghao Xia}, \bibinfo{person}{Xubin Ren}, \bibinfo{person}{Jiabin Tang}, \bibinfo{person}{Tianyi Chen}, \bibinfo{person}{Yong Xu}, {and} \bibinfo{person}{Chao Huang}.} \bibinfo{year}{2025}\natexlab{}.
\newblock \showarticletitle{Urban Computing in the Era of Large Language Models}.
\newblock \bibinfo{journal}{\emph{arXiv preprint arXiv:2504.02009}} (\bibinfo{year}{2025}).
\newblock


\bibitem[Li et~al\mbox{.}(2024f)]%
        {li2024urbangpt}
\bibfield{author}{\bibinfo{person}{Zhonghang Li}, \bibinfo{person}{Lianghao Xia}, \bibinfo{person}{Jiabin Tang}, \bibinfo{person}{Yong Xu}, \bibinfo{person}{Lei Shi}, \bibinfo{person}{Long Xia}, \bibinfo{person}{Dawei Yin}, {and} \bibinfo{person}{Chao Huang}.} \bibinfo{year}{2024}\natexlab{f}.
\newblock \showarticletitle{Urbangpt: Spatio-temporal large language models}. In \bibinfo{booktitle}{\emph{Proceedings of the 30th ACM SIGKDD Conference on Knowledge Discovery and Data Mining}}. \bibinfo{pages}{5351--5362}.
\newblock


\bibitem[Liang et~al\mbox{.}(2025)]%
        {liang2025foundation}
\bibfield{author}{\bibinfo{person}{Yuxuan Liang}, \bibinfo{person}{Haomin Wen}, \bibinfo{person}{Yutong Xia}, \bibinfo{person}{Ming Jin}, \bibinfo{person}{Bin Yang}, \bibinfo{person}{Flora Salim}, \bibinfo{person}{Qingsong Wen}, \bibinfo{person}{Shirui Pan}, {and} \bibinfo{person}{Gao Cong}.} \bibinfo{year}{2025}\natexlab{}.
\newblock \showarticletitle{Foundation Models for Spatio-Temporal Data Science: A Tutorial and Survey}.
\newblock \bibinfo{journal}{\emph{arXiv preprint arXiv:2503.13502}} (\bibinfo{year}{2025}).
\newblock


\bibitem[Lin et~al\mbox{.}(2023)]%
        {lin2023mcu}
\bibfield{author}{\bibinfo{person}{Haowei Lin}, \bibinfo{person}{Zihao Wang}, \bibinfo{person}{Jianzhu Ma}, {and} \bibinfo{person}{Yitao Liang}.} \bibinfo{year}{2023}\natexlab{}.
\newblock \showarticletitle{Mcu: A task-centric framework for open-ended agent evaluation in minecraft}.
\newblock \bibinfo{journal}{\emph{arXiv preprint arXiv:2310.08367}} (\bibinfo{year}{2023}).
\newblock


\bibitem[Lin et~al\mbox{.}(2018)]%
        {lin2018efficient}
\bibfield{author}{\bibinfo{person}{Kaixiang Lin}, \bibinfo{person}{Renyu Zhao}, \bibinfo{person}{Zhe Xu}, {and} \bibinfo{person}{Jiayu Zhou}.} \bibinfo{year}{2018}\natexlab{}.
\newblock \showarticletitle{Efficient large-scale fleet management via multi-agent deep reinforcement learning}. In \bibinfo{booktitle}{\emph{Proceedings of the 24th ACM SIGKDD international conference on knowledge discovery \& data mining}}. \bibinfo{pages}{1774--1783}.
\newblock


\bibitem[Liu et~al\mbox{.}(2024c)]%
        {liu2024adversarial}
\bibfield{author}{\bibinfo{person}{Fan Liu}, \bibinfo{person}{Zhao Xu}, {and} \bibinfo{person}{Hao Liu}.} \bibinfo{year}{2024}\natexlab{c}.
\newblock \showarticletitle{Adversarial tuning: Defending against jailbreak attacks for llms}.
\newblock \bibinfo{journal}{\emph{arXiv preprint arXiv:2406.06622}} (\bibinfo{year}{2024}).
\newblock


\bibitem[Liu et~al\mbox{.}(2023e)]%
        {liu2023robust}
\bibfield{author}{\bibinfo{person}{Fan Liu}, \bibinfo{person}{Weijia Zhang}, {and} \bibinfo{person}{Hao Liu}.} \bibinfo{year}{2023}\natexlab{e}.
\newblock \showarticletitle{Robust spatiotemporal traffic forecasting with reinforced dynamic adversarial training}. In \bibinfo{booktitle}{\emph{Proceedings of the 29th ACM SIGKDD Conference on Knowledge Discovery and Data Mining}}. \bibinfo{pages}{1417--1428}.
\newblock


\bibitem[Liu et~al\mbox{.}(2025a)]%
        {liu2025embodied}
\bibfield{author}{\bibinfo{person}{Huaping Liu}, \bibinfo{person}{Di Guo}, {and} \bibinfo{person}{Angelo Cangelosi}.} \bibinfo{year}{2025}\natexlab{a}.
\newblock \showarticletitle{Embodied Intelligence: A Synergy of Morphology, Action, Perception and Learning}.
\newblock \bibinfo{journal}{\emph{Comput. Surveys}} (\bibinfo{year}{2025}).
\newblock


\bibitem[Liu et~al\mbox{.}(2020)]%
        {liu2020multi}
\bibfield{author}{\bibinfo{person}{Hao Liu}, \bibinfo{person}{Jindong Han}, \bibinfo{person}{Yanjie Fu}, \bibinfo{person}{Jingbo Zhou}, \bibinfo{person}{Xinjiang Lu}, {and} \bibinfo{person}{Hui Xiong}.} \bibinfo{year}{2020}\natexlab{}.
\newblock \showarticletitle{Multi-modal transportation recommendation with unified route representation learning}.
\newblock \bibinfo{journal}{\emph{Proceedings of the VLDB Endowment}} \bibinfo{volume}{14}, \bibinfo{number}{3} (\bibinfo{year}{2020}), \bibinfo{pages}{342--350}.
\newblock


\bibitem[Liu et~al\mbox{.}(2025b)]%
        {liu2025large}
\bibfield{author}{\bibinfo{person}{Mingzhe Liu}, \bibinfo{person}{Liang Zhang}, \bibinfo{person}{Jianli Chen}, \bibinfo{person}{Wei-An Chen}, \bibinfo{person}{Zhiyao Yang}, \bibinfo{person}{L~James Lo}, \bibinfo{person}{Jin Wen}, {and} \bibinfo{person}{Zheng O’Neill}.} \bibinfo{year}{2025}\natexlab{b}.
\newblock \showarticletitle{Large language models for building energy applications: Opportunities and challenges}. In \bibinfo{booktitle}{\emph{Building Simulation}}. Springer, \bibinfo{pages}{1--10}.
\newblock


\bibitem[Liu et~al\mbox{.}(2024a)]%
        {liu2024aligning}
\bibfield{author}{\bibinfo{person}{Yang Liu}, \bibinfo{person}{Weixing Chen}, \bibinfo{person}{Yongjie Bai}, \bibinfo{person}{Xiaodan Liang}, \bibinfo{person}{Guanbin Li}, \bibinfo{person}{Wen Gao}, {and} \bibinfo{person}{Liang Lin}.} \bibinfo{year}{2024}\natexlab{a}.
\newblock \showarticletitle{Aligning cyber space with physical world: A comprehensive survey on embodied ai}.
\newblock \bibinfo{journal}{\emph{arXiv preprint arXiv:2407.06886}} (\bibinfo{year}{2024}).
\newblock


\bibitem[Liu et~al\mbox{.}(2023a)]%
        {liu2023prompt}
\bibfield{author}{\bibinfo{person}{Yi Liu}, \bibinfo{person}{Gelei Deng}, \bibinfo{person}{Yuekang Li}, \bibinfo{person}{Kailong Wang}, \bibinfo{person}{Zihao Wang}, \bibinfo{person}{Xiaofeng Wang}, \bibinfo{person}{Tianwei Zhang}, \bibinfo{person}{Yepang Liu}, \bibinfo{person}{Haoyu Wang}, \bibinfo{person}{Yan Zheng}, {et~al\mbox{.}}} \bibinfo{year}{2023}\natexlab{a}.
\newblock \showarticletitle{Prompt Injection attack against LLM-integrated Applications}.
\newblock \bibinfo{journal}{\emph{arXiv preprint arXiv:2306.05499}} (\bibinfo{year}{2023}).
\newblock


\bibitem[Liu et~al\mbox{.}(2023b)]%
        {liu2023urbankg}
\bibfield{author}{\bibinfo{person}{Yu Liu}, \bibinfo{person}{Jingtao Ding}, \bibinfo{person}{Yanjie Fu}, {and} \bibinfo{person}{Yong Li}.} \bibinfo{year}{2023}\natexlab{b}.
\newblock \showarticletitle{Urbankg: An urban knowledge graph system}.
\newblock \bibinfo{journal}{\emph{ACM Transactions on Intelligent Systems and Technology}} \bibinfo{volume}{14}, \bibinfo{number}{4} (\bibinfo{year}{2023}), \bibinfo{pages}{1--25}.
\newblock


\bibitem[Liu et~al\mbox{.}(2024b)]%
        {liu2024formalizing}
\bibfield{author}{\bibinfo{person}{Yupei Liu}, \bibinfo{person}{Yuqi Jia}, \bibinfo{person}{Runpeng Geng}, \bibinfo{person}{Jinyuan Jia}, {and} \bibinfo{person}{Neil~Zhenqiang Gong}.} \bibinfo{year}{2024}\natexlab{b}.
\newblock \showarticletitle{Formalizing and benchmarking prompt injection attacks and defenses}. In \bibinfo{booktitle}{\emph{33rd USENIX Security Symposium (USENIX Security 24)}}. \bibinfo{pages}{1831--1847}.
\newblock


\bibitem[Liu et~al\mbox{.}(2023c)]%
        {liu2023can}
\bibfield{author}{\bibinfo{person}{Yang Liu}, \bibinfo{person}{Fanyou Wu}, \bibinfo{person}{Zhiyuan Liu}, \bibinfo{person}{Kai Wang}, \bibinfo{person}{Feiyue Wang}, {and} \bibinfo{person}{Xiaobo Qu}.} \bibinfo{year}{2023}\natexlab{c}.
\newblock \showarticletitle{Can language models be used for real-world urban-delivery route optimization?}
\newblock \bibinfo{journal}{\emph{The Innovation}} \bibinfo{volume}{4}, \bibinfo{number}{6} (\bibinfo{year}{2023}).
\newblock


\bibitem[Liu et~al\mbox{.}(2023d)]%
        {liu2023trustworthy}
\bibfield{author}{\bibinfo{person}{Yang Liu}, \bibinfo{person}{Yuanshun Yao}, \bibinfo{person}{Jean-Francois Ton}, \bibinfo{person}{Xiaoying Zhang}, \bibinfo{person}{Ruocheng Guo}, \bibinfo{person}{Hao Cheng}, \bibinfo{person}{Yegor Klochkov}, \bibinfo{person}{Muhammad~Faaiz Taufiq}, {and} \bibinfo{person}{Hang Li}.} \bibinfo{year}{2023}\natexlab{d}.
\newblock \showarticletitle{Trustworthy llms: a survey and guideline for evaluating large language models' alignment}.
\newblock \bibinfo{journal}{\emph{arXiv preprint arXiv:2308.05374}} (\bibinfo{year}{2023}).
\newblock


\bibitem[Lowrie(1990)]%
        {lowrie1990scats}
\bibfield{author}{\bibinfo{person}{P Lowrie}.} \bibinfo{year}{1990}\natexlab{}.
\newblock \showarticletitle{Scats-a traffic responsive method of controlling urban traffic}.
\newblock \bibinfo{journal}{\emph{Sales information brochure published by Roads \& Traffic Authority, Sydney, Australia}} (\bibinfo{year}{1990}).
\newblock


\bibitem[Lu et~al\mbox{.}(2024)]%
        {lu2024ai}
\bibfield{author}{\bibinfo{person}{Chris Lu}, \bibinfo{person}{Cong Lu}, \bibinfo{person}{Robert~Tjarko Lange}, \bibinfo{person}{Jakob Foerster}, \bibinfo{person}{Jeff Clune}, {and} \bibinfo{person}{David Ha}.} \bibinfo{year}{2024}\natexlab{}.
\newblock \showarticletitle{The ai scientist: Towards fully automated open-ended scientific discovery}.
\newblock \bibinfo{journal}{\emph{arXiv preprint arXiv:2408.06292}} (\bibinfo{year}{2024}).
\newblock


\bibitem[Lu et~al\mbox{.}(2019)]%
        {lu2019learning}
\bibfield{author}{\bibinfo{person}{Hao Lu}, \bibinfo{person}{Xingwen Zhang}, {and} \bibinfo{person}{Shuang Yang}.} \bibinfo{year}{2019}\natexlab{}.
\newblock \showarticletitle{A learning-based iterative method for solving vehicle routing problems}. In \bibinfo{booktitle}{\emph{International conference on learning representations}}.
\newblock


\bibitem[Lyu et~al\mbox{.}(2023)]%
        {lyu2023if}
\bibfield{author}{\bibinfo{person}{Yan Lyu}, \bibinfo{person}{Hangxin Lu}, \bibinfo{person}{Min~Kyung Lee}, \bibinfo{person}{Gerhard Schmitt}, {and} \bibinfo{person}{Brian~Y Lim}.} \bibinfo{year}{2023}\natexlab{}.
\newblock \showarticletitle{IF-City: Intelligible fair city planning to measure, explain and mitigate inequality}.
\newblock \bibinfo{journal}{\emph{IEEE Transactions on Visualization and Computer Graphics}} (\bibinfo{year}{2023}).
\newblock


\bibitem[Madaan et~al\mbox{.}(2023)]%
        {madaan2023self}
\bibfield{author}{\bibinfo{person}{Aman Madaan}, \bibinfo{person}{Niket Tandon}, \bibinfo{person}{Prakhar Gupta}, \bibinfo{person}{Skyler Hallinan}, \bibinfo{person}{Luyu Gao}, \bibinfo{person}{Sarah Wiegreffe}, \bibinfo{person}{Uri Alon}, \bibinfo{person}{Nouha Dziri}, \bibinfo{person}{Shrimai Prabhumoye}, \bibinfo{person}{Yiming Yang}, {et~al\mbox{.}}} \bibinfo{year}{2023}\natexlab{}.
\newblock \showarticletitle{Self-refine: Iterative refinement with self-feedback}.
\newblock \bibinfo{journal}{\emph{Advances in Neural Information Processing Systems}}  \bibinfo{volume}{36} (\bibinfo{year}{2023}), \bibinfo{pages}{46534--46594}.
\newblock


\bibitem[Mai et~al\mbox{.}(2024)]%
        {mai2024opportunities}
\bibfield{author}{\bibinfo{person}{Gengchen Mai}, \bibinfo{person}{Weiming Huang}, \bibinfo{person}{Jin Sun}, \bibinfo{person}{Suhang Song}, \bibinfo{person}{Deepak Mishra}, \bibinfo{person}{Ninghao Liu}, \bibinfo{person}{Song Gao}, \bibinfo{person}{Tianming Liu}, \bibinfo{person}{Gao Cong}, \bibinfo{person}{Yingjie Hu}, {et~al\mbox{.}}} \bibinfo{year}{2024}\natexlab{}.
\newblock \showarticletitle{On the opportunities and challenges of foundation models for geoai (vision paper)}.
\newblock \bibinfo{journal}{\emph{ACM Transactions on Spatial Algorithms and Systems}} \bibinfo{volume}{10}, \bibinfo{number}{2} (\bibinfo{year}{2024}), \bibinfo{pages}{1--46}.
\newblock


\bibitem[Manivannan et~al\mbox{.}(2024)]%
        {manivannan2024climaqa}
\bibfield{author}{\bibinfo{person}{Veeramakali~Vignesh Manivannan}, \bibinfo{person}{Yasaman Jafari}, \bibinfo{person}{Srikar Eranky}, \bibinfo{person}{Spencer Ho}, \bibinfo{person}{Rose Yu}, \bibinfo{person}{Duncan Watson-Parris}, \bibinfo{person}{Yian Ma}, \bibinfo{person}{Leon Bergen}, {and} \bibinfo{person}{Taylor Berg-Kirkpatrick}.} \bibinfo{year}{2024}\natexlab{}.
\newblock \showarticletitle{ClimaQA: An Automated Evaluation Framework for Climate Foundation Models}.
\newblock \bibinfo{journal}{\emph{arXiv preprint arXiv:2410.16701}} (\bibinfo{year}{2024}).
\newblock


\bibitem[Mao et~al\mbox{.}(2023)]%
        {mao2023detecting}
\bibfield{author}{\bibinfo{person}{Jinzhu Mao}, \bibinfo{person}{Liu Cao}, \bibinfo{person}{Chen Gao}, \bibinfo{person}{Huandong Wang}, \bibinfo{person}{Hangyu Fan}, \bibinfo{person}{Depeng Jin}, {and} \bibinfo{person}{Yong Li}.} \bibinfo{year}{2023}\natexlab{}.
\newblock \showarticletitle{Detecting vulnerable nodes in urban infrastructure interdependent network}. In \bibinfo{booktitle}{\emph{Proceedings of the 29th ACM SIGKDD conference on knowledge discovery and data mining}}. \bibinfo{pages}{4617--4627}.
\newblock


\bibitem[McLeod(2007)]%
        {mcleod2007maslow}
\bibfield{author}{\bibinfo{person}{Saul McLeod}.} \bibinfo{year}{2007}\natexlab{}.
\newblock \showarticletitle{Maslow's hierarchy of needs}.
\newblock \bibinfo{journal}{\emph{Simply psychology}} \bibinfo{volume}{1}, \bibinfo{number}{1-18} (\bibinfo{year}{2007}).
\newblock


\bibitem[Mou et~al\mbox{.}(2024)]%
        {mou2024individual}
\bibfield{author}{\bibinfo{person}{Xinyi Mou}, \bibinfo{person}{Xuanwen Ding}, \bibinfo{person}{Qi He}, \bibinfo{person}{Liang Wang}, \bibinfo{person}{Jingcong Liang}, \bibinfo{person}{Xinnong Zhang}, \bibinfo{person}{Libo Sun}, \bibinfo{person}{Jiayu Lin}, \bibinfo{person}{Jie Zhou}, \bibinfo{person}{Xuanjing Huang}, {et~al\mbox{.}}} \bibinfo{year}{2024}\natexlab{}.
\newblock \showarticletitle{From Individual to Society: A Survey on Social Simulation Driven by Large Language Model-based Agents}.
\newblock \bibinfo{journal}{\emph{arXiv preprint arXiv:2412.03563}} (\bibinfo{year}{2024}).
\newblock


\bibitem[Ni et~al\mbox{.}(2025)]%
        {ni2025tp}
\bibfield{author}{\bibinfo{person}{Hang Ni}, \bibinfo{person}{Fan Liu}, \bibinfo{person}{Xinyu Ma}, \bibinfo{person}{Lixin Su}, \bibinfo{person}{Shuaiqiang Wang}, \bibinfo{person}{Dawei Yin}, \bibinfo{person}{Hui Xiong}, {and} \bibinfo{person}{Hao Liu}.} \bibinfo{year}{2025}\natexlab{}.
\newblock \showarticletitle{TP-RAG: Benchmarking Retrieval-Augmented Large Language Model Agents for Spatiotemporal-Aware Travel Planning}.
\newblock \bibinfo{journal}{\emph{arXiv preprint arXiv:2504.08694}} (\bibinfo{year}{2025}).
\newblock


\bibitem[Ni et~al\mbox{.}(2024)]%
        {ni2024planning}
\bibfield{author}{\bibinfo{person}{Hang Ni}, \bibinfo{person}{Yuzhi Wang}, {and} \bibinfo{person}{Hao Liu}.} \bibinfo{year}{2024}\natexlab{}.
\newblock \showarticletitle{Planning, Living and Judging: A Multi-agent LLM-based Framework for Cyclical Urban Planning}.
\newblock \bibinfo{journal}{\emph{arXiv preprint arXiv:2412.20505}} (\bibinfo{year}{2024}).
\newblock


\bibitem[Nie et~al\mbox{.}(2025)]%
        {nie2025joint}
\bibfield{author}{\bibinfo{person}{Tong Nie}, \bibinfo{person}{Junlin He}, \bibinfo{person}{Yuewen Mei}, \bibinfo{person}{Guoyang Qin}, \bibinfo{person}{Guilong Li}, \bibinfo{person}{Jian Sun}, {and} \bibinfo{person}{Wei Ma}.} \bibinfo{year}{2025}\natexlab{}.
\newblock \showarticletitle{Joint estimation and prediction of city-wide delivery demand: A large language model empowered graph-based learning approach}.
\newblock \bibinfo{journal}{\emph{Transportation Research Part E: Logistics and Transportation Review}}  \bibinfo{volume}{197} (\bibinfo{year}{2025}), \bibinfo{pages}{104075}.
\newblock


\bibitem[Ning et~al\mbox{.}(2024)]%
        {ning2024llm}
\bibfield{author}{\bibinfo{person}{Huan Ning}, \bibinfo{person}{Zhenlong Li}, \bibinfo{person}{Temitope Akinboyewa}, {and} \bibinfo{person}{M Naser~Lessani}.} \bibinfo{year}{2024}\natexlab{}.
\newblock \showarticletitle{LLM-Find: An Autonomous GIS Agent Framework for Geospatial Data Retrieval}.
\newblock \bibinfo{journal}{\emph{arXiv e-prints}} (\bibinfo{year}{2024}), \bibinfo{pages}{arXiv--2407}.
\newblock


\bibitem[Ning et~al\mbox{.}(2025)]%
        {ning2025dima}
\bibfield{author}{\bibinfo{person}{Yansong Ning}, \bibinfo{person}{Shuowei Cai}, \bibinfo{person}{Wei Li}, \bibinfo{person}{Jun Fang}, \bibinfo{person}{Naiqiang Tan}, \bibinfo{person}{Hua Chai}, {and} \bibinfo{person}{Hao Liu}.} \bibinfo{year}{2025}\natexlab{}.
\newblock \showarticletitle{DiMA: An LLM-Powered Ride-Hailing Assistant at DiDi}.
\newblock \bibinfo{journal}{\emph{arXiv preprint arXiv:2503.04768}} (\bibinfo{year}{2025}).
\newblock


\bibitem[Ning and Liu(2024)]%
        {ning2024urbankgent}
\bibfield{author}{\bibinfo{person}{Yansong Ning} {and} \bibinfo{person}{Hao Liu}.} \bibinfo{year}{2024}\natexlab{}.
\newblock \showarticletitle{UrbanKGent: A Unified Large Language Model Agent Framework for Urban Knowledge Graph Construction}.
\newblock \bibinfo{journal}{\emph{arXiv preprint arXiv:2402.06861}} (\bibinfo{year}{2024}).
\newblock


\bibitem[Ning et~al\mbox{.}(2023)]%
        {ning2023uukg}
\bibfield{author}{\bibinfo{person}{Yansong Ning}, \bibinfo{person}{Hao Liu}, \bibinfo{person}{Hao Wang}, \bibinfo{person}{Zhenyu Zeng}, {and} \bibinfo{person}{Hui Xiong}.} \bibinfo{year}{2023}\natexlab{}.
\newblock \showarticletitle{UUKG: Unified urban knowledge graph dataset for urban spatiotemporal prediction}.
\newblock \bibinfo{journal}{\emph{Advances in Neural Information Processing Systems}}  \bibinfo{volume}{36} (\bibinfo{year}{2023}), \bibinfo{pages}{62442--62456}.
\newblock


\bibitem[Otal et~al\mbox{.}(2024)]%
        {otal2024llm}
\bibfield{author}{\bibinfo{person}{Hakan~T Otal}, \bibinfo{person}{Eric Stern}, {and} \bibinfo{person}{M~Abdullah Canbaz}.} \bibinfo{year}{2024}\natexlab{}.
\newblock \showarticletitle{Llm-assisted crisis management: Building advanced llm platforms for effective emergency response and public collaboration}. In \bibinfo{booktitle}{\emph{2024 IEEE Conference on Artificial Intelligence (CAI)}}. IEEE, \bibinfo{pages}{851--859}.
\newblock


\bibitem[Ouyang et~al\mbox{.}(2022)]%
        {ouyang2022training}
\bibfield{author}{\bibinfo{person}{Long Ouyang}, \bibinfo{person}{Jeffrey Wu}, \bibinfo{person}{Xu Jiang}, \bibinfo{person}{Diogo Almeida}, \bibinfo{person}{Carroll Wainwright}, \bibinfo{person}{Pamela Mishkin}, \bibinfo{person}{Chong Zhang}, \bibinfo{person}{Sandhini Agarwal}, \bibinfo{person}{Katarina Slama}, \bibinfo{person}{Alex Ray}, {et~al\mbox{.}}} \bibinfo{year}{2022}\natexlab{}.
\newblock \showarticletitle{Training language models to follow instructions with human feedback}.
\newblock \bibinfo{journal}{\emph{Advances in neural information processing systems}}  \bibinfo{volume}{35} (\bibinfo{year}{2022}), \bibinfo{pages}{27730--27744}.
\newblock


\bibitem[Pan et~al\mbox{.}(2024)]%
        {pan2024survey}
\bibfield{author}{\bibinfo{person}{James~Jie Pan}, \bibinfo{person}{Jianguo Wang}, {and} \bibinfo{person}{Guoliang Li}.} \bibinfo{year}{2024}\natexlab{}.
\newblock \showarticletitle{Survey of vector database management systems}.
\newblock \bibinfo{journal}{\emph{The VLDB Journal}} \bibinfo{volume}{33}, \bibinfo{number}{5} (\bibinfo{year}{2024}), \bibinfo{pages}{1591--1615}.
\newblock


\bibitem[Pang et~al\mbox{.}(2024)]%
        {pang2024illm}
\bibfield{author}{\bibinfo{person}{Aoyu Pang}, \bibinfo{person}{Maonan Wang}, \bibinfo{person}{Man-On Pun}, \bibinfo{person}{Chung~Shue Chen}, {and} \bibinfo{person}{Xi Xiong}.} \bibinfo{year}{2024}\natexlab{}.
\newblock \showarticletitle{iLLM-TSC: Integration reinforcement learning and large language model for traffic signal control policy improvement}.
\newblock \bibinfo{journal}{\emph{arXiv preprint arXiv:2407.06025}} (\bibinfo{year}{2024}).
\newblock


\bibitem[Park et~al\mbox{.}(2023)]%
        {park2023generative}
\bibfield{author}{\bibinfo{person}{Joon~Sung Park}, \bibinfo{person}{Joseph O'Brien}, \bibinfo{person}{Carrie~Jun Cai}, \bibinfo{person}{Meredith~Ringel Morris}, \bibinfo{person}{Percy Liang}, {and} \bibinfo{person}{Michael~S Bernstein}.} \bibinfo{year}{2023}\natexlab{}.
\newblock \showarticletitle{Generative agents: Interactive simulacra of human behavior}. In \bibinfo{booktitle}{\emph{Proceedings of the 36th annual acm symposium on user interface software and technology}}. \bibinfo{pages}{1--22}.
\newblock


\bibitem[Park et~al\mbox{.}(2024)]%
        {park2024generative}
\bibfield{author}{\bibinfo{person}{Joon~Sung Park}, \bibinfo{person}{Carolyn~Q Zou}, \bibinfo{person}{Aaron Shaw}, \bibinfo{person}{Benjamin~Mako Hill}, \bibinfo{person}{Carrie Cai}, \bibinfo{person}{Meredith~Ringel Morris}, \bibinfo{person}{Robb Willer}, \bibinfo{person}{Percy Liang}, {and} \bibinfo{person}{Michael~S Bernstein}.} \bibinfo{year}{2024}\natexlab{}.
\newblock \showarticletitle{Generative agent simulations of 1,000 people}.
\newblock \bibinfo{journal}{\emph{arXiv preprint arXiv:2411.10109}} (\bibinfo{year}{2024}).
\newblock


\bibitem[Parker et~al\mbox{.}(2003)]%
        {parker2003multi}
\bibfield{author}{\bibinfo{person}{Dawn~C Parker}, \bibinfo{person}{Steven~M Manson}, \bibinfo{person}{Marco~A Janssen}, \bibinfo{person}{Matthew~J Hoffmann}, {and} \bibinfo{person}{Peter Deadman}.} \bibinfo{year}{2003}\natexlab{}.
\newblock \showarticletitle{Multi-agent systems for the simulation of land-use and land-cover change: a review}.
\newblock \bibinfo{journal}{\emph{Annals of the association of American Geographers}} \bibinfo{volume}{93}, \bibinfo{number}{2} (\bibinfo{year}{2003}), \bibinfo{pages}{314--337}.
\newblock


\bibitem[Perez et~al\mbox{.}(2022)]%
        {perez2022red}
\bibfield{author}{\bibinfo{person}{Ethan Perez}, \bibinfo{person}{Saffron Huang}, \bibinfo{person}{Francis Song}, \bibinfo{person}{Trevor Cai}, \bibinfo{person}{Roman Ring}, \bibinfo{person}{John Aslanides}, \bibinfo{person}{Amelia Glaese}, \bibinfo{person}{Nat McAleese}, {and} \bibinfo{person}{Geoffrey Irving}.} \bibinfo{year}{2022}\natexlab{}.
\newblock \showarticletitle{Red Teaming Language Models with Language Models}. In \bibinfo{booktitle}{\emph{Proceedings of the 2022 Conference on Empirical Methods in Natural Language Processing}}. \bibinfo{pages}{3419--3448}.
\newblock


\bibitem[Piao et~al\mbox{.}(2025)]%
        {piao2025agentsociety}
\bibfield{author}{\bibinfo{person}{Jinghua Piao}, \bibinfo{person}{Yuwei Yan}, \bibinfo{person}{Jun Zhang}, \bibinfo{person}{Nian Li}, \bibinfo{person}{Junbo Yan}, \bibinfo{person}{Xiaochong Lan}, \bibinfo{person}{Zhihong Lu}, \bibinfo{person}{Zhiheng Zheng}, \bibinfo{person}{Jing~Yi Wang}, \bibinfo{person}{Di Zhou}, {et~al\mbox{.}}} \bibinfo{year}{2025}\natexlab{}.
\newblock \showarticletitle{AgentSociety: Large-Scale Simulation of LLM-Driven Generative Agents Advances Understanding of Human Behaviors and Society}.
\newblock \bibinfo{journal}{\emph{arXiv preprint arXiv:2502.08691}} (\bibinfo{year}{2025}).
\newblock


\bibitem[Qiu et~al\mbox{.}(2024)]%
        {qiu2024ef}
\bibfield{author}{\bibinfo{person}{Zihang Qiu}, \bibinfo{person}{Chaojie Li}, \bibinfo{person}{Zhongyang Wang}, \bibinfo{person}{Renyou Xie}, \bibinfo{person}{Borui Zhang}, \bibinfo{person}{Huadong Mo}, \bibinfo{person}{Guo Chen}, {and} \bibinfo{person}{Zhaoyang Dong}.} \bibinfo{year}{2024}\natexlab{}.
\newblock \showarticletitle{EF-LLM: Energy Forecasting LLM with AI-assisted Automation, Enhanced Sparse Prediction, Hallucination Detection}.
\newblock \bibinfo{journal}{\emph{arXiv preprint arXiv:2411.00852}} (\bibinfo{year}{2024}).
\newblock


\bibitem[Radford et~al\mbox{.}(2019)]%
        {radford2019language}
\bibfield{author}{\bibinfo{person}{Alec Radford}, \bibinfo{person}{Jeffrey Wu}, \bibinfo{person}{Rewon Child}, \bibinfo{person}{David Luan}, \bibinfo{person}{Dario Amodei}, \bibinfo{person}{Ilya Sutskever}, {et~al\mbox{.}}} \bibinfo{year}{2019}\natexlab{}.
\newblock \showarticletitle{Language models are unsupervised multitask learners}.
\newblock \bibinfo{journal}{\emph{OpenAI blog}} \bibinfo{volume}{1}, \bibinfo{number}{8} (\bibinfo{year}{2019}), \bibinfo{pages}{9}.
\newblock


\bibitem[Raffel et~al\mbox{.}(2020)]%
        {raffel2020exploring}
\bibfield{author}{\bibinfo{person}{Colin Raffel}, \bibinfo{person}{Noam Shazeer}, \bibinfo{person}{Adam Roberts}, \bibinfo{person}{Katherine Lee}, \bibinfo{person}{Sharan Narang}, \bibinfo{person}{Michael Matena}, \bibinfo{person}{Yanqi Zhou}, \bibinfo{person}{Wei Li}, {and} \bibinfo{person}{Peter~J Liu}.} \bibinfo{year}{2020}\natexlab{}.
\newblock \showarticletitle{Exploring the limits of transfer learning with a unified text-to-text transformer}.
\newblock \bibinfo{journal}{\emph{Journal of machine learning research}} \bibinfo{volume}{21}, \bibinfo{number}{140} (\bibinfo{year}{2020}), \bibinfo{pages}{1--67}.
\newblock


\bibitem[Reza et~al\mbox{.}({[n.\,d.]})]%
        {rezacrix}
\bibfield{author}{\bibinfo{person}{Muhammad~Ashar Reza}, \bibinfo{person}{Aaditya Bisaria}, \bibinfo{person}{S Advaitha}, \bibinfo{person}{Alekhya Ponnekanti}, {and} \bibinfo{person}{Arti Arya}.} \bibinfo{year}{[n.\,d.]}\natexlab{}.
\newblock \showarticletitle{CriX: Intersection of Crime, Demographics and Explainable AI}.
\newblock  (\bibinfo{year}{[n.\,d.]}).
\newblock


\bibitem[Ribeiro et~al\mbox{.}(2016)]%
        {ribeiro2016should}
\bibfield{author}{\bibinfo{person}{Marco~Tulio Ribeiro}, \bibinfo{person}{Sameer Singh}, {and} \bibinfo{person}{Carlos Guestrin}.} \bibinfo{year}{2016}\natexlab{}.
\newblock \showarticletitle{" Why should i trust you?" Explaining the predictions of any classifier}. In \bibinfo{booktitle}{\emph{Proceedings of the 22nd ACM SIGKDD international conference on knowledge discovery and data mining}}. \bibinfo{pages}{1135--1144}.
\newblock


\bibitem[Rizvi et~al\mbox{.}(2024)]%
        {rizvi2024sparc}
\bibfield{author}{\bibinfo{person}{Md~Imbesat Rizvi}, \bibinfo{person}{Xiaodan Zhu}, {and} \bibinfo{person}{Iryna Gurevych}.} \bibinfo{year}{2024}\natexlab{}.
\newblock \showarticletitle{SpaRC and SpaRP: Spatial Reasoning Characterization and Path Generation for Understanding Spatial Reasoning Capability of Large Language Models}. In \bibinfo{booktitle}{\emph{Proceedings of the 62nd Annual Meeting of the Association for Computational Linguistics (Volume 1: Long Papers)}}. \bibinfo{pages}{4750--4767}.
\newblock


\bibitem[Rolnick et~al\mbox{.}(2022)]%
        {rolnick2022tackling}
\bibfield{author}{\bibinfo{person}{David Rolnick}, \bibinfo{person}{Priya~L Donti}, \bibinfo{person}{Lynn~H Kaack}, \bibinfo{person}{Kelly Kochanski}, \bibinfo{person}{Alexandre Lacoste}, \bibinfo{person}{Kris Sankaran}, \bibinfo{person}{Andrew~Slavin Ross}, \bibinfo{person}{Nikola Milojevic-Dupont}, \bibinfo{person}{Natasha Jaques}, \bibinfo{person}{Anna Waldman-Brown}, {et~al\mbox{.}}} \bibinfo{year}{2022}\natexlab{}.
\newblock \showarticletitle{Tackling climate change with machine learning}.
\newblock \bibinfo{journal}{\emph{ACM Computing Surveys (CSUR)}} \bibinfo{volume}{55}, \bibinfo{number}{2} (\bibinfo{year}{2022}), \bibinfo{pages}{1--96}.
\newblock


\bibitem[Russell and Norvig(2016)]%
        {russell2016artificial}
\bibfield{author}{\bibinfo{person}{Stuart~J Russell} {and} \bibinfo{person}{Peter Norvig}.} \bibinfo{year}{2016}\natexlab{}.
\newblock \bibinfo{booktitle}{\emph{Artificial intelligence: a modern approach}}.
\newblock \bibinfo{publisher}{pearson}.
\newblock


\bibitem[Salkham et~al\mbox{.}(2008)]%
        {salkham2008collaborative}
\bibfield{author}{\bibinfo{person}{As'~ad Salkham}, \bibinfo{person}{Raymond Cunningham}, \bibinfo{person}{Anurag Garg}, {and} \bibinfo{person}{Vinny Cahill}.} \bibinfo{year}{2008}\natexlab{}.
\newblock \showarticletitle{A collaborative reinforcement learning approach to urban traffic control optimization}. In \bibinfo{booktitle}{\emph{2008 IEEE/WIC/ACM International Conference on Web Intelligence and Intelligent Agent Technology}}, Vol.~\bibinfo{volume}{2}. IEEE, \bibinfo{pages}{560--566}.
\newblock


\bibitem[Schick et~al\mbox{.}(2023)]%
        {schick2023toolformer}
\bibfield{author}{\bibinfo{person}{Timo Schick}, \bibinfo{person}{Jane Dwivedi-Yu}, \bibinfo{person}{Roberto Dess{\`\i}}, \bibinfo{person}{Roberta Raileanu}, \bibinfo{person}{Maria Lomeli}, \bibinfo{person}{Eric Hambro}, \bibinfo{person}{Luke Zettlemoyer}, \bibinfo{person}{Nicola Cancedda}, {and} \bibinfo{person}{Thomas Scialom}.} \bibinfo{year}{2023}\natexlab{}.
\newblock \showarticletitle{Toolformer: Language models can teach themselves to use tools}.
\newblock \bibinfo{journal}{\emph{Advances in Neural Information Processing Systems}}  \bibinfo{volume}{36} (\bibinfo{year}{2023}), \bibinfo{pages}{68539--68551}.
\newblock


\bibitem[Shahid et~al\mbox{.}(2024)]%
        {shahid2024watchovergpt}
\bibfield{author}{\bibinfo{person}{Abdur~R Shahid}, \bibinfo{person}{Syed~Mhamudul Hasan}, \bibinfo{person}{Malithi~Wanniarachchi Kankanamge}, \bibinfo{person}{Md~Zarif Hossain}, {and} \bibinfo{person}{Ahmed Imteaj}.} \bibinfo{year}{2024}\natexlab{}.
\newblock \showarticletitle{WatchOverGPT: A Framework for Real-Time Crime Detection and Response Using Wearable Camera and Large Language Model}. In \bibinfo{booktitle}{\emph{2024 IEEE 48th Annual Computers, Software, and Applications Conference (COMPSAC)}}. IEEE, \bibinfo{pages}{2189--2194}.
\newblock


\bibitem[Shao et~al\mbox{.}(2024)]%
        {shao2024chain}
\bibfield{author}{\bibinfo{person}{Chenyang Shao}, \bibinfo{person}{Fengli Xu}, \bibinfo{person}{Bingbing Fan}, \bibinfo{person}{Jingtao Ding}, \bibinfo{person}{Yuan Yuan}, \bibinfo{person}{Meng Wang}, {and} \bibinfo{person}{Yong Li}.} \bibinfo{year}{2024}\natexlab{}.
\newblock \showarticletitle{Chain-of-planned-behaviour workflow elicits few-shot mobility generation in LLMs}.
\newblock \bibinfo{journal}{\emph{arXiv preprint arXiv:2402.09836}} (\bibinfo{year}{2024}).
\newblock


\bibitem[Sharkey et~al\mbox{.}(2024)]%
        {sharkey2024causal}
\bibfield{author}{\bibinfo{person}{Lee Sharkey}, \bibinfo{person}{Cl{\'\i}odhna~N{\'\i} Ghuidhir}, \bibinfo{person}{Dan Braun}, \bibinfo{person}{J{\'e}r{\'e}my Scheurer}, \bibinfo{person}{Mikita Balesni}, \bibinfo{person}{Lucius Bushnaq}, \bibinfo{person}{Charlotte Stix}, {and} \bibinfo{person}{Marius Hobbhahn}.} \bibinfo{year}{2024}\natexlab{}.
\newblock \showarticletitle{A causal framework for AI regulation and auditing}.
\newblock \bibinfo{journal}{\emph{Publisher: Preprints}} (\bibinfo{year}{2024}).
\newblock


\bibitem[Shinn et~al\mbox{.}(2023)]%
        {shinn2023reflexion}
\bibfield{author}{\bibinfo{person}{Noah Shinn}, \bibinfo{person}{Federico Cassano}, \bibinfo{person}{Ashwin Gopinath}, \bibinfo{person}{Karthik Narasimhan}, {and} \bibinfo{person}{Shunyu Yao}.} \bibinfo{year}{2023}\natexlab{}.
\newblock \showarticletitle{Reflexion: Language agents with verbal reinforcement learning}.
\newblock \bibinfo{journal}{\emph{Advances in Neural Information Processing Systems}}  \bibinfo{volume}{36} (\bibinfo{year}{2023}), \bibinfo{pages}{8634--8652}.
\newblock


\bibitem[Singla et~al\mbox{.}(2024)]%
        {singla2024adaptive}
\bibfield{author}{\bibinfo{person}{Pratham Singla}, \bibinfo{person}{Ayush Singh}, \bibinfo{person}{Adesh Gupta}, {and} \bibinfo{person}{Shivank Garg}.} \bibinfo{year}{2024}\natexlab{}.
\newblock \showarticletitle{Adaptive Urban Planning: A Hybrid Framework for Balanced City Development}.
\newblock \bibinfo{journal}{\emph{arXiv preprint arXiv:2412.15349}} (\bibinfo{year}{2024}).
\newblock


\bibitem[Snell et~al\mbox{.}(2024)]%
        {snell2024scaling}
\bibfield{author}{\bibinfo{person}{Charlie Snell}, \bibinfo{person}{Jaehoon Lee}, \bibinfo{person}{Kelvin Xu}, {and} \bibinfo{person}{Aviral Kumar}.} \bibinfo{year}{2024}\natexlab{}.
\newblock \showarticletitle{Scaling llm test-time compute optimally can be more effective than scaling model parameters}.
\newblock \bibinfo{journal}{\emph{arXiv preprint arXiv:2408.03314}} (\bibinfo{year}{2024}).
\newblock


\bibitem[Sofuoglu and Aviyente(2022)]%
        {sofuoglu2022gloss}
\bibfield{author}{\bibinfo{person}{Seyyid~Emre Sofuoglu} {and} \bibinfo{person}{Selin Aviyente}.} \bibinfo{year}{2022}\natexlab{}.
\newblock \showarticletitle{Gloss: Tensor-based anomaly detection in spatiotemporal urban traffic data}.
\newblock \bibinfo{journal}{\emph{Signal Processing}}  \bibinfo{volume}{192} (\bibinfo{year}{2022}), \bibinfo{pages}{108370}.
\newblock


\bibitem[Steinke(2022)]%
        {steinke2022composition}
\bibfield{author}{\bibinfo{person}{Thomas Steinke}.} \bibinfo{year}{2022}\natexlab{}.
\newblock \showarticletitle{Composition of differential privacy \& privacy amplification by subsampling}.
\newblock \bibinfo{journal}{\emph{arXiv preprint arXiv:2210.00597}} (\bibinfo{year}{2022}).
\newblock


\bibitem[Su et~al\mbox{.}(2024)]%
        {su2024living}
\bibfield{author}{\bibinfo{person}{Zhaochen Su}, \bibinfo{person}{Juntao Li}, \bibinfo{person}{Jun Zhang}, \bibinfo{person}{Tong Zhu}, \bibinfo{person}{Xiaoye Qu}, \bibinfo{person}{Pan Zhou}, \bibinfo{person}{Yan Bowen}, \bibinfo{person}{Yu Cheng}, {et~al\mbox{.}}} \bibinfo{year}{2024}\natexlab{}.
\newblock \showarticletitle{Living in the Moment: Can Large Language Models Grasp Co-Temporal Reasoning?}
\newblock \bibinfo{journal}{\emph{arXiv preprint arXiv:2406.09072}} (\bibinfo{year}{2024}).
\newblock


\bibitem[Su et~al\mbox{.}({[n.\,d.]})]%
        {sutimo}
\bibfield{author}{\bibinfo{person}{Zhaochen Su}, \bibinfo{person}{Jun Zhang}, \bibinfo{person}{Tong Zhu}, \bibinfo{person}{Xiaoye Qu}, \bibinfo{person}{Juntao Li}, \bibinfo{person}{Yu Cheng}, {et~al\mbox{.}}} \bibinfo{year}{[n.\,d.]}\natexlab{}.
\newblock \showarticletitle{Timo: Towards Better Temporal Reasoning for Language Models}. In \bibinfo{booktitle}{\emph{First Conference on Language Modeling}}.
\newblock


\bibitem[Sun et~al\mbox{.}(2022)]%
        {sun2022adversarial}
\bibfield{author}{\bibinfo{person}{Lichao Sun}, \bibinfo{person}{Yingtong Dou}, \bibinfo{person}{Carl Yang}, \bibinfo{person}{Kai Zhang}, \bibinfo{person}{Ji Wang}, \bibinfo{person}{Philip~S Yu}, \bibinfo{person}{Lifang He}, {and} \bibinfo{person}{Bo Li}.} \bibinfo{year}{2022}\natexlab{}.
\newblock \showarticletitle{Adversarial attack and defense on graph data: A survey}.
\newblock \bibinfo{journal}{\emph{IEEE Transactions on Knowledge and Data Engineering}} \bibinfo{volume}{35}, \bibinfo{number}{8} (\bibinfo{year}{2022}), \bibinfo{pages}{7693--7711}.
\newblock


\bibitem[Sun et~al\mbox{.}(2024a)]%
        {sun2024trustllm}
\bibfield{author}{\bibinfo{person}{Lichao Sun}, \bibinfo{person}{Yue Huang}, \bibinfo{person}{Haoran Wang}, \bibinfo{person}{Siyuan Wu}, \bibinfo{person}{Qihui Zhang}, \bibinfo{person}{Chujie Gao}, \bibinfo{person}{Yixin Huang}, \bibinfo{person}{Wenhan Lyu}, \bibinfo{person}{Yixuan Zhang}, \bibinfo{person}{Xiner Li}, {et~al\mbox{.}}} \bibinfo{year}{2024}\natexlab{a}.
\newblock \showarticletitle{Trustllm: Trustworthiness in large language models}.
\newblock \bibinfo{journal}{\emph{arXiv preprint arXiv:2401.05561}}  \bibinfo{volume}{3} (\bibinfo{year}{2024}).
\newblock


\bibitem[Sun et~al\mbox{.}(2024b)]%
        {sun2024crosslight}
\bibfield{author}{\bibinfo{person}{Qian Sun}, \bibinfo{person}{Rui Zha}, \bibinfo{person}{Le Zhang}, \bibinfo{person}{Jingbo Zhou}, \bibinfo{person}{Yu Mei}, \bibinfo{person}{Zhiling Li}, {and} \bibinfo{person}{Hui Xiong}.} \bibinfo{year}{2024}\natexlab{b}.
\newblock \showarticletitle{CrossLight: Offline-to-Online Reinforcement Learning for Cross-City Traffic Signal Control}. In \bibinfo{booktitle}{\emph{Proceedings of the 30th ACM SIGKDD Conference on Knowledge Discovery and Data Mining}}. \bibinfo{pages}{2765--2774}.
\newblock


\bibitem[Sun et~al\mbox{.}(2023)]%
        {sun2023unleashing}
\bibfield{author}{\bibinfo{person}{Yimin Sun}, \bibinfo{person}{Chao Wang}, {and} \bibinfo{person}{Yan Peng}.} \bibinfo{year}{2023}\natexlab{}.
\newblock \showarticletitle{Unleashing the potential of large language model: Zero-shot vqa for flood disaster scenario}. In \bibinfo{booktitle}{\emph{Proceedings of the 4th International Conference on Artificial Intelligence and Computer Engineering}}. \bibinfo{pages}{368--373}.
\newblock


\bibitem[Tan et~al\mbox{.}(2023)]%
        {tan2023towards}
\bibfield{author}{\bibinfo{person}{Qingyu Tan}, \bibinfo{person}{Hwee~Tou Ng}, {and} \bibinfo{person}{Lidong Bing}.} \bibinfo{year}{2023}\natexlab{}.
\newblock \showarticletitle{Towards Benchmarking and Improving the Temporal Reasoning Capability of Large Language Models}. In \bibinfo{booktitle}{\emph{Proceedings of the 61st Annual Meeting of the Association for Computational Linguistics (Volume 1: Long Papers)}}. \bibinfo{pages}{14820--14835}.
\newblock


\bibitem[Tang et~al\mbox{.}(2016)]%
        {tang2016locationspark}
\bibfield{author}{\bibinfo{person}{Mingjie Tang}, \bibinfo{person}{Yongyang Yu}, \bibinfo{person}{Qutaibah~M Malluhi}, \bibinfo{person}{Mourad Ouzzani}, {and} \bibinfo{person}{Walid~G Aref}.} \bibinfo{year}{2016}\natexlab{}.
\newblock \showarticletitle{Locationspark: A distributed in-memory data management system for big spatial data}.
\newblock \bibinfo{journal}{\emph{Proceedings of the VLDB Endowment}} \bibinfo{volume}{9}, \bibinfo{number}{13} (\bibinfo{year}{2016}), \bibinfo{pages}{1565--1568}.
\newblock


\bibitem[Tang et~al\mbox{.}(2024a)]%
        {tang2024large}
\bibfield{author}{\bibinfo{person}{Yiqing Tang}, \bibinfo{person}{Xingyuan Dai}, \bibinfo{person}{Chen Zhao}, \bibinfo{person}{Qi Cheng}, {and} \bibinfo{person}{Yisheng Lv}.} \bibinfo{year}{2024}\natexlab{a}.
\newblock \showarticletitle{Large language model-driven urban traffic signal control}. In \bibinfo{booktitle}{\emph{2024 Australian \& New Zealand Control Conference (ANZCC)}}. IEEE, \bibinfo{pages}{67--71}.
\newblock


\bibitem[Tang et~al\mbox{.}(2024b)]%
        {tang2024itinera}
\bibfield{author}{\bibinfo{person}{Yihong Tang}, \bibinfo{person}{Zhaokai Wang}, \bibinfo{person}{Ao Qu}, \bibinfo{person}{Yihao Yan}, \bibinfo{person}{Zhaofeng Wu}, \bibinfo{person}{Dingyi Zhuang}, \bibinfo{person}{Jushi Kai}, \bibinfo{person}{Kebing Hou}, \bibinfo{person}{Xiaotong Guo}, \bibinfo{person}{Jinhua Zhao}, {et~al\mbox{.}}} \bibinfo{year}{2024}\natexlab{b}.
\newblock \showarticletitle{ItiNera: Integrating Spatial Optimization with Large Language Models for Open-domain Urban Itinerary Planning}. In \bibinfo{booktitle}{\emph{Proceedings of the 2024 Conference on Empirical Methods in Natural Language Processing: Industry Track}}. \bibinfo{pages}{1413--1432}.
\newblock


\bibitem[Thompson(1980)]%
        {thompson1980moral}
\bibfield{author}{\bibinfo{person}{Dennis~F Thompson}.} \bibinfo{year}{1980}\natexlab{}.
\newblock \showarticletitle{Moral responsibility of public officials: The problem of many hands}.
\newblock \bibinfo{journal}{\emph{American Political Science Review}} \bibinfo{volume}{74}, \bibinfo{number}{4} (\bibinfo{year}{1980}), \bibinfo{pages}{905--916}.
\newblock


\bibitem[Thulke et~al\mbox{.}(2024)]%
        {thulke2024climategpt}
\bibfield{author}{\bibinfo{person}{David Thulke}, \bibinfo{person}{Yingbo Gao}, \bibinfo{person}{Petrus Pelser}, \bibinfo{person}{Rein Brune}, \bibinfo{person}{Rricha Jalota}, \bibinfo{person}{Floris Fok}, \bibinfo{person}{Michael Ramos}, \bibinfo{person}{Ian van Wyk}, \bibinfo{person}{Abdallah Nasir}, \bibinfo{person}{Hayden Goldstein}, {et~al\mbox{.}}} \bibinfo{year}{2024}\natexlab{}.
\newblock \showarticletitle{Climategpt: Towards ai synthesizing interdisciplinary research on climate change}.
\newblock \bibinfo{journal}{\emph{arXiv preprint arXiv:2401.09646}} (\bibinfo{year}{2024}).
\newblock


\bibitem[Ullah et~al\mbox{.}(2020)]%
        {ullah2020applications}
\bibfield{author}{\bibinfo{person}{Zaib Ullah}, \bibinfo{person}{Fadi Al-Turjman}, \bibinfo{person}{Leonardo Mostarda}, {and} \bibinfo{person}{Roberto Gagliardi}.} \bibinfo{year}{2020}\natexlab{}.
\newblock \showarticletitle{Applications of artificial intelligence and machine learning in smart cities}.
\newblock \bibinfo{journal}{\emph{Computer Communications}}  \bibinfo{volume}{154} (\bibinfo{year}{2020}), \bibinfo{pages}{313--323}.
\newblock


\bibitem[Vaghefi et~al\mbox{.}(2023)]%
        {vaghefi2023chatclimate}
\bibfield{author}{\bibinfo{person}{Saeid~Ashraf Vaghefi}, \bibinfo{person}{Dominik Stammbach}, \bibinfo{person}{Veruska Muccione}, \bibinfo{person}{Julia Bingler}, \bibinfo{person}{Jingwei Ni}, \bibinfo{person}{Mathias Kraus}, \bibinfo{person}{Simon Allen}, \bibinfo{person}{Chiara Colesanti-Senni}, \bibinfo{person}{Tobias Wekhof}, \bibinfo{person}{Tobias Schimanski}, {et~al\mbox{.}}} \bibinfo{year}{2023}\natexlab{}.
\newblock \showarticletitle{ChatClimate: Grounding conversational AI in climate science}.
\newblock \bibinfo{journal}{\emph{Communications Earth \& Environment}} \bibinfo{volume}{4}, \bibinfo{number}{1} (\bibinfo{year}{2023}), \bibinfo{pages}{480}.
\newblock


\bibitem[Vaswani et~al\mbox{.}(2017)]%
        {vaswani2017attention}
\bibfield{author}{\bibinfo{person}{Ashish Vaswani}, \bibinfo{person}{Noam Shazeer}, \bibinfo{person}{Niki Parmar}, \bibinfo{person}{Jakob Uszkoreit}, \bibinfo{person}{Llion Jones}, \bibinfo{person}{Aidan~N Gomez}, \bibinfo{person}{{\L}ukasz Kaiser}, {and} \bibinfo{person}{Illia Polosukhin}.} \bibinfo{year}{2017}\natexlab{}.
\newblock \showarticletitle{Attention is all you need}.
\newblock \bibinfo{journal}{\emph{Advances in neural information processing systems}}  \bibinfo{volume}{30} (\bibinfo{year}{2017}).
\newblock


\bibitem[Verma et~al\mbox{.}(2023)]%
        {verma2023generative}
\bibfield{author}{\bibinfo{person}{Deepank Verma}, \bibinfo{person}{Olaf Mumm}, {and} \bibinfo{person}{Vanessa~Miriam Carlow}.} \bibinfo{year}{2023}\natexlab{}.
\newblock \showarticletitle{Generative agents in the streets: Exploring the use of Large Language Models (LLMs) in collecting urban perceptions}.
\newblock \bibinfo{journal}{\emph{arXiv preprint arXiv:2312.13126}} (\bibinfo{year}{2023}).
\newblock


\bibitem[von Wahl et~al\mbox{.}(2022)]%
        {von2022reinforcement}
\bibfield{author}{\bibinfo{person}{Leonie von Wahl}, \bibinfo{person}{Nicolas Tempelmeier}, \bibinfo{person}{Ashutosh Sao}, {and} \bibinfo{person}{Elena Demidova}.} \bibinfo{year}{2022}\natexlab{}.
\newblock \showarticletitle{Reinforcement learning-based placement of charging stations in urban road networks}. In \bibinfo{booktitle}{\emph{Proceedings of the 28th ACM SIGKDD Conference on Knowledge Discovery and Data Mining}}. \bibinfo{pages}{3992--4000}.
\newblock


\bibitem[Wang et~al\mbox{.}(2024a)]%
        {wang2024traffic}
\bibfield{author}{\bibinfo{person}{Bingzhang Wang}, \bibinfo{person}{Zhiyu Cai}, \bibinfo{person}{Muhammad~Monjurul Karim}, \bibinfo{person}{Chenxi Liu}, {and} \bibinfo{person}{Yinhai Wang}.} \bibinfo{year}{2024}\natexlab{a}.
\newblock \showarticletitle{Traffic performance gpt (tp-gpt): Real-time data informed intelligent chatbot for transportation surveillance and management}.
\newblock \bibinfo{journal}{\emph{arXiv preprint arXiv:2405.03076}} (\bibinfo{year}{2024}).
\newblock


\bibitem[Wang et~al\mbox{.}(2025)]%
        {wang2025chattime}
\bibfield{author}{\bibinfo{person}{Chengsen Wang}, \bibinfo{person}{Qi Qi}, \bibinfo{person}{Jingyu Wang}, \bibinfo{person}{Haifeng Sun}, \bibinfo{person}{Zirui Zhuang}, \bibinfo{person}{Jinming Wu}, \bibinfo{person}{Lei Zhang}, {and} \bibinfo{person}{Jianxin Liao}.} \bibinfo{year}{2025}\natexlab{}.
\newblock \showarticletitle{Chattime: A unified multimodal time series foundation model bridging numerical and textual data}. In \bibinfo{booktitle}{\emph{Proceedings of the AAAI Conference on Artificial Intelligence}}, Vol.~\bibinfo{volume}{39}. \bibinfo{pages}{12694--12702}.
\newblock


\bibitem[Wang et~al\mbox{.}(2024d)]%
        {wang2024survey}
\bibfield{author}{\bibinfo{person}{Lei Wang}, \bibinfo{person}{Chen Ma}, \bibinfo{person}{Xueyang Feng}, \bibinfo{person}{Zeyu Zhang}, \bibinfo{person}{Hao Yang}, \bibinfo{person}{Jingsen Zhang}, \bibinfo{person}{Zhiyuan Chen}, \bibinfo{person}{Jiakai Tang}, \bibinfo{person}{Xu Chen}, \bibinfo{person}{Yankai Lin}, {et~al\mbox{.}}} \bibinfo{year}{2024}\natexlab{d}.
\newblock \showarticletitle{A survey on large language model based autonomous agents}.
\newblock \bibinfo{journal}{\emph{Frontiers of Computer Science}} \bibinfo{volume}{18}, \bibinfo{number}{6} (\bibinfo{year}{2024}), \bibinfo{pages}{186345}.
\newblock


\bibitem[Wang et~al\mbox{.}(2024e)]%
        {wang2024llm}
\bibfield{author}{\bibinfo{person}{Maonan Wang}, \bibinfo{person}{Aoyu Pang}, \bibinfo{person}{Yuheng Kan}, \bibinfo{person}{Man-On Pun}, \bibinfo{person}{Chung~Shue Chen}, {and} \bibinfo{person}{Bo Huang}.} \bibinfo{year}{2024}\natexlab{e}.
\newblock \showarticletitle{LLM-assisted light: Leveraging large language model capabilities for human-mimetic traffic signal control in complex urban environments}.
\newblock \bibinfo{journal}{\emph{arXiv preprint arXiv:2403.08337}} (\bibinfo{year}{2024}).
\newblock


\bibitem[Wang et~al\mbox{.}(2023b)]%
        {wang2023would}
\bibfield{author}{\bibinfo{person}{Xinglei Wang}, \bibinfo{person}{Meng Fang}, \bibinfo{person}{Zichao Zeng}, {and} \bibinfo{person}{Tao Cheng}.} \bibinfo{year}{2023}\natexlab{b}.
\newblock \showarticletitle{Where would i go next? large language models as human mobility predictors}.
\newblock \bibinfo{journal}{\emph{arXiv preprint arXiv:2308.15197}} (\bibinfo{year}{2023}).
\newblock


\bibitem[Wang et~al\mbox{.}(2024c)]%
        {wang2024news}
\bibfield{author}{\bibinfo{person}{Xinlei Wang}, \bibinfo{person}{Maike Feng}, \bibinfo{person}{Jing Qiu}, \bibinfo{person}{Jinjin Gu}, {and} \bibinfo{person}{Junhua Zhao}.} \bibinfo{year}{2024}\natexlab{c}.
\newblock \showarticletitle{From news to forecast: Integrating event analysis in llm-based time series forecasting with reflection}.
\newblock \bibinfo{journal}{\emph{Advances in Neural Information Processing Systems}}  \bibinfo{volume}{37} (\bibinfo{year}{2024}), \bibinfo{pages}{58118--58153}.
\newblock


\bibitem[Wang et~al\mbox{.}(2024b)]%
        {wang2024simulating}
\bibfield{author}{\bibinfo{person}{Yiding Wang}, \bibinfo{person}{Yuxuan Chen}, \bibinfo{person}{Fangwei Zhong}, \bibinfo{person}{Long Ma}, {and} \bibinfo{person}{Yizhou Wang}.} \bibinfo{year}{2024}\natexlab{b}.
\newblock \showarticletitle{Simulating Human-like Daily Activities with Desire-driven Autonomy}.
\newblock \bibinfo{journal}{\emph{arXiv preprint arXiv:2412.06435}} (\bibinfo{year}{2024}).
\newblock


\bibitem[Wang and Zhao(2024)]%
        {wang2024tram}
\bibfield{author}{\bibinfo{person}{Yuqing Wang} {and} \bibinfo{person}{Yun Zhao}.} \bibinfo{year}{2024}\natexlab{}.
\newblock \showarticletitle{TRAM: Benchmarking Temporal Reasoning for Large Language Models}. In \bibinfo{booktitle}{\emph{Findings of the Association for Computational Linguistics ACL 2024}}. \bibinfo{pages}{6389--6415}.
\newblock


\bibitem[Wang et~al\mbox{.}(2023a)]%
        {wang2023humanoid}
\bibfield{author}{\bibinfo{person}{Zhilin Wang}, \bibinfo{person}{Yu~Ying Chiu}, {and} \bibinfo{person}{Yu~Cheung Chiu}.} \bibinfo{year}{2023}\natexlab{a}.
\newblock \showarticletitle{Humanoid Agents: Platform for Simulating Human-like Generative Agents}. In \bibinfo{booktitle}{\emph{Proceedings of the 2023 Conference on Empirical Methods in Natural Language Processing: System Demonstrations}}. \bibinfo{pages}{167--176}.
\newblock


\bibitem[Webersinke et~al\mbox{.}(2021)]%
        {webersinke2021climatebert}
\bibfield{author}{\bibinfo{person}{Nicolas Webersinke}, \bibinfo{person}{Mathias Kraus}, \bibinfo{person}{Julia~Anna Bingler}, {and} \bibinfo{person}{Markus Leippold}.} \bibinfo{year}{2021}\natexlab{}.
\newblock \showarticletitle{Climatebert: A pretrained language model for climate-related text}.
\newblock \bibinfo{journal}{\emph{arXiv preprint arXiv:2110.12010}} (\bibinfo{year}{2021}).
\newblock


\bibitem[Wei et~al\mbox{.}(2019)]%
        {wei2019colight}
\bibfield{author}{\bibinfo{person}{Hua Wei}, \bibinfo{person}{Nan Xu}, \bibinfo{person}{Huichu Zhang}, \bibinfo{person}{Guanjie Zheng}, \bibinfo{person}{Xinshi Zang}, \bibinfo{person}{Chacha Chen}, \bibinfo{person}{Weinan Zhang}, \bibinfo{person}{Yanmin Zhu}, \bibinfo{person}{Kai Xu}, {and} \bibinfo{person}{Zhenhui Li}.} \bibinfo{year}{2019}\natexlab{}.
\newblock \showarticletitle{Colight: Learning network-level cooperation for traffic signal control}. In \bibinfo{booktitle}{\emph{Proceedings of the 28th ACM international conference on information and knowledge management}}. \bibinfo{pages}{1913--1922}.
\newblock


\bibitem[Wei et~al\mbox{.}(2021)]%
        {wei2021finetuned}
\bibfield{author}{\bibinfo{person}{Jason Wei}, \bibinfo{person}{Maarten Bosma}, \bibinfo{person}{Vincent~Y Zhao}, \bibinfo{person}{Kelvin Guu}, \bibinfo{person}{Adams~Wei Yu}, \bibinfo{person}{Brian Lester}, \bibinfo{person}{Nan Du}, \bibinfo{person}{Andrew~M Dai}, {and} \bibinfo{person}{Quoc~V Le}.} \bibinfo{year}{2021}\natexlab{}.
\newblock \showarticletitle{Finetuned language models are zero-shot learners}.
\newblock \bibinfo{journal}{\emph{arXiv preprint arXiv:2109.01652}} (\bibinfo{year}{2021}).
\newblock


\bibitem[Wei et~al\mbox{.}(2022)]%
        {wei2022chain}
\bibfield{author}{\bibinfo{person}{Jason Wei}, \bibinfo{person}{Xuezhi Wang}, \bibinfo{person}{Dale Schuurmans}, \bibinfo{person}{Maarten Bosma}, \bibinfo{person}{Fei Xia}, \bibinfo{person}{Ed Chi}, \bibinfo{person}{Quoc~V Le}, \bibinfo{person}{Denny Zhou}, {et~al\mbox{.}}} \bibinfo{year}{2022}\natexlab{}.
\newblock \showarticletitle{Chain-of-thought prompting elicits reasoning in large language models}.
\newblock \bibinfo{journal}{\emph{Advances in neural information processing systems}}  \bibinfo{volume}{35} (\bibinfo{year}{2022}), \bibinfo{pages}{24824--24837}.
\newblock


\bibitem[Wickramasekara et~al\mbox{.}(2025)]%
        {wickramasekara2025exploring}
\bibfield{author}{\bibinfo{person}{Akila Wickramasekara}, \bibinfo{person}{Frank Breitinger}, {and} \bibinfo{person}{Mark Scanlon}.} \bibinfo{year}{2025}\natexlab{}.
\newblock \showarticletitle{Exploring the potential of large language models for improving digital forensic investigation efficiency}.
\newblock \bibinfo{journal}{\emph{Forensic Science International: Digital Investigation}}  \bibinfo{volume}{52} (\bibinfo{year}{2025}), \bibinfo{pages}{301859}.
\newblock


\bibitem[Wiering et~al\mbox{.}(2004)]%
        {wiering2004intelligent}
\bibfield{author}{\bibinfo{person}{Marco Wiering}, \bibinfo{person}{Jelle Van~Veenen}, \bibinfo{person}{Jilles Vreeken}, {and} \bibinfo{person}{Arne Koopman}.} \bibinfo{year}{2004}\natexlab{}.
\newblock \showarticletitle{Intelligent traffic light control}.
\newblock \bibinfo{journal}{\emph{Institute of Information and Computing Sciences. Utrecht University}} (\bibinfo{year}{2004}).
\newblock


\bibitem[Wiering et~al\mbox{.}(2000)]%
        {wiering2000multi}
\bibfield{author}{\bibinfo{person}{Marco~A Wiering} {et~al\mbox{.}}} \bibinfo{year}{2000}\natexlab{}.
\newblock \showarticletitle{Multi-agent reinforcement learning for traffic light control}. In \bibinfo{booktitle}{\emph{Machine Learning: Proceedings of the Seventeenth International Conference (ICML'2000)}}. \bibinfo{pages}{1151--1158}.
\newblock


\bibitem[Williams et~al\mbox{.}(2023)]%
        {williams2023epidemic}
\bibfield{author}{\bibinfo{person}{Ross Williams}, \bibinfo{person}{Niyousha Hosseinichimeh}, \bibinfo{person}{Aritra Majumdar}, {and} \bibinfo{person}{Navid Ghaffarzadegan}.} \bibinfo{year}{2023}\natexlab{}.
\newblock \showarticletitle{Epidemic modeling with generative agents}.
\newblock \bibinfo{journal}{\emph{arXiv preprint arXiv:2307.04986}} (\bibinfo{year}{2023}).
\newblock


\bibitem[Wooldridge and Jennings(1995)]%
        {wooldridge1995intelligent}
\bibfield{author}{\bibinfo{person}{Michael Wooldridge} {and} \bibinfo{person}{Nicholas~R Jennings}.} \bibinfo{year}{1995}\natexlab{}.
\newblock \showarticletitle{Intelligent agents: Theory and practice}.
\newblock \bibinfo{journal}{\emph{The knowledge engineering review}} \bibinfo{volume}{10}, \bibinfo{number}{2} (\bibinfo{year}{1995}), \bibinfo{pages}{115--152}.
\newblock


\bibitem[Wu et~al\mbox{.}(2024b)]%
        {wu2024lade}
\bibfield{author}{\bibinfo{person}{Lixia Wu}, \bibinfo{person}{Haomin Wen}, \bibinfo{person}{Haoyuan Hu}, \bibinfo{person}{Xiaowei Mao}, \bibinfo{person}{Yutong Xia}, \bibinfo{person}{Ergang Shan}, \bibinfo{person}{Jianbin Zheng}, \bibinfo{person}{Junhong Lou}, \bibinfo{person}{Yuxuan Liang}, \bibinfo{person}{Liuqing Yang}, {et~al\mbox{.}}} \bibinfo{year}{2024}\natexlab{b}.
\newblock \showarticletitle{LaDe: The First Comprehensive Last-mile Express Dataset from Industry}. In \bibinfo{booktitle}{\emph{Proceedings of the 30th ACM SIGKDD Conference on Knowledge Discovery and Data Mining}}. \bibinfo{pages}{5991--6002}.
\newblock


\bibitem[Wu et~al\mbox{.}(2024a)]%
        {wu2024mind}
\bibfield{author}{\bibinfo{person}{Wenshan Wu}, \bibinfo{person}{Shaoguang Mao}, \bibinfo{person}{Yadong Zhang}, \bibinfo{person}{Yan Xia}, \bibinfo{person}{Li Dong}, \bibinfo{person}{Lei Cui}, {and} \bibinfo{person}{Furu Wei}.} \bibinfo{year}{2024}\natexlab{a}.
\newblock \showarticletitle{Mind's Eye of LLMs: Visualization-of-Thought Elicits Spatial Reasoning in Large Language Models}. In \bibinfo{booktitle}{\emph{The Thirty-eighth Annual Conference on Neural Information Processing Systems}}.
\newblock


\bibitem[Xi et~al\mbox{.}(2025)]%
        {xi2025rise}
\bibfield{author}{\bibinfo{person}{Zhiheng Xi}, \bibinfo{person}{Wenxiang Chen}, \bibinfo{person}{Xin Guo}, \bibinfo{person}{Wei He}, \bibinfo{person}{Yiwen Ding}, \bibinfo{person}{Boyang Hong}, \bibinfo{person}{Ming Zhang}, \bibinfo{person}{Junzhe Wang}, \bibinfo{person}{Senjie Jin}, \bibinfo{person}{Enyu Zhou}, {et~al\mbox{.}}} \bibinfo{year}{2025}\natexlab{}.
\newblock \showarticletitle{The rise and potential of large language model based agents: A survey}.
\newblock \bibinfo{journal}{\emph{Science China Information Sciences}} \bibinfo{volume}{68}, \bibinfo{number}{2} (\bibinfo{year}{2025}), \bibinfo{pages}{121101}.
\newblock


\bibitem[Xia et~al\mbox{.}(2024)]%
        {xia2024question}
\bibfield{author}{\bibinfo{person}{Yongqi Xia}, \bibinfo{person}{Yi Huang}, \bibinfo{person}{Qianqian Qiu}, \bibinfo{person}{Xueying Zhang}, \bibinfo{person}{Lizhi Miao}, {and} \bibinfo{person}{Yixiang Chen}.} \bibinfo{year}{2024}\natexlab{}.
\newblock \showarticletitle{A question and answering service of typhoon disasters based on the t5 large language model}.
\newblock \bibinfo{journal}{\emph{ISPRS International Journal of Geo-Information}} \bibinfo{volume}{13}, \bibinfo{number}{5} (\bibinfo{year}{2024}), \bibinfo{pages}{165}.
\newblock


\bibitem[Xia et~al\mbox{.}(2025)]%
        {xia2025reimagining}
\bibfield{author}{\bibinfo{person}{Yutong Xia}, \bibinfo{person}{Ao Qu}, \bibinfo{person}{Yunhan Zheng}, \bibinfo{person}{Yihong Tang}, \bibinfo{person}{Dingyi Zhuang}, \bibinfo{person}{Yuxuan Liang}, \bibinfo{person}{Cathy Wu}, \bibinfo{person}{Roger Zimmermann}, {and} \bibinfo{person}{Jinhua Zhao}.} \bibinfo{year}{2025}\natexlab{}.
\newblock \showarticletitle{Reimagining Urban Science: Scaling Causal Inference with Large Language Models}.
\newblock \bibinfo{journal}{\emph{arXiv preprint arXiv:2504.12345}} (\bibinfo{year}{2025}).
\newblock


\bibitem[Xiao et~al\mbox{.}(2024)]%
        {xiao2024refound}
\bibfield{author}{\bibinfo{person}{Congxi Xiao}, \bibinfo{person}{Jingbo Zhou}, \bibinfo{person}{Yixiong Xiao}, \bibinfo{person}{Jizhou Huang}, {and} \bibinfo{person}{Hui Xiong}.} \bibinfo{year}{2024}\natexlab{}.
\newblock \showarticletitle{ReFound: Crafting a Foundation Model for Urban Region Understanding upon Language and Visual Foundations}. In \bibinfo{booktitle}{\emph{Proceedings of the 30th ACM SIGKDD Conference on Knowledge Discovery and Data Mining}}. \bibinfo{pages}{3527--3538}.
\newblock


\bibitem[Xu et~al\mbox{.}(2023)]%
        {xu2023urban}
\bibfield{author}{\bibinfo{person}{Fengli Xu}, \bibinfo{person}{Jun Zhang}, \bibinfo{person}{Chen Gao}, \bibinfo{person}{Jie Feng}, {and} \bibinfo{person}{Yong Li}.} \bibinfo{year}{2023}\natexlab{}.
\newblock \showarticletitle{Urban generative intelligence (ugi): A foundational platform for agents in embodied city environment}.
\newblock \bibinfo{journal}{\emph{arXiv preprint arXiv:2312.11813}} (\bibinfo{year}{2023}).
\newblock


\bibitem[Xu et~al\mbox{.}(2025)]%
        {xu2025mm}
\bibfield{author}{\bibinfo{person}{Ronghui Xu}, \bibinfo{person}{Hanyin Cheng}, \bibinfo{person}{Chenjuan Guo}, \bibinfo{person}{Hongfan Gao}, \bibinfo{person}{Jilin Hu}, \bibinfo{person}{Sean~Bin Yang}, {and} \bibinfo{person}{Bin Yang}.} \bibinfo{year}{2025}\natexlab{}.
\newblock \showarticletitle{Mm-path: Multi-modal, multi-granularity path representation learning}. In \bibinfo{booktitle}{\emph{Proceedings of the 31st ACM SIGKDD Conference on Knowledge Discovery and Data Mining V. 1}}. \bibinfo{pages}{1703--1714}.
\newblock


\bibitem[Yan(2021)]%
        {yan2021fairness}
\bibfield{author}{\bibinfo{person}{An Yan}.} \bibinfo{year}{2021}\natexlab{}.
\newblock \bibinfo{booktitle}{\emph{Fairness-aware Spatio-temporal Prediction for Cities}}.
\newblock \bibinfo{publisher}{University of Washington}.
\newblock


\bibitem[Yan and Howe(2019)]%
        {yan2019fairst}
\bibfield{author}{\bibinfo{person}{An Yan} {and} \bibinfo{person}{Bill Howe}.} \bibinfo{year}{2019}\natexlab{}.
\newblock \showarticletitle{Fairst: Equitable spatial and temporal demand prediction for new mobility systems}. In \bibinfo{booktitle}{\emph{Proceedings of the 27th ACM SIGSPATIAL International Conference on Advances in Geographic Information Systems}}. \bibinfo{pages}{552--555}.
\newblock


\bibitem[Yan et~al\mbox{.}(2024a)]%
        {yan2024urbanclip}
\bibfield{author}{\bibinfo{person}{Yibo Yan}, \bibinfo{person}{Haomin Wen}, \bibinfo{person}{Siru Zhong}, \bibinfo{person}{Wei Chen}, \bibinfo{person}{Haodong Chen}, \bibinfo{person}{Qingsong Wen}, \bibinfo{person}{Roger Zimmermann}, {and} \bibinfo{person}{Yuxuan Liang}.} \bibinfo{year}{2024}\natexlab{a}.
\newblock \showarticletitle{Urbanclip: Learning text-enhanced urban region profiling with contrastive language-image pretraining from the web}. In \bibinfo{booktitle}{\emph{Proceedings of the ACM Web Conference 2024}}. \bibinfo{pages}{4006--4017}.
\newblock


\bibitem[Yan et~al\mbox{.}(2024b)]%
        {yan2024opencity}
\bibfield{author}{\bibinfo{person}{Yuwei Yan}, \bibinfo{person}{Qingbin Zeng}, \bibinfo{person}{Zhiheng Zheng}, \bibinfo{person}{Jingzhe Yuan}, \bibinfo{person}{Jie Feng}, \bibinfo{person}{Jun Zhang}, \bibinfo{person}{Fengli Xu}, {and} \bibinfo{person}{Yong Li}.} \bibinfo{year}{2024}\natexlab{b}.
\newblock \showarticletitle{OpenCity: A Scalable Platform to Simulate Urban Activities with Massive LLM Agents}.
\newblock \bibinfo{journal}{\emph{arXiv preprint arXiv:2410.21286}} (\bibinfo{year}{2024}).
\newblock


\bibitem[Yang et~al\mbox{.}(2024b)]%
        {yang2024thinking}
\bibfield{author}{\bibinfo{person}{Jihan Yang}, \bibinfo{person}{Shusheng Yang}, \bibinfo{person}{Anjali~W Gupta}, \bibinfo{person}{Rilyn Han}, \bibinfo{person}{Li Fei-Fei}, {and} \bibinfo{person}{Saining Xie}.} \bibinfo{year}{2024}\natexlab{b}.
\newblock \showarticletitle{Thinking in space: How multimodal large language models see, remember, and recall spaces}.
\newblock \bibinfo{journal}{\emph{arXiv preprint arXiv:2412.14171}} (\bibinfo{year}{2024}).
\newblock


\bibitem[Yang et~al\mbox{.}(2024a)]%
        {yang2024llm}
\bibfield{author}{\bibinfo{person}{Joshua~C Yang}, \bibinfo{person}{Damian Dalisan}, \bibinfo{person}{Marcin Korecki}, \bibinfo{person}{Carina~I Hausladen}, {and} \bibinfo{person}{Dirk Helbing}.} \bibinfo{year}{2024}\natexlab{a}.
\newblock \showarticletitle{LLM Voting: Human Choices and AI Collective Decision-Making}. In \bibinfo{booktitle}{\emph{Proceedings of the AAAI/ACM Conference on AI, Ethics, and Society}}, Vol.~\bibinfo{volume}{7}. \bibinfo{pages}{1696--1708}.
\newblock


\bibitem[Yang et~al\mbox{.}(2019)]%
        {yang2019federated}
\bibfield{author}{\bibinfo{person}{Qiang Yang}, \bibinfo{person}{Yang Liu}, \bibinfo{person}{Tianjian Chen}, {and} \bibinfo{person}{Yongxin Tong}.} \bibinfo{year}{2019}\natexlab{}.
\newblock \showarticletitle{Federated machine learning: Concept and applications}.
\newblock \bibinfo{journal}{\emph{ACM Transactions on Intelligent Systems and Technology (TIST)}} \bibinfo{volume}{10}, \bibinfo{number}{2} (\bibinfo{year}{2019}), \bibinfo{pages}{1--19}.
\newblock


\bibitem[Yi et~al\mbox{.}(2024)]%
        {yi2024filternet}
\bibfield{author}{\bibinfo{person}{Kun Yi}, \bibinfo{person}{Jingru Fei}, \bibinfo{person}{Qi Zhang}, \bibinfo{person}{Hui He}, \bibinfo{person}{Shufeng Hao}, \bibinfo{person}{Defu Lian}, {and} \bibinfo{person}{Wei Fan}.} \bibinfo{year}{2024}\natexlab{}.
\newblock \showarticletitle{Filternet: Harnessing frequency filters for time series forecasting}.
\newblock \bibinfo{journal}{\emph{Advances in Neural Information Processing Systems}}  \bibinfo{volume}{37} (\bibinfo{year}{2024}), \bibinfo{pages}{55115--55140}.
\newblock


\bibitem[Yin et~al\mbox{.}(2024)]%
        {yin2024survey}
\bibfield{author}{\bibinfo{person}{Shukang Yin}, \bibinfo{person}{Chaoyou Fu}, \bibinfo{person}{Sirui Zhao}, \bibinfo{person}{Ke Li}, \bibinfo{person}{Xing Sun}, \bibinfo{person}{Tong Xu}, {and} \bibinfo{person}{Enhong Chen}.} \bibinfo{year}{2024}\natexlab{}.
\newblock \showarticletitle{A survey on multimodal large language models}.
\newblock \bibinfo{journal}{\emph{National Science Review}} \bibinfo{volume}{11}, \bibinfo{number}{12} (\bibinfo{year}{2024}).
\newblock


\bibitem[Yu et~al\mbox{.}(2025)]%
        {yu2025spatial}
\bibfield{author}{\bibinfo{person}{Dazhou Yu}, \bibinfo{person}{Riyang Bao}, \bibinfo{person}{Gengchen Mai}, {and} \bibinfo{person}{Liang Zhao}.} \bibinfo{year}{2025}\natexlab{}.
\newblock \showarticletitle{Spatial-rag: Spatial retrieval augmented generation for real-world spatial reasoning questions}.
\newblock \bibinfo{journal}{\emph{arXiv preprint arXiv:2502.18470}} (\bibinfo{year}{2025}).
\newblock


\bibitem[Yu et~al\mbox{.}(2015)]%
        {yu2015geospark}
\bibfield{author}{\bibinfo{person}{Jia Yu}, \bibinfo{person}{Jinxuan Wu}, {and} \bibinfo{person}{Mohamed Sarwat}.} \bibinfo{year}{2015}\natexlab{}.
\newblock \showarticletitle{Geospark: A cluster computing framework for processing large-scale spatial data}. In \bibinfo{booktitle}{\emph{Proceedings of the 23rd SIGSPATIAL international conference on advances in geographic information systems}}. \bibinfo{pages}{1--4}.
\newblock


\bibitem[Yuan et~al\mbox{.}(2024b)]%
        {yuan2024back}
\bibfield{author}{\bibinfo{person}{Chenhan Yuan}, \bibinfo{person}{Qianqian Xie}, \bibinfo{person}{Jimin Huang}, {and} \bibinfo{person}{Sophia Ananiadou}.} \bibinfo{year}{2024}\natexlab{b}.
\newblock \showarticletitle{Back to the future: Towards explainable temporal reasoning with large language models}. In \bibinfo{booktitle}{\emph{Proceedings of the ACM Web Conference 2024}}. \bibinfo{pages}{1963--1974}.
\newblock


\bibitem[Yuan et~al\mbox{.}(2024a)]%
        {yuan2024unist}
\bibfield{author}{\bibinfo{person}{Yuan Yuan}, \bibinfo{person}{Jingtao Ding}, \bibinfo{person}{Jie Feng}, \bibinfo{person}{Depeng Jin}, {and} \bibinfo{person}{Yong Li}.} \bibinfo{year}{2024}\natexlab{a}.
\newblock \showarticletitle{Unist: A prompt-empowered universal model for urban spatio-temporal prediction}. In \bibinfo{booktitle}{\emph{Proceedings of the 30th ACM SIGKDD Conference on Knowledge Discovery and Data Mining}}. \bibinfo{pages}{4095--4106}.
\newblock


\bibitem[Yuan et~al\mbox{.}(2025)]%
        {yuan2025collmlight}
\bibfield{author}{\bibinfo{person}{Zirui Yuan}, \bibinfo{person}{Siqi Lai}, {and} \bibinfo{person}{Hao Liu}.} \bibinfo{year}{2025}\natexlab{}.
\newblock \showarticletitle{CoLLMLight: Cooperative Large Language Model Agents for Network-Wide Traffic Signal Control}.
\newblock \bibinfo{journal}{\emph{arXiv preprint arXiv:2503.11739}} (\bibinfo{year}{2025}).
\newblock


\bibitem[Zhai et~al\mbox{.}(2025)]%
        {zhai2025heterogeneous}
\bibfield{author}{\bibinfo{person}{Xuehao Zhai}, \bibinfo{person}{Junqi Jiang}, \bibinfo{person}{Adam Dejl}, \bibinfo{person}{Antonio Rago}, \bibinfo{person}{Fangce Guo}, \bibinfo{person}{Francesca Toni}, {and} \bibinfo{person}{Aruna Sivakumar}.} \bibinfo{year}{2025}\natexlab{}.
\newblock \showarticletitle{Heterogeneous graph neural networks with post-hoc explanations for multi-modal and explainable land use inference}.
\newblock \bibinfo{journal}{\emph{Information Fusion}} (\bibinfo{year}{2025}), \bibinfo{pages}{103057}.
\newblock


\bibitem[Zhan et~al\mbox{.}(2022)]%
        {zhan2022deepthermal}
\bibfield{author}{\bibinfo{person}{Xianyuan Zhan}, \bibinfo{person}{Haoran Xu}, \bibinfo{person}{Yue Zhang}, \bibinfo{person}{Xiangyu Zhu}, \bibinfo{person}{Honglei Yin}, {and} \bibinfo{person}{Yu Zheng}.} \bibinfo{year}{2022}\natexlab{}.
\newblock \showarticletitle{Deepthermal: Combustion optimization for thermal power generating units using offline reinforcement learning}. In \bibinfo{booktitle}{\emph{Proceedings of the AAAI Conference on Artificial Intelligence}}, Vol.~\bibinfo{volume}{36}. \bibinfo{pages}{4680--4688}.
\newblock


\bibitem[Zhang et~al\mbox{.}(2017b)]%
        {zhang2017deep}
\bibfield{author}{\bibinfo{person}{Junbo Zhang}, \bibinfo{person}{Yu Zheng}, {and} \bibinfo{person}{Dekang Qi}.} \bibinfo{year}{2017}\natexlab{b}.
\newblock \showarticletitle{Deep spatio-temporal residual networks for citywide crowd flows prediction}. In \bibinfo{booktitle}{\emph{Proceedings of the AAAI conference on artificial intelligence}}, Vol.~\bibinfo{volume}{31}.
\newblock


\bibitem[Zhang et~al\mbox{.}(2017a)]%
        {zhang2017security}
\bibfield{author}{\bibinfo{person}{Kuan Zhang}, \bibinfo{person}{Jianbing Ni}, \bibinfo{person}{Kan Yang}, \bibinfo{person}{Xiaohui Liang}, \bibinfo{person}{Ju Ren}, {and} \bibinfo{person}{Xuemin~Sherman Shen}.} \bibinfo{year}{2017}\natexlab{a}.
\newblock \showarticletitle{Security and privacy in smart city applications: Challenges and solutions}.
\newblock \bibinfo{journal}{\emph{IEEE communications magazine}} \bibinfo{volume}{55}, \bibinfo{number}{1} (\bibinfo{year}{2017}), \bibinfo{pages}{122--129}.
\newblock


\bibitem[Zhang et~al\mbox{.}(2024a)]%
        {zhang2024trafficgpt}
\bibfield{author}{\bibinfo{person}{Siyao Zhang}, \bibinfo{person}{Daocheng Fu}, \bibinfo{person}{Wenzhe Liang}, \bibinfo{person}{Zhao Zhang}, \bibinfo{person}{Bin Yu}, \bibinfo{person}{Pinlong Cai}, {and} \bibinfo{person}{Baozhen Yao}.} \bibinfo{year}{2024}\natexlab{a}.
\newblock \showarticletitle{Trafficgpt: Viewing, processing and interacting with traffic foundation models}.
\newblock \bibinfo{journal}{\emph{Transport Policy}}  \bibinfo{volume}{150} (\bibinfo{year}{2024}), \bibinfo{pages}{95--105}.
\newblock


\bibitem[Zhang et~al\mbox{.}(2024b)]%
        {zhang2024meta}
\bibfield{author}{\bibinfo{person}{Weijia Zhang}, \bibinfo{person}{Jindong Han}, \bibinfo{person}{Hao Liu}, \bibinfo{person}{Wei Fan}, \bibinfo{person}{Hao Wang}, {and} \bibinfo{person}{Hui Xiong}.} \bibinfo{year}{2024}\natexlab{b}.
\newblock \showarticletitle{Meta-Transfer Learning Empowered Temporal Graph Networks for Cross-City Real Estate Appraisal}.
\newblock \bibinfo{journal}{\emph{arXiv preprint arXiv:2410.08947}} (\bibinfo{year}{2024}).
\newblock


\bibitem[Zhang et~al\mbox{.}(2024c)]%
        {zhang2024towards}
\bibfield{author}{\bibinfo{person}{Weijia Zhang}, \bibinfo{person}{Jindong Han}, \bibinfo{person}{Zhao Xu}, \bibinfo{person}{Hang Ni}, \bibinfo{person}{Hao Liu}, {and} \bibinfo{person}{Hui Xiong}.} \bibinfo{year}{2024}\natexlab{c}.
\newblock \showarticletitle{Towards urban general intelligence: A review and outlook of urban foundation models}.
\newblock \bibinfo{journal}{\emph{arXiv preprint arXiv:2402.01749}} (\bibinfo{year}{2024}).
\newblock


\bibitem[Zhang et~al\mbox{.}(2022)]%
        {zhang2022multi}
\bibfield{author}{\bibinfo{person}{Weijia Zhang}, \bibinfo{person}{Hao Liu}, \bibinfo{person}{Jindong Han}, \bibinfo{person}{Yong Ge}, {and} \bibinfo{person}{Hui Xiong}.} \bibinfo{year}{2022}\natexlab{}.
\newblock \showarticletitle{Multi-agent graph convolutional reinforcement learning for dynamic electric vehicle charging pricing}. In \bibinfo{booktitle}{\emph{Proceedings of the 28th ACM SIGKDD conference on knowledge discovery and data mining}}. \bibinfo{pages}{2471--2481}.
\newblock


\bibitem[Zhao et~al\mbox{.}(2016)]%
        {zhao2016towards}
\bibfield{author}{\bibinfo{person}{Kaiqi Zhao}, \bibinfo{person}{Yiding Liu}, \bibinfo{person}{Quan Yuan}, \bibinfo{person}{Lisi Chen}, \bibinfo{person}{Zhida Chen}, {and} \bibinfo{person}{Gao Cong}.} \bibinfo{year}{2016}\natexlab{}.
\newblock \showarticletitle{Towards personalized maps: mining user preferences from geo-textual data}.
\newblock \bibinfo{journal}{\emph{Proceedings of the VLDB Endowment}} \bibinfo{volume}{9}, \bibinfo{number}{13} (\bibinfo{year}{2016}), \bibinfo{pages}{1545--1548}.
\newblock


\bibitem[Zheng(2015)]%
        {zheng2015trajectory}
\bibfield{author}{\bibinfo{person}{Yu Zheng}.} \bibinfo{year}{2015}\natexlab{}.
\newblock \showarticletitle{Trajectory data mining: an overview}.
\newblock \bibinfo{journal}{\emph{ACM Transactions on Intelligent Systems and Technology (TIST)}} \bibinfo{volume}{6}, \bibinfo{number}{3} (\bibinfo{year}{2015}), \bibinfo{pages}{1--41}.
\newblock


\bibitem[Zheng et~al\mbox{.}(2014a)]%
        {zheng2014urban}
\bibfield{author}{\bibinfo{person}{Yu Zheng}, \bibinfo{person}{Licia Capra}, \bibinfo{person}{Ouri Wolfson}, {and} \bibinfo{person}{Hai Yang}.} \bibinfo{year}{2014}\natexlab{a}.
\newblock \showarticletitle{Urban computing: concepts, methodologies, and applications}.
\newblock \bibinfo{journal}{\emph{ACM Transactions on Intelligent Systems and Technology (TIST)}} \bibinfo{volume}{5}, \bibinfo{number}{3} (\bibinfo{year}{2014}), \bibinfo{pages}{1--55}.
\newblock


\bibitem[Zheng et~al\mbox{.}(2024)]%
        {zheng2024survey}
\bibfield{author}{\bibinfo{person}{Yu Zheng}, \bibinfo{person}{Qianyue Hao}, \bibinfo{person}{Jingwei Wang}, \bibinfo{person}{Changzheng Gao}, \bibinfo{person}{Jinwei Chen}, \bibinfo{person}{Depeng Jin}, {and} \bibinfo{person}{Yong Li}.} \bibinfo{year}{2024}\natexlab{}.
\newblock \showarticletitle{A Survey of Machine Learning for Urban Decision Making: Applications in Planning, Transportation, and Healthcare}.
\newblock \bibinfo{journal}{\emph{Comput. Surveys}} \bibinfo{volume}{57}, \bibinfo{number}{4} (\bibinfo{year}{2024}), \bibinfo{pages}{1--41}.
\newblock


\bibitem[Zheng et~al\mbox{.}(2023)]%
        {zheng2023spatial}
\bibfield{author}{\bibinfo{person}{Yu Zheng}, \bibinfo{person}{Yuming Lin}, \bibinfo{person}{Liang Zhao}, \bibinfo{person}{Tinghai Wu}, \bibinfo{person}{Depeng Jin}, {and} \bibinfo{person}{Yong Li}.} \bibinfo{year}{2023}\natexlab{}.
\newblock \showarticletitle{Spatial planning of urban communities via deep reinforcement learning}.
\newblock \bibinfo{journal}{\emph{Nature Computational Science}} \bibinfo{volume}{3}, \bibinfo{number}{9} (\bibinfo{year}{2023}), \bibinfo{pages}{748--762}.
\newblock


\bibitem[Zheng et~al\mbox{.}({[n.\,d.]})]%
        {zhengurbanplanbench}
\bibfield{author}{\bibinfo{person}{Yu Zheng}, \bibinfo{person}{Longyi Liu}, \bibinfo{person}{Yuming Lin}, \bibinfo{person}{Jie Feng}, \bibinfo{person}{Guozhen Zhang}, \bibinfo{person}{Depeng Jin}, {and} \bibinfo{person}{Yong Li}.} \bibinfo{year}{[n.\,d.]}\natexlab{}.
\newblock \showarticletitle{UrbanPlanBench: A Comprehensive Assessment of Urban Planning Abilities in Large Language Models}.
\newblock  (\bibinfo{year}{[n.\,d.]}).
\newblock


\bibitem[Zheng et~al\mbox{.}(2025)]%
        {urbanplanbench2025}
\bibfield{author}{\bibinfo{person}{Yu Zheng}, \bibinfo{person}{Longyi Liu}, \bibinfo{person}{Yuming Lin}, \bibinfo{person}{Jie Feng}, \bibinfo{person}{Guozhen Zhang}, \bibinfo{person}{Depeng Jin}, {and} \bibinfo{person}{Yong Li}.} \bibinfo{year}{2025}\natexlab{}.
\newblock \bibinfo{title}{UrbanPlanBench: A Comprehensive Assessment of Urban Planning Abilities in Large Language Models}.
\newblock
\newblock
\urldef\tempurl%
\url{https://openreview.net/forum?id=Dl5JaX7zoN}
\showURL{%
\tempurl}


\bibitem[Zheng et~al\mbox{.}(2014b)]%
        {zheng2014diagnosing}
\bibfield{author}{\bibinfo{person}{Yu Zheng}, \bibinfo{person}{Tong Liu}, \bibinfo{person}{Yilun Wang}, \bibinfo{person}{Yanmin Zhu}, \bibinfo{person}{Yanchi Liu}, {and} \bibinfo{person}{Eric Chang}.} \bibinfo{year}{2014}\natexlab{b}.
\newblock \showarticletitle{Diagnosing New York city's noises with ubiquitous data}. In \bibinfo{booktitle}{\emph{Proceedings of the 2014 ACM international joint conference on pervasive and ubiquitous computing}}. \bibinfo{pages}{715--725}.
\newblock


\bibitem[Zhong et~al\mbox{.}(2025)]%
        {zhong2025time}
\bibfield{author}{\bibinfo{person}{Siru Zhong}, \bibinfo{person}{Weilin Ruan}, \bibinfo{person}{Ming Jin}, \bibinfo{person}{Huan Li}, \bibinfo{person}{Qingsong Wen}, {and} \bibinfo{person}{Yuxuan Liang}.} \bibinfo{year}{2025}\natexlab{}.
\newblock \showarticletitle{Time-VLM: Exploring Multimodal Vision-Language Models for Augmented Time Series Forecasting}.
\newblock \bibinfo{journal}{\emph{arXiv preprint arXiv:2502.04395}} (\bibinfo{year}{2025}).
\newblock


\bibitem[Zhou et~al\mbox{.}(2025)]%
        {urbench2025}
\bibfield{author}{\bibinfo{person}{Baichuan Zhou}, \bibinfo{person}{Haote Yang}, \bibinfo{person}{Dairong Chen}, \bibinfo{person}{Junyan Ye}, \bibinfo{person}{Tianyi Bai}, \bibinfo{person}{Jinhua Yu}, \bibinfo{person}{Songyang Zhang}, \bibinfo{person}{Dahua Lin}, \bibinfo{person}{Conghui He}, {and} \bibinfo{person}{Weijia Li}.} \bibinfo{year}{2025}\natexlab{}.
\newblock \showarticletitle{Urbench: A comprehensive benchmark for evaluating large multimodal models in multi-view urban scenarios}. In \bibinfo{booktitle}{\emph{Proceedings of the AAAI Conference on Artificial Intelligence}}, Vol.~\bibinfo{volume}{39}. \bibinfo{pages}{10707--10715}.
\newblock


\bibitem[Zhou et~al\mbox{.}(2021)]%
        {zhou2021informer}
\bibfield{author}{\bibinfo{person}{Haoyi Zhou}, \bibinfo{person}{Shanghang Zhang}, \bibinfo{person}{Jieqi Peng}, \bibinfo{person}{Shuai Zhang}, \bibinfo{person}{Jianxin Li}, \bibinfo{person}{Hui Xiong}, {and} \bibinfo{person}{Wancai Zhang}.} \bibinfo{year}{2021}\natexlab{}.
\newblock \showarticletitle{Informer: Beyond efficient transformer for long sequence time-series forecasting}. In \bibinfo{booktitle}{\emph{Proceedings of the AAAI conference on artificial intelligence}}, Vol.~\bibinfo{volume}{35}. \bibinfo{pages}{11106--11115}.
\newblock


\bibitem[Zhou et~al\mbox{.}(2019)]%
        {zhou2019cyber}
\bibfield{author}{\bibinfo{person}{Yuchen Zhou}, \bibinfo{person}{F~Richard Yu}, \bibinfo{person}{Jian Chen}, {and} \bibinfo{person}{Yonghong Kuo}.} \bibinfo{year}{2019}\natexlab{}.
\newblock \showarticletitle{Cyber-physical-social systems: A state-of-the-art survey, challenges and opportunities}.
\newblock \bibinfo{journal}{\emph{IEEE Communications Surveys \& Tutorials}} \bibinfo{volume}{22}, \bibinfo{number}{1} (\bibinfo{year}{2019}), \bibinfo{pages}{389--425}.
\newblock


\bibitem[Zhou et~al\mbox{.}(2024)]%
        {zhou2024large}
\bibfield{author}{\bibinfo{person}{Zhilun Zhou}, \bibinfo{person}{Yuming Lin}, \bibinfo{person}{Depeng Jin}, {and} \bibinfo{person}{Yong Li}.} \bibinfo{year}{2024}\natexlab{}.
\newblock \showarticletitle{Large language model for participatory urban planning}.
\newblock \bibinfo{journal}{\emph{arXiv preprint arXiv:2402.17161}} (\bibinfo{year}{2024}).
\newblock


\bibitem[Zhou and Yu(2024)]%
        {zhou2024can}
\bibfield{author}{\bibinfo{person}{Zihao Zhou} {and} \bibinfo{person}{Rose Yu}.} \bibinfo{year}{2024}\natexlab{}.
\newblock \showarticletitle{Can LLMs Understand Time Series Anomalies?}
\newblock \bibinfo{journal}{\emph{arXiv preprint arXiv:2410.05440}} (\bibinfo{year}{2024}).
\newblock


\bibitem[Zhou(2022)]%
        {zhou2022rehearsal}
\bibfield{author}{\bibinfo{person}{Zhi-Hua Zhou}.} \bibinfo{year}{2022}\natexlab{}.
\newblock \showarticletitle{Rehearsal: learning from prediction to decision}.
\newblock \bibinfo{journal}{\emph{Frontiers of Computer Science}} \bibinfo{volume}{16}, \bibinfo{number}{4} (\bibinfo{year}{2022}), \bibinfo{pages}{164352}.
\newblock


\bibitem[Zhu et~al\mbox{.}(2024)]%
        {zhu2024plangpt}
\bibfield{author}{\bibinfo{person}{He Zhu}, \bibinfo{person}{Wenjia Zhang}, \bibinfo{person}{Nuoxian Huang}, \bibinfo{person}{Boyang Li}, \bibinfo{person}{Luyao Niu}, \bibinfo{person}{Zipei Fan}, \bibinfo{person}{Tianle Lun}, \bibinfo{person}{Yicheng Tao}, \bibinfo{person}{Junyou Su}, \bibinfo{person}{Zhaoya Gong}, {et~al\mbox{.}}} \bibinfo{year}{2024}\natexlab{}.
\newblock \showarticletitle{PlanGPT: Enhancing urban planning with tailored language model and efficient retrieval}.
\newblock \bibinfo{journal}{\emph{arXiv preprint arXiv:2402.19273}} (\bibinfo{year}{2024}).
\newblock


\bibitem[Zou et~al\mbox{.}(2025)]%
        {zou2025deep}
\bibfield{author}{\bibinfo{person}{Xingchen Zou}, \bibinfo{person}{Yibo Yan}, \bibinfo{person}{Xixuan Hao}, \bibinfo{person}{Yuehong Hu}, \bibinfo{person}{Haomin Wen}, \bibinfo{person}{Erdong Liu}, \bibinfo{person}{Junbo Zhang}, \bibinfo{person}{Yong Li}, \bibinfo{person}{Tianrui Li}, \bibinfo{person}{Yu Zheng}, {et~al\mbox{.}}} \bibinfo{year}{2025}\natexlab{}.
\newblock \showarticletitle{Deep learning for cross-domain data fusion in urban computing: Taxonomy, advances, and outlook}.
\newblock \bibinfo{journal}{\emph{Information Fusion}}  \bibinfo{volume}{113} (\bibinfo{year}{2025}), \bibinfo{pages}{102606}.
\newblock


\end{thebibliography}

% \appendix

\end{document}